\documentclass[longauth]{aa}

\usepackage{graphicx, xcolor}
\usepackage{subcaption}
\newcommand\newsubcap[1]{\phantomcaption%
       \caption*{\figurename~\thefigure\thesubfigure: #1}}
\definecolor{cerulean}{rgb}{0.0, 0.48, 0.65}
\usepackage[colorlinks, citecolor=cerulean]{hyperref}
\usepackage{pdflscape}
 \usepackage{lscape}
 \usepackage{tablefootnote}
\usepackage{graphicx}

\def\kms{km\,s$^{-1}$}

\def\Msun{$M_{\odot}$}

\def\halpha{H${\alpha}$}

\def\bloem{BLOeM}
\newcommand{\giraffe}{{\sc giraffe}}
\newcommand{\flames}{{\sc flames}}

\defcitealias{Shenar2024}{Paper~I}

\usepackage{txfonts}
\begin{document}

   \title{Binarity at Low Metallicity (BLOeM)
          \thanks{Based on observations collected at the European Southern Observatory under ESO program ID 112.25W2.}}
    \subtitle{Multiplicity properties of Oe and Be stars}

    \author{
   J.~Bodensteiner\inst{\ref{inst:antonpannekoek},\ref{inst:ESO}}
   \and T.~Shenar\inst{\ref{inst:TelAv}}
   \and H.~Sana\inst{\ref{inst:kul},\ref{lgi}}
   % from here on multiplicity team & other contributions in alphabetical order
   \and N.~Britavskiy \inst{\ref{inst:rob}}
   \and P.~A.~Crowther\inst{\ref{inst:sheffield}}
   \and N.~Langer\inst{\ref{inst:bonn}}
   \and D.~J.~Lennon\inst{\ref{inst:iac}, \ref{inst:ull}}
   \and L.~Mahy \inst{\ref{inst:rob}}
   \and L.~R.~Patrick\inst{\ref{inst:cab}} 
   \and J.~I.~Villase\~{n}or\inst{\ref{inst:mpia}}
   % from here on all alphabetical
   \and M.~Abdul-Masih\inst{\ref{inst:iac}, \ref{inst:ull}}
   \and D.~M.~Bowman\inst{\ref{inst:newcastle}, \ref{inst:kul}}
   \and A.~de~Koter \inst{\ref{inst:antonpannekoek}, \ref{inst:kul}} 
   \and S.~E.~de Mink \inst{\ref{inst:mpa}}
   \and K.~Deshmukh \inst{\ref{inst:kul}}
   \and M.~Fabry \inst{\ref{inst:kul}, \ref{inst:villa}}
   \and A.~Gilkis \inst{\ref{inst:cambridge}}
   \and Y.~G{\"o}tberg \inst{\ref{inst:ista}}
   \and G.~Holgado \inst{\ref{inst:iac}}
   \and R.~G.~Izzard\inst{\ref{inst:surr}}
   \and S.~Janssens\inst{\ref{inst:kul}}
   \and V.~M.~Kalari\inst{\ref{inst:gemini}}
   \and Z.~Keszthelyi\inst{\ref{inst:naoj}}
   \and J.~Kub\'at\inst{\ref{inst:ondrejov}}
   \and I.~Mandel\inst{\ref{inst:monash},\ref{inst:ozgrav}}
   \and G.~Maravelias\inst{\ref{inst:noa},\ref{inst:forth}}
   \and L.~M.~Oskinova\inst{\ref{inst:up}}
   \and D.~Pauli\inst{\ref{inst:up}}
   \and V.~Ramachandran\inst{\ref{inst:ari}}
   \and D.~F.~Rocha\inst{\ref{inst:ON_Br}}
   \and M.~Renzo\inst{\ref{inst:steward}}
   \and A.~A.~C.~Sander\inst{\ref{inst:ari}}
   \and F.~R.~N.~Schneider\inst{\ref{inst:hits},\ref{inst:ari}}
   \and A.~Schootemeijer\inst{\ref{inst:bonn}}
   \and K.~Sen\inst{\ref{inst:umk}}
   \and M.~Stoop\inst{\ref{inst:antonpannekoek}}
   \and S.~Toonen\inst{\ref{inst:antonpannekoek}}
   \and J.~Th.~van~Loon\inst{\ref{inst:keele}}
   \and R.~Valli\inst{\ref{inst:mpa}}
   \and A. Vigna-G\'omez\inst{\ref{inst:mpa}}
   \and J.~S.~Vink\inst{\ref{inst:armagh}} 
   \and C.~Wang\inst{\ref{inst:mpa}}
   \and X.-T.~Xu\inst{\ref{inst:bonn}}
    }
   \institute{
   {{Anton Pannekoek Institute for Astronomy, University of Amsterdam, Science Park 904, 1098 XH Amsterdam, the Netherlands} \label{inst:antonpannekoek}} \\ \email{j.bodensteiner@uva.nl}
   \and {ESO - European Southern Observatory, Karl-Schwarzschild-Strasse 2, 85748 Garching bei M\"unchen, Germany\label{inst:ESO}}  
    \and {The School of Physics and Astronomy, Tel Aviv University, Tel Aviv 6997801, Israel\label{inst:TelAv}}
    \and {Institute of Astronomy, KU Leuven, Celestijnenlaan 200D, 3001 Leuven, Belgium\label{inst:kul}}
    \and
    Leuven Gravity Institute, KU Leuven, Celestijnenlaan 200D box 2415, 3001 Leuven, Belgium\label{lgi}
    \and {Royal Observatory of Belgium, Avenue Circulaire/Ringlaan 3, B-1180 Brussels, Belgium} \label{inst:rob}
    \and {Department of Physics \& Astronomy, Hounsfield Road, University of Sheffield, Sheffield, S3 7RH, United Kingdom\label{inst:sheffield}} 
    \and {Argelander-Institut f\"{u}r Astronomie, Universit\"{a}t Bonn, Auf dem H\"{u}gel 71, 53121 Bonn, Germany\label{inst:bonn}}
    \and {Instituto de Astrof\'isica de Canarias, C. V\'ia L\'actea, s/n, 38205 La Laguna, Santa Cruz de Tenerife, Spain\label{inst:iac}}
    \and {Universidad de La Laguna, Dpto. Astrof\'isica, Av.\ Astrof\'sico Francisco S\'anchez, 38206 La Laguna, Santa Cruz de Tenerife, Spain\label{inst:ull}}
    \and {Centro de Astrobiolog\'ia (CSIC-INTA), Ctra.\ Torrej\'on a Ajalvir km 4, 28850 Torrej\'on de Ardoz, Spain\label{inst:cab}}
    \and {Max-Planck-Institut f\"{u}r Astronomie, K\"{o}nigstuhl 17, D-69117 Heidelberg, Germany\label{inst:mpia}}
    \and {School of Mathematics, Statistics and Physics, Newcastle University, Newcastle upon Tyne, NE1 7RU, UK\label{inst:newcastle}}
    \and {Max-Planck-Institute for Astrophysics, Karl-Schwarzschild-Strasse 1, 85748 Garching, Germany\label{inst:mpa}}
    \and {Department of Astrophysics and Planetary Science, Villanova University, 800 E Lancaster Ave., PA 19085, USA\label{inst:villa}}
    \and {Institute of Astronomy, University of Cambridge, Madingley Road, Cambridge CB3 0HA, United Kingdom} \label{inst:cambridge}
    \and {{Institute of Science and Technology Austria (ISTA), Am Campus 1, 3400 Klosterneuburg, Austria}\label{inst:ista}}
    \and {{Astrophysics Group, Department of Physics, University of Surrey, Guildford GU2 7XH, UK}\label{inst:surr}}
    \and {Gemini Observatory/NSF's NOIRLab, Casilla 603, La Serena, Chile}\label{inst:gemini}
    \and {Center for Computational Astrophysics, Division of Science, National Astronomical Observatory of Japan, 2-21-1, Osawa, Mitaka, Tokyo 181-8588, Japan\label{inst:naoj}}
    \and {Astronomical Institute, Academy of Sciences of the Czech Republic, Fri\v{c}ova 298, CZ-251 65 Ond\v{r}ejov, Czech Republic}\label{inst:ondrejov}
    \and {{School of Physics and Astronomy, Monash University, Clayton VIC 3800, Australia}\label{inst:monash}}
    \and {{ARC Centre of Excellence for Gravitational-wave Discovery (OzGrav), Melbourne, Australia}\label{inst:ozgrav}}
    \and {IAASARS, National Observatory of Athens, GR-15236, Penteli, Greece}\label{inst:noa}
    \and {Institute of Astrophysics, FORTH, GR-71110, Heraklion, Greece}\label{inst:forth} 
    \and {Institut f\"ur Physik und Astronomie, Universit\"at Potsdam, Karl-Liebknecht-Str. 24/25, 14476 Potsdam, Germany\label{inst:up}}
    \and {Observat\'orio Nacional, R. Gen. Jos\'e Cristino, 77 - Vasco da Gama, Rio de Janeiro - RJ, 20921-400, Brazil\label{inst:ON_Br}}
    \and {Department of Astronomy \& Steward Observatory, 933 N. Cherry Ave., Tucson, AZ 85721, USA\label{inst:steward}}
    \and {Zentrum f\"ur Astronomie der Universit\"at Heidelberg, Astronomisches Rechen-Institut, M\"onchhofstr. 12-14, 69120 Heidelberg, Germany\label{inst:ari}}
    \and Heidelberger Institut f{\"u}r Theoretische Studien, Schloss-Wolfsbrunnenweg 35, 69118 Heidelberg, Germany\label{inst:hits}
    \and {{Institute of Astronomy, Faculty of Physics, Astronomy and Informatics, Nicolaus Copernicus University, Grudziadzka 5	, 87-100 Torun, Poland}\label{inst:umk}}
    \and {Lennard-Jones Laboratories, Keele University, ST5 5BG, UK\label{inst:keele}}
    \and {Armagh Observatory, College Hill, Armagh, BT61 9DG, Northern Ireland, UK\label{inst:armagh}}
    }
   \date{Received xx Month Year; accepted xx Month Year}

  \abstract
  % context heading (optional)
   {Rapidly rotating classical OBe stars have been proposed as the products of binary interactions, and %observations of peculiar post-interaction binary systems have demonstrated
   the fraction of Be stars with compact companions implies that at least some are. However, to constrain the interaction physics spinning up the OBe stars,
   a large sample of homogeneously analysed OBe stars with well-determined binary characteristics and orbital parameters are required.}
  % aims heading (mandatory)
   {We investigate the multiplicity properties of a sample of 18 Oe, 62 Be, and two Of?p stars observed within the \bloem\ survey in the Small Magellanic Cloud. We analyse the first nine epochs of spectroscopic observations obtained over approximately three months in 2023.}
  % methods heading (mandatory)
   {Radial velocities (RVs) of all stars are measured using cross-correlation based on different sets of absorption and emission lines. Applying commonly-used binarity criteria we classify objects as binaries, binary candidates, and apparently single (RV stable) objects. We further inspect the spectra for double-lined spectroscopic binaries and cross-match with catalogues of X-ray sources and photometric binaries.}
  % results heading (mandatory)
   {We classify 14 OBe stars as binaries, and an additional 11 as binary candidates. The two Of?p stars are apparently single. We find two more objects that are most likely currently interacting binaries. Without those, the observed binary fraction for the remaining OBe sample of 78 stars is f$^\mathrm{OBe}_\mathrm{obs}=0.18\,\pm\,0.04$ (f$^\mathrm{OBe}_\mathrm{obs+cand}=0.32\pm$0.05 including candidates). This binary fraction is less than half of that measured for OB stars in \bloem. Combined with the lower fraction of SB2s, this suggests that OBe stars have indeed fundamentally different present-day binary properties than OB stars. We find no evidence for OBe binaries with massive compact companions, in contrast to expectations from binary population synthesis.} %We further do not detect any obvious OBe binaries with massive compact companions, in contrast to predictions by binary population synthesis.}
  % conclusions heading (optional)
   {Our results support the binary scenario as an important formation channel for OBe stars, as post-interaction binaries may have been disrupted or the stripped companions of OBe stars are harder to detect. Further observations are required to characterize the detected binaries, their orbital parameters, and the nature of their companions.}


    \titlerunning{BLOeM - Multiplicity of Oe and Be stars}
   \authorrunning{Bodensteiner et al.}

   \maketitle
%
%--------------------------------------------------------------------

%%%%%%%%%%%%%%%%%%%%%%%%%%%%%%%%%%%%%%%%%%%%%%%%%%%%%%%%%%%%%%%%%%%%%
%%%%%%%%%%%%%%%%%%%%%%%%%%%%%%%%%%%%%%%%%%%%%%%%%%%%%%%%%%%%%%%%%%%%%
%%%%%%%%%%%%%%%%%%%%%%%%%%%%%%%%%%%%%%%%%%%%%%%%%%%%%%%%%%%%%%%%%%%%%
\section{Introduction}\label{sec:intro}
Observations have undoubtedly demonstrated that most massive stars live their lives in binary or higher-order multiple systems \citep[e.g.,][]{Abt1984, Sana2014, Kobulnicky2014, Dunstall2015, Moe2017, Offner2023, Bordier2024}. A majority of those stars interact with their companion at some point of their evolution \citep{Sana2012}. This drastically changes the evolutionary path of both binary components, and implies that there should be a large number of post-interaction systems that, depending on the interaction pathway they undergo (stable or unstable mass transfer, or merger), have different orbital and physical characteristics and probe different interaction physics \citep[e.g.][]{Paczynski1967, Podsiadlowski1992, Wellstein2001, Langer2012, deMink2014, DeMarco2017, Eldridge2022}. Given the large uncertainties that binary evolution is still subject to, detecting and characterizing more post-interaction binaries yields crucial new constraints that will help improve our understanding \citep[e.g.][]{Marchant2024, ChenX2024}.

However, the characteristics of interaction products, and how to tell them apart from stars that truly evolve as isolated stars, remain unclear and strongly depend on the type of interaction that may have occurred. In addition to peculiar chemical surface abundances, magnetic fields, or apparent younger ages compared to a reference star or population \citep[e.g.,][]{Schneider2019, Irrgang2022, Sen2022, Shenar2023, Frost2024}, a star's rotational velocity has been proposed as important characteristic to distinguish single- and binary-interaction channels. While stars stripped of their hydrogen-rich envelopes are often predicted to be slow rotators \citep[e.g.,][]{Kippenhahn1967, Schurmann2022, Sen2023}, mass gainers are expected to rotate close to or at their critical spin \citep[e.g.][]{Packet1981, Blaauw1993, deMink2013, Ramirez-Agudelo2013, Ramirez-Agudelo2015, Mahy2020a, Renzo2021}. 

In this context, classical OBe stars are of interest. These are rapidly rotating, non-radially pulsating, non-supergiant OB-type stars that are surrounded by a circumstellar gaseous decretion disk, causing characteristic emission lines in their spectra \citep[e.g.,][]{Rivinius2013}. The origin of the rapid rotation of classical OBe stars is often linked to mass transfer in binary interactions. Theoretical works have shown that the mass gainers in binaries can become rapid rotators \citep[e.g.,][]{Pols1991} and population synthesis computations illustrate that the binary channel can produce a large number of post-interaction rapid rotators that observationally may look like OBe stars \citep{Shao2014, Shao2021, WangC2020}. However, \citet{Hastings2021} pointed out that the large OBe fraction in star clusters in the Large and Small Magellanic Cloud (LMC, SMC) can only be reproduced assuming that a large fraction of binaries undergo stable mass transfer without merging. Alternatively, the rapid rotation of classical OBe stars is explained by single-star evolution. They may be born as rapid rotators \citep[e.g.,][]{Bodenheimer1995}, or efficient angular-momentum transport from the stellar core to the envelope might bring them close to their critical velocity towards the end of the main-sequence (MS) evolution \citep[e.g.,][]{Ekstrom2012, Hastings2020}. In both single-star scenarios, the multiplicity properties of classical OBe stars are expected to be similar to their OB counterparts, that is many of them are expected in pre-interaction systems with MS companions \citep[e.g.,][]{Sana2011, Banyard2021}. Contrarily, the binary channel predicts a lack of OBe\,+\,MS binaries, and most OBe stars should have stripped or compact companions \citep[e.g.,][]{WangC2024}.

Observations in the Milky Way indicate a lack of massive OBe stars in close binary systems with MS companions \citep{Bodensteiner2020b}. The detection of post-interaction OBe binaries demonstrates that at least some are indeed interaction products. First, there are OBe binaries with hot-subdwarf O and B \citep[sdOB; e.g.,][]{Heber2009} companions. Those are mostly detected in the UV \citep[e.g.,][]{Gies1998, Koubsky2012, WangL2021, WangL2023}, or recently also using interferometry \citep[e.g.,][]{Klement2022, Klement2024}. Secondly, some Be binaries further in their evolution have compact companions. In those Be X-ray binaries (BeXRBs), transient X-ray emission can occur when the compact object (usually a neutron star; NS) periodically moves through the disk of the Be star during the course of its orbit \citep[e.g.,][]{Reig2011, Coe2015, Haberl2016}. They can provide important constraints on uncertain binary physics, such as the mass transfer stability and efficiency \citep[e.g.,][]{Vinciguerra2020, Rocha2024}. Some single OBe stars were also interpreted as products of binary interaction, in which the binary system was disrupted in the supernova explosion of the mass donor \citep[e.g.,][]{Boubert2018, Neuhauser2020, Renzo2021}. Recently, another group of post-interaction Be binaries in a phase relatively soon after the mass transfer phase was proposed \citep[with LB-1 and HR\,6819 as proto-types,][]{Shenar2020, Bodensteiner2020c, ElBadry2020a, Frost2022}. These systems are detected as double-lined spectroscopic binaries (SB2s). The (partially or fully) stripped donor is thermally relaxing, thus still large and over-luminous for its mass. It is typically cooler than an sdOB star and due,  to its slow rotation, can be detected based on narrow absorption lines in the spectrum.
The rapidly-rotating mass gainer is often initially only detected due to OBe-typical emission lines arising in the circumstellar disk. Recently, three such systems were reported in the Small Magellanic Cloud \citep[SMC; ][]{Ramachandran2023, Ramachandran2024}. 

Overall, only few post-interaction Be-binary systems have been detected, and even less have a fully characterized set of orbital and stellar parameters \citep[see][for an overview of Be+sdOB binaries]{WangL2023}. Reasons for this are on the one hand, it is difficult to identify such systems observationally, in particular the Be\,+\,sdOBs, in which the sdOB components are optically faint and contribute little to the overall flux \citep[e.g.,][]{Gotberg2018}. Given their masses below 1\Msun, they only induce small radial-velocity (RV) variations on the more massive Be companion. On the other hand, the Be phenomenon is known to be transient, with emission lines appearing and disappearing on timescales of months, years, and decades \citep[e.g.,][]{Townsend2004}, so not all of these objects might show the characteristic OBe emission. Indeed, several binary systems were reported to contain a recently stripped star, in which the mass gainer shows no emission lines typical for OBe stars, for example NGC\,1850\,BH1 \citep[e.g.,][]{Saracino2022, ElBadry2022, Saracino2023} and VFTS\,291 \citep{Villasenor2023} in the LMC, and AzV\,476 \citep{Pauli2022} in the SMC.

Observationally, classical OBe stars can be confused with other objects that have a similar signature but a different physical nature \citep[e.g.,][]{Rivinius2013}. More generally, OBe stars are defined as OB stars that show (or have shown) emission lines in their spectra. These include classical OBe stars, but also other objects like magnetic stars, interacting binaries with an accretion disk causing the emission \citep[e.g.,][]{Kriz1975}, supergiants that show emission lines due to their stellar winds, or young stellar objects. %Especially in the case of Oe stars, it remains uncertain up to which mass a disk can be sustained \citep[e.g.,][]{Vink2009}. %So far, few large-scale spectroscopic surveys of classical OBe stars with homogeneous data sets exist.

With a metallicity (Z) of 0.2 solar \citep[e.g.][]{Russell1990}, the SMC is a prime target for investigating OBe stars, as stellar winds are expected to be weaker and stars spin-down less easily \citep[e.g.,][]{Vink2001, Langer2012, Smith2014}. \citet{Schootemeijer2022} reported the SMC to host a higher fraction of OBe stars (f$_\mathrm{OBe}\sim31\%$) than what is observed in our Galaxy and the LMC \citep[see also][]{Viera2021}. The OBe fractions are especially higher in clusters, reaching $\sim40\%$ \citep[e.g.,][]{Grebel1992b, Iqbal2013}. There are also many BeXRBs in the SMC \citep[][]{Coe2015, Haberl2016, McBride2017}, outnumbering those reported in the LMC, despite the LMC having a larger population of (massive) stars \citep{Antoniou2016}. %This could also be due to the recent SMC star formation history rather than its metallicity \citep{Vinciguerra2020}.

This paper is part of a series investigating the multiplicity properties of massive stars across the SMC observed in the Binarity at LOw Metallicity (\bloem) survey. \bloem\ is a multi-epoch spectroscopic survey of almost 1000 massive stars, described in detail in \citet[][hereafter \citetalias{Shenar2024}]{Shenar2024}. %Analysing the first 9 of 25 epochs, the sample is divided in O-type stars \citep{Sana2024}, B-type dwarfs and giants \citep{Villasenor2024}, early-B supergiants \citep{Britavskiy2024} as well as late-B, A- and F-type supergiants \citep{Patrick2024}.
Here, we focus on the 82 OBe stars observed in \bloem. This is not only the largest homogeneous spectroscopic dataset of OBe stars, especially at low Z, but also allows for a direct comparison to the \bloem\ OB stars. This manuscript is organized as follows: we briefly summarize the sample selection and observations in Sect.\,\ref{sec:sample}, and describe the RV measurements and binary classification criteria in Sect.\,\ref{sec:bin}. The observed binary fractions of OBe stars in \bloem\ are presented in Sect.\,\ref{sec:results}, and discussed and compared to other BLOeM subsamples and previous works in Sect.\,\ref{sec:discussion}. We summarize the results and conclude with Sect.\,\ref{sec:conclusion}.

%%%%%%%%%%%%%%%%%%%%%%%%%%%%%%%%%%%%%%%%%%%%%%%%%%%%%%%%%%%%%%%%%%%%%
%%%%%%%%%%%%%%%%%%%%%%%%%%%%%%%%%%%%%%%%%%%%%%%%%%%%%%%%%%%%%%%%%%%%%
%%%%%%%%%%%%%%%%%%%%%%%%%%%%%%%%%%%%%%%%%%%%%%%%%%%%%%%%%%%%%%%%%%%%%

\section{BLOeM sample selection and observations}\label{sec:sample}
The \bloem\, survey uses the \flames\, multi-object instrument at the VLT in Paranal, Chile, with the \giraffe\, {\'e}chelle spectrograph \citep{Pasquini2002}, allowing for a simultaneous observation of almost 130 stars per field. Covering eight fields over the SMC, 929 stars were observed. After ESO Period P112, nine epochs are available that were taken over the course of approximately four months (September to December 2023). Those were spread unevenly, with a minimum time of one day between the observations, resulting in the eight fields having different time samplings. Due to technical issues with fibres, a handful of stars could only be observed four times. 

The \bloem\ targets are, by design, selected without any prior knowledge of their types. To ensure the observation of massive stars, targets are chosen based on their \textit{Gaia} DR3 \citep{Gaia2023} magnitudes and colors in comparison to evolutionary tracks from the extended grid of \citet{Schootemeijer2019} with mass-dependent overshooting and SMC metallicity \citep[described in App. B of][]{Hastings2021}. In particular, objects were selected that should have initial masses above 8\,\Msun\ (Fig.\,\ref{fig:bin_cmd}). Foreground objects were removed based on their \textit{Gaia} distances and proper motions. Stars in clusters and other crowded regions that \flames\ cannot resolve were avoided. Apart from the technical limitations of the spectrograph and the fibre allocation (i.e., the minimum distance between two objects to be able to allocate fibres), only the star's magnitude and no additional information went into our target selection.

All observations are reduced, sky corrected and normalized following the same automated procedure. The final products after data reduction are 1D normalized, sky-subtracted spectra covering the wavelength range 3960 to 4565\,\r{A} for each epoch. The resolving power of the given \flames\ setup is approximately 6000, and the individual spectra have typical signal-to-noise ratios (S/Ns) of 50 per pixel, with a few spectra with S/Ns of only around 25, and others above 100. We refer to \citetalias{Shenar2024} for a more detailed description of the target selection, the observations and data reduction, as well as an observing log for each of the fields.

Here, we focus on the 82 OBe stars in the \bloem\ sample of 929 stars in total. The total fraction of OBe stars in \bloem\ is 11\% (including the extra eight stars with large H$\alpha$ EWs in \textit{Gaia}). Dividing by spectral types, the Oe star fraction is 13\% while the Be fraction is 10\%. Their spectral identification is described in detail in \citetalias{Shenar2024}. In short: they were identified in the \bloem\ sample based on emission lines in their spectra, mainly in the Balmer lines H$\gamma$ and H$\delta$ (H$\epsilon$ is at the edge of the wavelength coverage and often not usable). To cross-check, in \citetalias{Shenar2024}, we also investigated the \textit{Gaia} low-resolution spectra for all stars (not only the ones classified as OBe), which cover the H$\alpha$ line that is commonly the strongest emission line in OBe spectra. Eight non-supergiant stars have significantly large H$\alpha$ equivalent widths (EWs) (indicative of emission, see figure 9 in \citetalias{Shenar2024}) but do not show emission in the \bloem\ spectra. Given the low number, and to ensure the usage of a consistent dataset, we here only investigate the stars identified as OBe stars in \bloem. 

To minimize contamination of the OBe sample, we visually investigated Digitized Sky Survey (DSS)-red images for the presence of large-scale nebulosities. The intrinsically narrow but instrumentally broadened nebular emission lines can mimic the spectrum of OBe stars in medium-resolution spectroscopy. Objects in regions with dense nebulosities, or objects that only show weak, narrow emission lines, were thus classified as contaminated by nebulae (`neb'). While some objects might still be misclassified as OBe stars, others might be missed due to the transient nature of the Be phenomenon. Our sample of OBe stars is thus most likely not complete and more OBe stars might be detected or removed from the sample with more observations, in particular covering \halpha. Because of their rapid rotation and an infrared excess caused by the cooler circumstellar disk, OBe stars are known to have redder colors than their OB counterparts. Given the target selection based on evolutionary tracks, especially at the faint end of the sample, redder OBe stars were not observed due to the color cut (see also Appendix\,\ref{app:xrbs}).

\begin{figure}
    \centering 
    \includegraphics[width=0.95\linewidth]{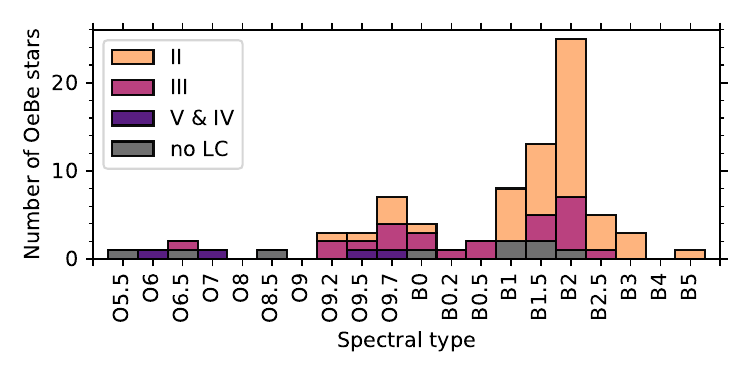}
    \caption{Distribution of spectral types in the OBe sample. The color indicates the assigned luminosity class of the stars (`V' and `IV' in purple, `III' in violet, `II' in ocher, and stars without classification in gray).% Note the uneven spacing between spectral sub-types on the x-axis.
    }
    \label{fig:spt_histo}
\end{figure}

Figure\,\ref{fig:spt_histo} provides an overview of the spectral types and luminosity classes of OBe stars, as reported in \citetalias{Shenar2024}. Most targets are formally classified as luminosity class II, however, we note that the luminosity-class definition for OBe stars is generally difficult as important diagnostic lines are often contaminated, or even fully dominated by emission infilling. While a majority of the OBe sample are B-type stars (62 objects), there are also 20 O-type stars in the sample. This number is large compared to the few Oe stars known in the Galaxy \citep[e.g., in the BeSS database;][]{Neiner2011}. There seem to be two peaks in the spectral type distribution: one around O9.7, and another one at B2. While the first peak resembles the one reported by \citet{GoldenMarx2016} based on the RIOTS survey \citep{Lamb2016} in the SMC, the peak in spectral types around B2 agrees with previous findings of OBe stars \citep[e.g.,][]{Neiner2011}, and BeXRBs in the Galaxy, the SMC and the LMC \citep[e.g.,][]{Maravelias2014}.

\begin{figure*}
    \centering
    \includegraphics[width=\linewidth]{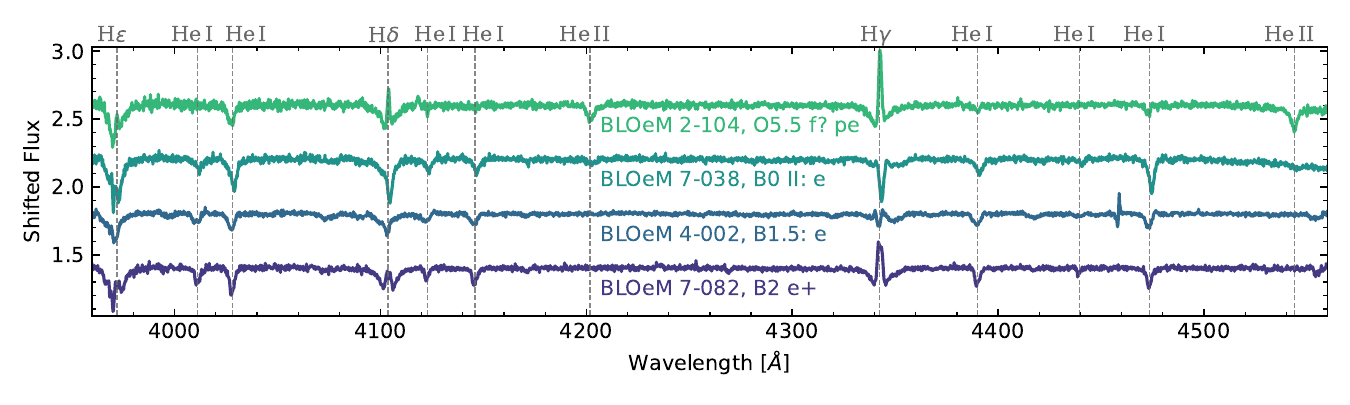} 
    \caption{Spectra of four BLOeM emission-line stars, from top to bottom BLOem 2-104 (an Of?p stars with spectral type O5.5\,f?\,pe), BLOeM 7-038 (with only weak emission infilling in the wings of the Balmer lines), BLOeM 4-002 (a B1.5:\,e star with broad absorption lines), and BLOeM\,7-082 (a B2\,e+ star with narrow absorption lines and strong double-peaked Balmer emission.}
    \label{fig:example_specs}
\end{figure*}

Two of the targets with the earliest spectral type, BLOeM\,2-104 and BLOeM\,4-039, were classified as `Of?p' stars in \citetalias{Shenar2024} (they appear without luminosity class in Fig.\,\ref{fig:spt_histo}). As all Galactic Of?p stars are strongly magnetic \citep[e.g.][]{Walborn1972, Naze2010, Grunhut2017, Keszthelyi2023}, they are most likely not classical OBe stars. They are included in this sample as they show emission in their spectra, however, they are not included in the OBe binary statistics (Sect.\,\ref{sec:bin_stat}) and their binary status is discussed separately. An example single-epoch spectrum of three OBe stars and one Of?p star is shown in Fig.\,\ref{fig:example_specs}.
%to illustrate the diversity in the width of the absorption lines as well as in the shape and strength of the emission lines.

\section{Binary classification}\label{sec:bin}
\subsection{Radial velocity measurement}\label{sec:RV_meas}
To assess the radial velocities (RVs) of our OBe sample, we use the cross-correlation method \citep[CCF;][]{Zucker2003}, which has previously been applied to OBe stars successfully \citep[e.g.,][Janssens et al. in prep.]{Bodensteiner2020c, Janssens2023}. We follow a two-step process: first, the highest S/N spectrum is used as the CCF template to measure RVs of each epoch. Each spectrum is then shifted to the same reference frame, and a new co-added template spectrum is produced. In a second step, this higher-S/N template is used as new template for the CCF. The measured RVs are thus not absolute RVs, but relative to the template spectrum, which differs from object to object. \citet{Patrick2024} 
investigate the overall stability of the wavelength calibration in the \bloem\ dataset using the late-type supergiants, which provide a large number of spectral lines for accurate RV determination. They find it to be stable within a few tens of \kms, which is lower than typical errors measured for the OBe stars.

One assumption of the CCF is that the spectra show no intrinsic variability apart from a shift in wavelength. While the photospheric absorption lines might be impacted by intrinsic stellar variability such as pulsations \citep[see review by][]{Bowman2020b}, the emission lines in OBe stars trace disk variability. In case of variability of the star or the disk, the formal errors on the RVs measured may be underestimated (Janssens et al. in prep.). One object, namely BLOeM\,6-034, shows clear signatures of being an SB2. Because of the large changes in the line shape, the CCF is not reliable and we measure the RVs by simultaneously fitting two Gaussian profiles to the spectra \citep[see e.g.,][]{Sana2013}. More information about this target is given in \href{10.5281/zenodo.14679548}{Appendix E}.

Generally, we employ two sets of spectral lines: first, we use all available absorption lines that are not impacted by emission. As the affected lines and the degree of infilling varies from star to star, we visually inspect the individual spectra and select emission-free lines. The main lines used here are \ion{He}{i} lines at $\lambda$$\lambda$4009, 4026, 4121, 4144, 4388, and 4471\,\r{A}. For the O- and B0 stars, we also use the \ion{He}{ii} lines at $\lambda$$\lambda$4200 and 4542\,$\r{A}$ (which are generally the least affected by the disk). In most stars, metal lines are too weak to provide reliable constraints. One exception is BLOeM\,5-071, for which we measure RVs from the strong, narrow \ion{Si}{IV} absorption line at $\lambda$ 4088\,\r{A}. %Depending on the broadening of the spectral lines, we adjusted the considered wavelength window for each line star by star. 

Secondly, we use Balmer emission lines (mainly H$\gamma$, but sometimes also H$\delta$) to independently constrain the RVs of all objects with significant emission. Those measurements likely suffer from larger uncertainties as emission lines are intrinsically variable, which is not taken into account by the CCF. Still the emission lines have important advantages. The profiles are sharper, and generally have a higher S/N compared to the broad, shallow absorption lines. They are also less susceptible to stellar pulsations.
In classical OBe stars, emission lines are thought to arise in the circumstellar disk. They thus trace the motion of the star generating the disk and help determine whether the absorption lines in the spectra are formed in the same star. In particular in Be binaries with recently stripped companions (see Sect.\,\ref{sec:intro}), the emission lines commonly do not follow the narrow absorption lines, indicating that there are two luminous stars present.
%Using the methodology described above we measured RVs for all the 82 objects in our sample. For a handful of objects, the emission lines were too weak to separate them from the absorption lines, in which case only absorption-line RVs were measured. 

\subsection{Binary criteria}\label{sec:bin_class}
To distinguish binaries from RV-stable objects in a homogeneous way, we use the same binary criteria recently applied to massive stars in the literature \citep[e.g.,][]{Sana2013, Dunstall2015, Bodensteiner2020a, Banyard2021, Mahy2022}. To be classified as binary, at least two individual epochs $i$ and $j$ have to simultaneously satisfy the following two criteria: 1) the difference in RVs measured in the two epochs ($\Delta \mathrm{RV}_{ij}$) must exceed 4$\sigma_{\mathrm{RV}ij}$, and 2) the RV variation $\Delta \mathrm{RV}_{ij}$ must exceed a given threshold $\mathrm{RV}_\mathrm{crit}$. Following previous works, and to be consistent with the analysis of O- and B-type stars in \bloem\, \citep[see][]{Sana2024, Villasenor2024}, we choose a conservative threshold value of 20\,\kms. This criterion is designed to minimize contamination by objects that appear RV variable due to other causes like stellar winds or pulsations. Stars that do not fulfill both criteria are classified as RV-stable (referred apparently single below). Those could either be truly single, or they are undetected binaries with either longer periods than the observing campaign, high eccentricity, or RV variations below the threshold of 20\,\kms\ including RV variables seen near pole-on. Indeed, several Galactic Be binaries with stripped companions have RV amplitudes below this conservative threshold \citep[see e.g.,][for a recent compilation]{WangL2023}. %, while objects that fulfill one of the criteria .

Adopting different values $\mathrm{RV}_\mathrm{crit}$, we find that the observed binary fraction does not strongly depend on the chosen $\mathrm{RV}_\mathrm{crit}$=20\,\kms\ (App.\,\ref{app:bin_crit}).
OBe stars might be more likely to exceed this simple threshold as they to pulsate \citep[e.g.,][]{Rivinius2013, Labadie2022}. This can cause line profile variations and mimic RV shifts of typically up to 20\,\kms. We thus carefully inspect all measured RVs visually and search for indications of line-profile variability typical of pulsations \citep[where, usually, the core of the line is variable, but the wings are stable, ][]{Aerts2009} and flag uncertain cases as such. 
%The further characterisation of these objects will benefit from additional epochs that will be taken in the framework of the \bloem\ survey. 

\subsection{Double-lined spectroscopic binaries}
In addition to the aforementioned binary criteria, we visually inspect the spectra and measured RVs for SB2 signatures. SB2s are usually defined as stars where two stellar components are visible in the spectra that move in anti-phase with each other. BLOeM\,6-034 (also fulfilling both RV criteria) is the only object that clearly shows this. 

Six additional objects show a similar signature. Their absorption lines move in antiphase with the emission lines while no second stellar component can be discerned. As the emission lines in anti-phase indirectly indicate the presence of a secondary, we soften the standard SB2 definition and classify those as such. Two of them (BLOeM\,3-031 and BLOeM\,5-071) were already detected as SB2 in \citetalias{Shenar2024}. Five of the six targets also fulfill both binary criteria, while BLOeM\,2-111 is only classified as binary because of the SB2 classification (the measured $\Delta$RV$_\mathrm{max}=7\pm3$\,\kms\ is below the threshold).

\subsection{Additional data}
Binaries can be classified by means other than spectroscopy. \citetalias{Shenar2024} provides a cross-match of the \bloem\ dataset with the OGLE catalogue of photometrically variable stars for the SMC \citep{Pawlak2016}, indicating eclipsing binaries (EB) and ellipsoidal variables, as well as with common X-ray catalogues. Here, we discuss the OBe stars among those.

Three stars (BLOeM\,3-018, BLOeM\,3-031, BLOeM\,6-034) are detected as eclipsing binaries in \citet{Pawlak2016}. According to the OGLE catalogue, BLOeM\,3-018 (\object{OGLE-SMC-ECL-6595}) shows ellipsoidal variability with a period of 1.44\,d. We classify it as RV-stable in \bloem\ (neither of the two criteria are fulfilled), which could be the case if the two sets of spectral lines never deblend. We thus consider it as apparently single in the following. BLOeM\,3-031 (\object{OGLE-SMC-ECL-1232}) and BLOeM\,6-034 (\object{OGLE-SMC-ECL-5838}) are classified as ``detached or semi-detached'' systems with periods of 4.37\,d and 6.43\,d, respectively. Using the OGLE orbital parameters, and estimating the RV semi-amplitudes from the \bloem\ data, we find that the RVs of BLOeM\,3-031 match the OGLE orbit. For BLOeM\,6-034, we find a decent match assuming half the OGLE period (see Appendix\,\ref{sec:ogle}). The short periods as well as the light curves illustrate that both systems are most likely semi-detached, interacting binaries and not classical OBe stars. They will thus be removed from the OBe binary fraction.

In \citetalias{Shenar2024}, we also cross-matched the \bloem\ targets with several catalogues of known X-Ray sources. Four OBe stars (BLOeM\,2-055, BLOeM\,2-082, BLOeM\,4-026, and BLOeM\,4-113) spatially coincide with high-mass X-Ray binaries (HMXBs), making them potenital BeXRBs. Only BLOeM\,2-082 fulfills both binary criteria and is classified as binary. BLOeM\,4-113 is classified as candidate binary as the measured RVs do not exceed the 20\,\kms\ threshold. BLOeM\,2-055 and BLOeM\,4-026 show no statistically significant RV variations, and are classified as RV-stable. For now, we stick with the \bloem\ binarity classification of these objects, and only count BLOeM\,2-082 as binary in the binary fraction (see below).

\section{Results}\label{sec:results}
In the previous section, we investigated the binary properties of all 82 OBe stars based on their RV variations, including additional data when available. We discuss the individual (candidate) binaries and give their RV curves and spectra at RV extremes in \href{10.5281/zenodo.14679548}{Appendices E and F}. In the upcoming section, we provide the binary statistics of the 78 stars still counting as classical OBe stars, setting aside the two Of?p stars (both with stable RVs) as well as the two stars classified as semi-detached systems based on OGLE photometry and RV behaviour.

\begin{figure}
    \centering
    \includegraphics[width=\linewidth]{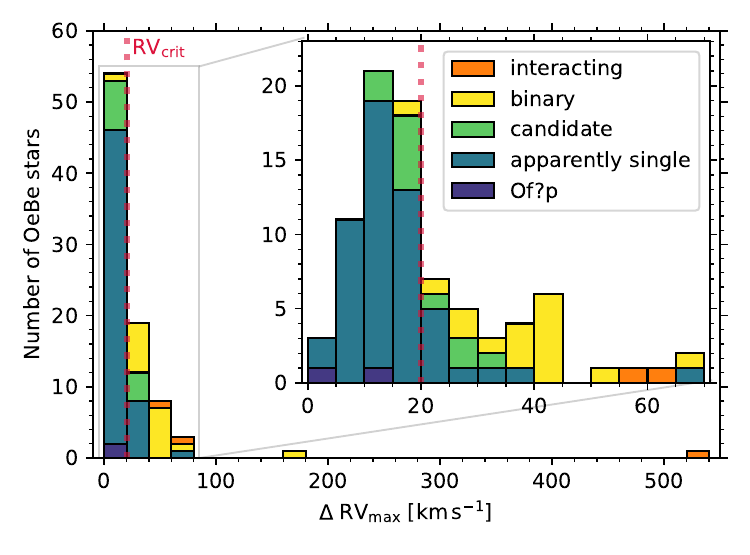}
    \caption{Maximum RV amplitude for all sample stars, with a zoom-in around lower RVs. Apparently single stars are shown in blue, binary candidates in green and binaries in yellow. Also indicated are the two Of?p stars (dark blue) and the two interacting binaries (orange). Both components of the SB2 BLOeM\,6-034 are included. Our RV threshold RV$_\mathrm{crit} = 20$\,\kms\ is given by red dotted lines.}
    \label{fig:bin_rvs}
\end{figure}

\subsection{OBe binary fraction}\label{sec:bin_stat}
Based on the introduced binary classification, we detect 14 binaries (13 from the RV criteria, and one SB2) among the 78 remaining OBe stars. We further find 11 candidate binaries that fulfill one of the two criteria, and passed our visual inspection as potential binaries. The remaining 53 stars are found to be RV stable and will be referred to as apparently single. The observed binary fraction of the OBe sample is $\mathrm{f}^\mathrm{OBe}_\mathrm{obs}=0.18\,\pm\,0.04$, with the error given from binomial statistics. %This provides a lower limit on the intrinsic binary fraction of OBe stars, as discussed in the subsequent section.
%The error quoted here is from binomial statistics.

Among the 14 systems classified as binaries, we find different behaviours of absorption and emission lines. In two objects, (BLOeM\,1-040 and BLOeM\,2-109), the emission and absorption lines follow a similar RV trend and no other component is visible (making them SB1s). In three objects, the emission lines remain stationary (BLOeM\,5-069, BLOeM\,5-071, BLOeM\,7-013), while in four objects they appear to move in anti-phase to the absorption lines (BLOeM\,1-113, BLOeM\,2-021, BLOeM\,2-111, BLOeM\,7-082). Four objects show no clear trend (BLOeM\,2-066, BLOeM\,2-082, BLOeM\,6-001, BLOeM\,7-045).
In one object (BLOeM\,7-114), the Balmer emission was too weak to measure RVs from the emission lines and only the absorption-line measurement is available. We discuss potential natures of these objects in Sect.\,\ref{sec:bin_nat}. 

Eleven additional objects are classified as candidate binaries. Either their maximum RV amplitudes are just below the 20\,\kms\ threshold but showed a continuous trend over time (as expected in a binary system, e.g., BLOeM\,5-035), or a pair of RVs satisfies both aforementioned criteria but visual inspection raised doubts that the object is indeed a binary system (e.g., BLOeM\,8-059). Additionally, in some cases, the binary status is only indicated by the RVs measured from the emission lines, while the absorption lines show no significant signal (e.g., BLOeM\,4-113). The extra observations that will be secured with \bloem\ will help classify these uncertain cases. Assuming all these candidates are indeed binaries increases the observed binary fraction to $\mathrm{f}^\mathrm{OBe}_\mathrm{obs+cand}=0.32\,\pm\,0.05$. 

The measured maximum RV amplitudes of all objects are shown in Fig.\,\ref{fig:bin_rvs}, also indicating the binary classification. Here, we also include the two Of?p stars, and the two objects classified as interacting binaries (for the SB2, we provide RVs for the primary and secondary), one of which stands out immediately due to the very high maximum RV amplitude of over 500\,\kms\ (BLOeM\,3-031). Similarly, with an RV amplitude of roughly 175\,\kms\, BLOeM\,5-071 stands out, which is another potential SB2 system. The other stars classified as binaries show RV amplitudes $\Delta\,\mathrm{RV}_\mathrm{max}$ between 20\,\kms\, (by definition) and 65\,\kms. The binary candidates have maximum RV amplitudes between 10 and 35\,\kms, with the low-amplitudes ones being classified as candidates due to a binary-like RV curve. Apart from one clear outlier (with $\Delta\,\mathrm{RV}_\mathrm{max}>65$\,\kms), the maximum RV amplitudes of the stars classified as ``RV constant'' peak around 10\,\kms\ with a few objects having RV amplitudes up to 40\,\kms. These have noisy spectra and typically broad absorption lines, and thus the measured RVs have large error bars of the order of the RV differences. They do not fulfill the significance criterion and show no trend that might indicate binarity. Some objects show RV variations below 5\,\kms, indicating that they are stable in terms of RVs but also show no other strong variability caused, for example, by pulsations. 

\begin{figure}
    \centering
    \includegraphics[width=\linewidth]{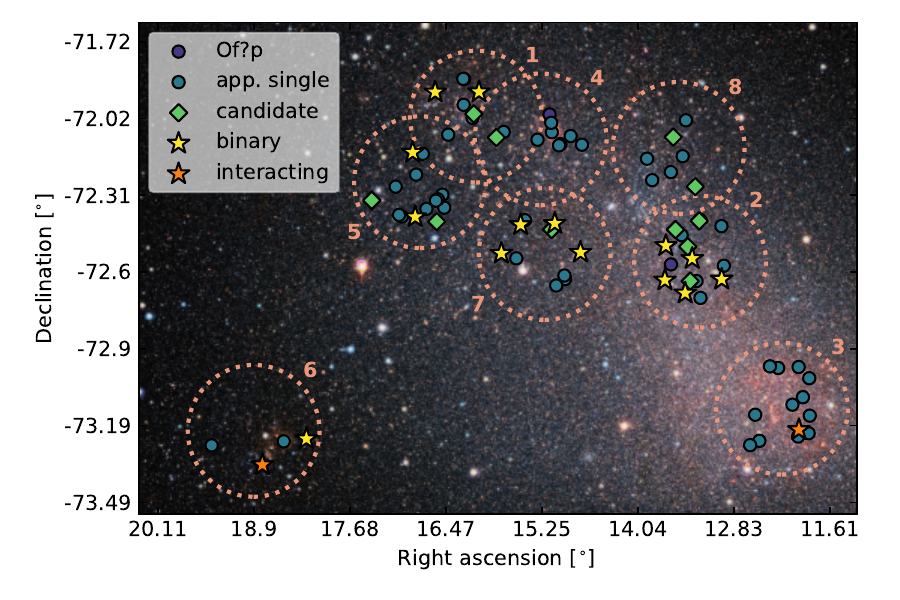}
    \caption{Location of detected binaries (yellow stars), binary candidates (green diamonds) and apparently single stars (blue circles) across the eight fields (orange dotted circles). Also indicated are the two Of?p stars (dark blue circles) and the two interacting binaries (orange stars). The background is a false-color (Y-J-K$_\mathrm{S}$) image from VISTA \citep[ESO/VISTA VMC,][]{Cioni2011}.}
    \label{fig:spatial}
\end{figure}

Figure\,\ref{fig:spatial} shows the distribution of binaries and binary candidates over the different fields in the SMC. Overall, field 6 has a surprisingly low number of OBe stars, one of which is the interacting binary BLOeM\,6-034. While most binaries are detected in fields 2 and 7, the only binary detected in field 3 is the second interacting binary. Given the overall low number of objects per field, however, the error bars on the respective binary fractions are large and no clear spatial trend across the SMC can be seen. We will re-investigate potential spatial differences, which might also be rooted in a difference in stellar masses and ages across the fields, in future work.

\begin{figure}
    \centering
    \includegraphics[width=0.95\linewidth]{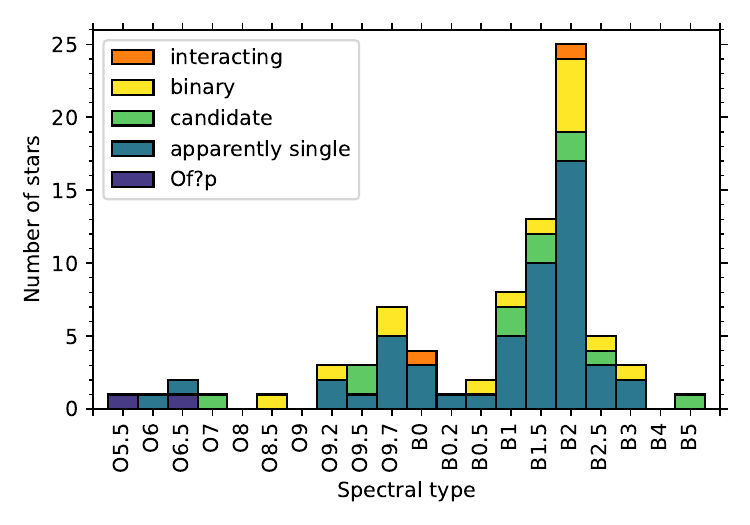}
    \caption{Binary statistics as a function of spectral type (see also Fig.\,\ref{fig:spt_histo}). The color coding is the same as in Fig.\,\ref{fig:spatial}.}
    \label{fig:bin_histo}
\end{figure}

Figure\,\ref{fig:bin_histo} shows the distribution of spectral types of the sample, now classified as apparently single, candidates binaries and binaries, with the two interacting binaries and Of?p stars also indicated. It illustrates that binaries are detected among both Oe and Be stars, with observed binary fractions of $\mathrm{f}^\mathrm{Oe}_\mathrm{obs}=0.22\,\pm\,0.10$ and $\mathrm{f}^\mathrm{Be}_\mathrm{obs}=0.17\,\pm\,0.05$, respectively. Those agree within the errors. As the Oe sample is relatively small (only 18 objects) the binomial uncertainty is fairly large. Notably, there are no binaries detected among the earliest spectral types (two of which are the Of?p stars). As the star with the latest spectral type, BLOeM\,2-097 (B5\,II:e), is classified as candidate binary, the spectral classification might be affected.

\begin{figure}
    \centering
    \includegraphics[width=\linewidth]{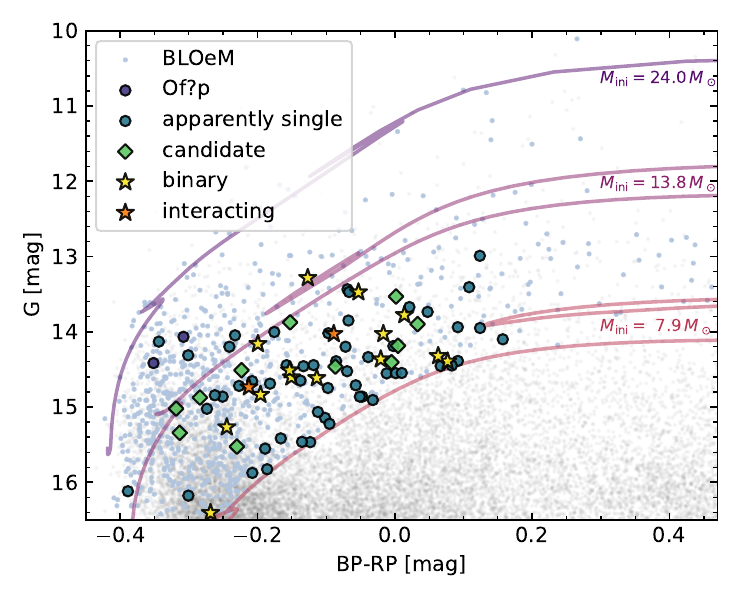}
    \caption{Binary statistics in the \textit{Gaia} CMD (same color coding as Fig.\,\ref{fig:spatial}). Over-plotted is the \bloem\,sample (blue dots, \citetalias{Shenar2024}), all \textit{Gaia} sources from which the sample was selected (grey dots), and evolutionary tracks from the extension of the grid of  \citet{Schootemeijer2019} with initial masses as indicated.}
    \label{fig:bin_cmd}
\end{figure}

Figure\,\ref{fig:bin_cmd} shows the location of the OBe sample and the stars classified as binary, candidate binary, and RV constant, in the color-magnitude diagram (CMD) constructed from \textit{Gaia} BP-RP colors and G-band magnitudes \citep{Gaia2023}. Overlaid are single-star evolutionary tracks from \citet{Schootemeijer2019} and adapted by \citet{Hastings2021}, adjusted for the SMC distance \citep[62\,kpc;][]{Graczyk2020} and average extinction \citep[assuming a reddening of $E_{\rm BP - RP} = 0.14\,$mag and an extinction of $A_G = 0.28\,$mag,][]{Schootemeijer2022}. In general, OBe stars mostly avoid the region around the zero-age-main-sequence (ZAMS) and are located towards or beyond the terminal-age main-sequence compared to the evolutionary tracks. Compared to the overall \bloem\ sample, they are redder than their non-emission counterparts, which could be caused by a combination of the rapid rotation and a contribution of their cooler disks. Fig.\,\ref{fig:bin_cmd} further shows that detected binaries and candidates are generally spread across the entire region populated by OBe stars in terms of color and brightness. However, there is a notable lack of detected binaries among the most massive stars in the sample, corresponding to the earliest spectral types. 

\subsection{Potential nature of detected binaries}\label{sec:bin_nat}
As mentioned in Sect.\,\ref{sec:bin_stat}, part of our detected OBe binaries have an SB1 signature, in which absorption and emission lines follow the same RV trend (BLOeM\,1-040 and BLOeM\,2-109). 
In other systems, the absorption lines move in anti-phase with the emission lines (four objects), or the emission lines appear stationary over moving absorption lines (three objects). We assign these three possibilities as Class 1, 2 and 3, respectively, and discuss them here in more detail.

{\bf Class 1 -- SB1 OBe binaries:}
In the systems classified as SB1 (BLOeM\,1-040 and BLOeM\,2-109), the emission traces the absorption lines and no additional component is visible. We postulate that the double-peaked emission arises in a circumstellar disk around the same object that produces the absorption lines, the OBe star. The companion could be a low-mass MS star, or an optically-faint object, such as a stripped star, a NS, or a black hole (BH). While unlikely, is can currently also not be excluded that the companion is a similarly bright, rapid rotator with broad lines difficult to discern in the complicated spectra. Both systems show RV variability with timescales of the order of the observing period, that is around two months. To explain the relatively large RV amplitudes with such an orbital period, in particular in the case of BLOeM\,2-109, a fairly massive object must be present. %Additional observations are required to investigate the nature of the companion. 
% >> veeery rough first estimate for BLOeM 2-109 gives f(m) = 0.36 Msun, assuming M1 of 7 Msun (B2 star) gives a minimum mass of the unseen object of ~3.5 Msun !

{\bf Class 2 -- SB2 OBe + absorption-line companions:}
In four objects, namely BLOeM\,1-113, BLOeM\,2-021, BLOeM\,2-111, and BLOeM\,7-082, we find that the absorption and emission lines move in anti-phase with each other. Such a signature can have two potential causes. The emission lines could be stationary, and the moving absorption lines can induce an apparent motion of the emission lines \citep[][]{AbdulMasih2020, ElBadry2020a}. A stationary emission line could for example be caused by a nebulosity or a circumbinary disk \citep[which has so far not been detected in massive binary systems, and is most likely hard to sustain in massive binaries,][]{Izzard2023}. Furthermore, the large RV amplitudes measured in those systems make this an unlikely interpretation. The same signature can also arise when two stars, causing the absorption and emission lines, are truly moving in anti-phase. These again can be subdivided into two separate cases: those containing ``true'' classical OBe stars, and those containing other emission-line objects. The latter could for example be interacting binary systems such as the two interacting binaries detected here in BLOeM, where the emission arises in a disk around the accreting star. Such objects can be misclassified as OBe stars, and are usually short-period systems that show a photometric signature of binarity as well \citep[][]{Waelkens1991}, similar to BLOeM\,3-031 and BLOeM\,6-034 . On the other hand, systems with anti-phase RVs could contain classical OBe stars with a luminous companion. The systems could either be pre-interaction binaries with OBe stars and MS companions, which so far have not been detected. In this case the OBe stars would be formed through a single-star channel. Alternatively, they could be post-interaction binaries in which both the mass gainer (the OBe component) and the stripped mass donor, are similarly bright (see below). The determination of the orbital period of these systems will help in understanding their nature, especially if they contain a classical OBe star or not.

{\bf Class 3 -- Stationary emission + absorption-line binary:}
In three objects (BLOeM\,5-069, BLOeM\,5-071, BLOeM\,7-013), the absorption lines show large RV amplitudes as in an SB1 binary, while the emission lines are basically stationary. In BLOeM\,5-071, the narrow absorption lines move with large RV amplitudes of $\Delta\,\mathrm{RV}_\mathrm{max}\sim 175$\,\kms. In BLOeM\,5-069, the emission component is difficult to measure as there is a central narrow absorption component which could indicate an OBe star viewed edge-on. Those systems also show the signature of two individual components, a star as well as a potential disk. A similar signature was recently observed in post-interaction binaries just after mass transfer such as LB-1 and HR 6819 \citep[e.g.,][]{AbdulMasih2020, Shenar2020, Bodensteiner2020c, ElBadry2020a}. This initially led to their mis-classification as systems containing BHs. 
%In those, the stripped companion, though visible in the optical, is much less massive than the Be star, which is surrounded by the disk causing the emission lines. Therefore, the RV shift of the stripped star shows large amplitudes, while the emission lines are barely moving (or appearing stationary). 
Contrary, such a signature could be caused by a truly stationary emission source such as a nebulosity, a circumbinary disk, or a massive unseen object like a BH surrounded by an accretion disk (as the initial interpretation of LB-1 proposed). The analysis of future epochs will tell if the emission lines are indeed stationary, shedding light on their nature.

%These three are thus indeed most likely interacting binaries where the emission arises in an accretion disk, or from circumstellar material. Whether this interpretation can be applied to all the stars with a similar spectroscopic signature remains to be investigated with further observations.

\section{Discussion}\label{sec:discussion}

\subsection{Observational biases}\label{sec:biases} 
The observed binary fraction of a population of stars can in principle be corrected for observational biases to obtain an intrinsic, bias corrected fraction \citep[see e.g.,][]{Sana2012, Sana2013, Kiminki2012}. This requires knowledge of the expected distribution of orbital parameters of the binary systems (most importantly the periods and mass ratios) or these need to be constrained simultaneously during the bias correction process. Given that these distributions were measured for pre-interaction systems and might be mass and metallicity independent \citep[see e.g.,][]{Almeida2017,Villasenor2021, Banyard2021}, populations of pre-interaction binaries such as the OB stars in \bloem\ can be corrected for observational biases in a straight-forward manner \citep[see][]{Sana2024, Villasenor2024}.

OBe stars, however, are interpreted as a population dominated by binary interaction products. Therefore, the pre-interaction distributions usually used in the bias correction are not valid here. The distributions of periods and mass ratios for post-interaction systems are not observationally constrained (yet). One possibility would be to use binary population synthesis models starting from the observed pre-interaction binaries and computing their evolution. However, there are still many uncertainties in the binary interaction physics, the assumption on supernova kicks, as well as which simulated stars would appear as OBe stars observationally. This implies that the simulations should be treated with great care and we therefore refrain from providing a formal bias correction here.

To nevertheless evaluate the types of binary systems that would be detectable, we simulate the detection probability of the \bloem\ OBe survey for an illustrative 10\Msun\ Be primary and a range of orbital periods ($P$) from 1 day to $\sim$~10 years and mass ratios ($q=M_2/M_\mathrm{Be}$) from 0.01 to 1. We simulate 10\,000 \bloem-like observing campaigns of Be stars. We randomly draw the eccentricity of the system from a flat eccentricity distribution between 0 and 0.9, assume a random orientation of the orbit in 3D space, and adopt a random time of periastron passage. We further assume typical errors of 5~\kms\ and adopt the temporal sampling of the  observations of OBe stars in \bloem. We finally adopt the same binary detection criteria as described in Sect.~\ref{sec:bin} and evaluate the detection probability in the $P-q$ plane. 

\begin{figure}
    \centering
    \includegraphics[width=\linewidth]{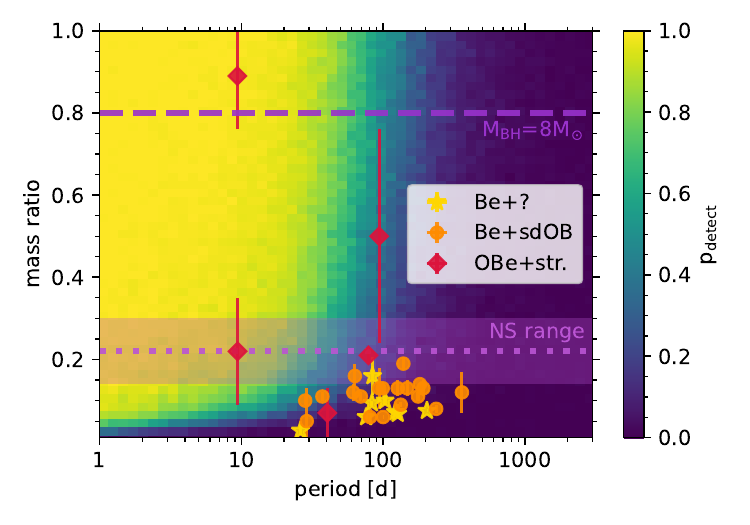}
    \caption{Binary detection probability p$_\mathrm{detect}$ as a function of period and mass ratio for a 10\,\Msun\,Be star. Overplotted are OBe binaries with known periods and mass ratios from \citet{Klement2024} and \citet[][Be+sdOBs in orange and Be binaries with companions of uncertain or debated nature in yellow]{WangL2023}, and recently stripped stars from \citet{Shenar2020}, \citet{Bodensteiner2020c}, \citet{Pauli2022} and \citet{Ramachandran2024}. Overplotted are also lines for potential Be+NS binaries (assuming NS masses between 1.4 and 3\,\Msun), and a Be binary with an 8\,\Msun\,BH companion.}
    \label{fig:pdetect}
\end{figure}

The results in Fig.\,\ref{fig:pdetect} are independent of the underlying distributions of periods and mass-ratios and only weakly depend on assumptions on the eccentricity \citep[e.g.,][]{Sana2013}.
Here, we also show known detected OBe binaries. Those include systems with sdOB companions, which are mostly detected and characterised using far-UV spectroscopy \citep[see][for a recent overview]{WangL2023} or interferometry \citep{Klement2024}. We also include so-called $\gamma$\,Cas-like binaries listed in \citet{WangL2023} for which the nature of the companion is still debated \citep[see also][]{Naze2022}. Finally, we plot OBe binaries with recently stripped companions: LB-1 \citep{Shenar2020}, HR\,6819 \citep{Bodensteiner2020c}, two systems from \citet[][namely \object{2dFS 2553} and \object{Sk -71 35}]{Ramachandran2024}, and \object{AzV 476} from \citet{Pauli2022}. Also indicated are the mass ratio range covered by typical Be+NS systems, assuming a 10~\Msun\ Be star and NS companion masses between 1.4 and 3.0\,\Msun, and a 10\,\Msun\ Be star with a BH companion of 8\,\Msun.

The simulations in Fig.\,\ref{fig:pdetect} show that our observational setup is sensitive to short-period systems of almost all mass ratios up to observing periods of about 3 months (approx.\ the duration of the observing campaign). It drops strongly for longer periods, and most systems with periods above 300\,d would not be detected with the given setup. While the detection probability does not vary strongly with mass ratio above $q=0.2$ in this period range, small mass ratios ($q<0.05$) are unlikely to be detected at any period due to the low reflex motion induced on the Be star. We repeated the same simulations for a 7\Msun\ primary Be star and find similar results. This is to be expected as they are not sensitive to the assumed primary mass (but on the mass ratio).

This numerical experiment demonstrates that the current observational campaign would have allowed us to detect OBe+MS binaries with mass ratios above 0.2 and periods below 100\,days. Those are expected to exist abundantly if OBe stars had similar binary properties than OB-type stars (see also next section). Furthermore, OBe binaries with BH companions, which are predicted by binary population synthesis based on dense grids of detailed binary evolution models \citep[e.g.][Xu et al. in prep.]{Langer2020b} would have been detected. Their lower-mass counterparts, NS companions, can be detected with a probability >90\% if their periods are below $\sim$30\,days. NS binaries with longer periods are unlikely detected as such, which can explain why two HMXBs appear as apparently single stars based on our RV campaign. Systems with even lower mass ratios, like sdOB or white dwarf binaries, cannot be detected with the given setup. The overall low observed binary fraction of the OBe sample thus indicates that there are few OBe+MS systems, and suggests that OBe stars mostly have stripped companions \citep[see e.g.][]{WangC2024}. Some of the binary systems detected here, in particular the SB1s, may be OBe+NS or OBe+BH systems.

As mentioned in Sect.\,\ref{sec:sample}, we recall here that our sample of OBe stars is incomplete. First, as described in \citetalias{Shenar2024}, given the observing strategy, the overall sample is incomplete in terms of massive stars. Notably, already the avoidance of crowded regions leads to missing many OBe objects: \citet{Dufton2019} report the presence of 73 Be stars in NGC\,346 that could not be observed in \bloem. Secondly, given the lack of the H$\alpha$ line in our observing range, which is usually the strongest emission line in OBe stars, stars with the weakest emission are potentially not identified as such in our sample. Thirdly, the transient nature of the Be phenomenon and the short time span covered by the observations lead to an additional incompleteness. On the other hand, several objects might be classified as OBe stars here, even if they are not classical OBe stars by nature. %Additionally, given the selection based on magnitude, binaries are more likely to be selected than single objects of similar intrinsic magnitude. 

\subsection{Comparison to OB stars in \bloem}
The \bloem\ survey allows for a direct comparison of the observed binary properties of OBe to OB stars. In particular, these objects are in the same %metallicity
environment, were observed with the same observational setup and timing, and the data were analysed in a similar way. These alleviate some of the common uncertainties when comparing observed properties of OBe stars with other objects, derived using different observing setups or techniques, or targeting stars located in different environments.

The RV analysis of the 139 O-type stars in \bloem\ yields an observed binary fraction of f$^\mathrm{O}_\mathrm{obs}=0.45\pm0.04$ \citep{Sana2024}.
The observed binary fraction measured in the 311 early B-type stars with luminosity classes III-V, similar to the stars in this sample, is f$^\mathrm{B}_\mathrm{obs}=0.49\pm0.03$ \citep{Villasenor2024}, 
and thus agrees with the O-star measurement. The binary fraction of the 78 classical OBe stars of f$^\mathrm{OBe}_\mathrm{obs}=0.18\,\pm\,0.04$ is less than half the value measured in OB stars, and thus differs significantly. Even when including all potential candidates (Sect.\,\ref{sec:bin_stat}), the observed binary fraction of OBe stars of f$^\mathrm{OBe}_\mathrm{obs+cand}=0.32\,\pm\,0.04$ remains significantly below that of OB stars. %It is further likely that some of those systems will be removed from the OBe sample with more observations, as they might be semi-detached binaries.   % problem, OBe stars are Luminosity class V-II, so they are not completely comparable to Jaime's sample (Nikolay has the LC II objects). But: LCs are very difficult to assess of OBe. 

Also the number of SB2s, which are in most cases pre-interaction binaries \citep{deMink2014}, detected among the OBe stars is much lower than for OB stars. Among the O stars, 22 out of 139 objects were classified as SB2, which is $0.16\pm0.03$ of the sample \citep{Sana2024}. Among the B stars, the SB2 fraction is $0.18\pm0.02$ \citep[55 out of 308,][]{Villasenor2024}) . In the OBe sample, we find that seven of the 78 systems, that is $0.09\pm0.03$, show an SB2-like signature, much lower than the fraction in OB stars. We have softened the SB2 definition here, as we include systems where absorption and emission lines move in anti-phase, indicating the presence of a companion star that is surrounded by a circumstellar disk, which is not directly detected. Some of those could also be interacting binaries like BLOeM\,3-031 and BLOeM\,6-034, and thus no classical OBe stars at all, further reducing the number SB2s. 

As discussed in Sect.\,\ref{sec:bin_class}, the spectra of OBe stars are impacted by additional variability, which might lead to a mis-classification of stable objects as binaries. Contrary, stellar variability could also mask the presence of binary companions, although only the ones with weak signatures. Obvious binaries -- those with similar masses, short periods and large RV amplitudes -- should have been detected among the OBe stars. Here, however, we find that the observed binary fraction and the number of detected SB2s in OBe stars are significantly lower than the ones measured for their OB counterparts. This implies that the underlying multiplicity properties of OBe stars  are different (likely longer orbital periods, or smaller mass ratios, or dimmer companions as would be expected if they are of a different nature than in OB+MS binaries). This is expected if OBe stars are indeed the products of binary interactions.%, as theory predicts their companions to be either stripped stars (in a short-lived optically bright phase, or a longer-lived hot sdOB like phase), or compact objects, if the system did not get unbound in a potential supernova explosion. 

\subsection{Comparison to OBe stars in the SMC}
% paragraph about RIOTS
The Runaways and Isolated O-Type Star Spectroscopic Survey of the SMC \citep[RIOTS4,][]{Lamb2016} targeted almost 400 isolated early-type stars with single- or multi-epoch spectroscopy (the latter for less than 10\% of the sample), and found that 42\% of their stars are classical OBe stars \citep{Lamb2016}. In particular, they find a large fraction of classical Oe stars, an order of magnitude higher than reported in the Galaxy \citep[][]{GoldenMarx2016}. \citet{Dallas2022} and \citet{Phillips2024} find that OBe stars are more isolated than OB stars, and that their spatial distribution and kinematics is similar to HMXBs. This leads them to interpret OBe stars as likely interaction products. 

Only ten targets are in common between RIOTS4 and the OBe sample of \bloem. The derived spectral types largely agree within one sub-type. One object (BLOeM\,7-051) did not show emission in RIOTS4 but appears as Be star in \bloem. Only one \bloem\ object (BLOeM\,2-082) was included in the multi-epoch monitoring of RIOTS4. Based on four available epochs, it was classified as binary in RIOTS4 \citep{Lamb2016}, and is also classified as binary in \bloem. 

% NGC 346 and NGC 330
Due to technical limitations, the \bloem\ survey avoids the dense cores of star clusters, mainly the young cluster NGC\,346 with an age below 5\,Myr \citep[e.g., ][]{Sabbi2008}, and NGC\,330 with a age of about 45\,Myr \citep{Patrick2020}. Those harbour many massive stars,in particular also OBe stars \citep[e.g.,][]{Dufton2019, Milone2018, Bodensteiner2020a}, and both were targeted by spectroscopic studies recently. 

\citet{Dufton2019} provided a stellar census of NGC\,346 based on VLT/FLAMES data of over 250 O- and early-B stars, detecting 70 Be stars among them. The Be star fraction of $0.27\pm0.03$ is thus larger than the one in \bloem\ \citep[which amounts to approximately 11\%,][]{Shenar2024}. This might be linked to the selection of targets and the incompleteness of the \bloem\ OBe sample discussed in Sect.\,\ref{sec:sample}. Given that only two to four epochs spanning 40 days or less were available, the detection probability for binaries is low, and no binary fraction for OBe stars was provided \citep[see][]{Dufton2022}. 

NGC\,330 was studied based on six epochs obtained over the course of approximately 1.5 years with the integral-field spectrograph MUSE at the VLT \citep[][]{Bodensteiner2020a, Bodensteiner2021, Bodensteiner2023}. While covering practically the entire optical wavelength range, MUSE only provides a spectral resolving power between 1700 and 4000, much lower than the FLAMES data used here. Given the age of the cluster, almost no O-type stars are present anymore. Among the roughly 330 B-type stars, 115 show emission lines typical of Be stars, resulting in a Be fraction of f$_\mathrm{Be}=0.35\pm0.03$, similar to the one measured in NGC~346. Adopting the same binary criteria invoked here, \citet{Bodensteiner2021} derived an observed binary fraction of the Be sample of only f$^\mathrm{Be}_\mathrm{obs}=7.5\pm2.7\%$. This is even lower than the fraction derived here, and could be linked to the lower binary detection probability in NGC~330, that comes from the lower resolving power and fewer epochs available. Still, obvious binary systems with large RV amplitudes should have been detected.

\subsection{Comparison to HMXBs in the SMC}
The SMC is known for its large number of HMXBs, in particular BeXRBs \citep[e.g.,][]{Coe2015, Antoniou2019}. However, none of the BeXRBs in \citet{Coe2015} and only three sources in \citet{Antoniou2019} overlap with our OBe sample. In addition to those three sources, there is one more HMXBs from other X-ray catalogues \citep[e.g., ][]{cscv2, Sturm2013}. Only one of those four XRBs are detected as binary in \bloem, and an additional one as candidate binary (the measured $\Delta$RV$_\mathrm{max}$ does not exceed 20\,\kms). If the other two objects are indeed BeXRBs, and not chance alignments between the optical and the X-ray source positions, their companions are most likely lower-mass NSs, or their periods would be longer than the current observing campaign. 

Overall, most HMXBs listed in the catalogue by \citet{Antoniou2019} and the BeXRBs from \citet{Coe2015} are significantly fainter and redder than the majority of the \bloem\ sample. Their redder colors place them mostly beyond our color cut based on the 8\Msun\ evolutionary track employed in the target selection. This color bias particularly affects the faint end of the \bloem\ sample. This is illustrated in Appendix \ref{app:xrbs}, where we compare the samples by \citet{Antoniou2019} and \citet{Coe2015} with the \bloem\ sample. It demonstrates again the incompleteness of the \bloem\ survey for faint, red Be stars, and explains the overall small number of BeXRBs in this large sample of classical OBe stars.

%\subsection{Comparison to OBe stars in the SMC}
%\begin{itemize}
%    \item BeXRB plot
%    \item Dufton+2019: NGC\,346 and surroundings, among those 70 Be stars
%    \item Dufton+2022: VFTS OBe stars
%    \item Bodensteiner+2021: multiplicity of NGC 330
%    \item BeXRBs in Vinciguerra?
    %\item comparison to literature (for example: Varsha's XShootU stars?) % DP: and maybe her star from the SGS? Ramachandran+2023, one candiadte Oe system  suspected to be a binary AzV 493, see Oey+2023, might also be in this sample
%\end{itemize}

\section{Conclusion}\label{sec:conclusion}
In this study, we investigate the binary properties of 82 OBe stars observed in the context of the \bloem\ survey. We measured RVs using the CCF technique in nine epochs obtained over the course of almost three months with the \flames\ multi-object spectrograph at the VLT, and searched for RV variations. Both the stellar absorption lines and the emission lines characteristic of these types of objects are investigated. All spectra were visually inspected for signatures of SB2s.

After excluding two Of?p stars and two interacting binaries, we detect fourteen binaries among our 78 Be objects as well as eleven candidate binaries. Considering only the robust detections, we measure an observed binary fraction of the OBe sample in \bloem\ of $\mathrm{f}^\mathrm{OBe}_\mathrm{obs} = 0.18\pm0.04$, which is less than half of the observed binary fraction measured for the O and the B stars in \bloem\ \citep{Sana2024, Villasenor2024}. 
%We refrain from applying the usual bias correction to the observations as it assumes pre-interaction orbital properties in the simulated binaries, while OBe stars are likely post-interaction objects and thus have altered orbital parameters because of their evolutionary history. 
An estimate of the detection probability of our campaign shows that systems with periods below 100\,d and mass ratios above 0.2 would most likely have been detected. The lower binary fraction among OBe stars relative to OB stars is most likely easily explained when assuming OBe stars are post-interaction products.

In some binaries, the RVs measured from emission lines follow the absorption lines, while in others, there seems to be an anti-phase motion, or the emission lines appear stationary (see Sect.\,\ref{sec:bin_nat}). These different signatures imply different properties of the companions. When following the same trend, the emission and absorption lines arise from the same object, and the companion is not detected in the optical. Those could be systems with massive compact object companions, or rapidly rotating MS stars. When moving in antiphase, two visible components are present, a star and the source of the emission lines, which might be an accretion or decretion disk around another star, or a stationary component that could arise from material located around the binary system (a circumbinary disk, or a nebulosity). Those systems could thus either be currently interacting binaries, or recently stripped stars with OBe companions, in which the stripped star is still puffed-up and bright.

Our observing campaign is most sensitive to detecting binaries with periods below 100\,d and mass ratios above 0.1. In particular the mass-ratio sensitivity implies that OBe stars with stripped stars or neutron-star companions, that would invoke only small semi-amplitudes in the OBe stars, would mostly appear as apparently single with the given setup. Contrary, we would have detected systems with similar mass-ratio companions in the aforementioned period range. Given the masses considered here, those could either be BHs or massive MS stars. The former are predicted to exist by populations synthesis calculations \citep[e.g.,][]{Langer2020b}, but are not observed here. The lack of the latter illustrates again the previously reported lack of OBe stars with MS companions.

%While the OBe sample was carefully selected (see \citetalias{Shenar2024}), 

%The OBe sample is likely incomplete, suffering from the magnitude and color cuts and the fact that the Be phenomenon is transient and thus likely some stars did not show emission during the observations so far. Similarly, some objects classified as OBe stars in \citetalias{Shenar2024} might not be classical OBe stars, which means that they cannot be interpreted as rapidly rotating stars with decretion disks. This holds for the two Of?p stars, as well as the two interacting binaries identified here, but most likely more such re-classification may become necessary in the future.
%A similar spectroscopic signature can arise from different physical origins, such as nebular contamination, interacting binaries, magnetic stars (as testified by the two Of?p stars in the sample) or SB2s in which the lines never fully deblend in wavelength space, thus mimic the appearance of emission in the core of the lines. Based on the information we currently have, we consider the stars in the sample as OBe stars, and future observations will help to refine the classification. 

The additional 16 epochs obtained in the \bloem\, survey will allow further constraints on the nature of our detected binaries, particularly the nature of the companions that could either be MS objects (if OBe stars could form as single stars), stripped stars or compact objects. The new observations, which in total cover two years of observing time, will also allow us to detect more Be binaries, especially objects with much longer periods. %than the initial phase of the survey.
Detecting additional OBe binaries and constraining their orbital and physical parameters is crucial to further constrain the evolutionary origin of classical OBe stars. If at least part of them turn out to be products of binary interactions, the characterization of additional systems will allow for direct constraints on the physics of the interaction that such systems experienced. %Those are valuable inputs in binary evolution calculations and population synthesis predictions with long-lasting consequences for many other fields.

\section*{Data availability} 
Appendices D, E and F are available via \href{10.5281/zenodo.14679548}{10.5281/zenodo.14679548}.

\begin{acknowledgements}
The research leading to these results has received funding from the European Research Council (ERC) under the European Union's Horizon 2020 and Horizon Europe research and innovation programme (grant agreement numbers 772225: MULTIPLES, and 945806) and is supported by the Deutsche Forschungsgemeinschaft (DFG, German Research Foundation) under Germany’s Excellence Strategy EXC 2181/1-390900948 (the Heidelberg STRUCTURES Excellence Cluster).
DMB gratefully acknowledges funding from  UK Research and Innovation (UKRI) in the form of a Frontier Research grant under the UK government's ERC Horizon Europe funding guarantee (SYMPHONY; grant number: EP/Y031059/1), and a Royal Society University Research Fellowship (URF; grant number: URF{\textbackslash}R1{\textbackslash}231631). RGI is funded by STFC grant ST/Y002350/1 as part of the BRIDGCE UK network, and thanks IReNA colleagues for stimulating discussions. Z.K. acknowledges support from JSPS Kakenhi Grant-in-Aid for Scientific Research (23K19071). IM acknowledges support from the Australian Research Council (ARC) Centre of Excellence for GravitationalWave Discovery (OzGrav), through project number CE230100016. DP acknowledges financial support by the Deutsches Zentrum f\"ur Luft und Raumfahrt (DLR) grant FKZ 50OR2005. DFR is thankful for the support of CAPES-Br and FAPERJ/DSC-10 (SEI-260003/001630/2023). 
AACS and VR are supported by the German Deutsche Forschungsgemeinschaft (DFG) under Project-ID 445674056 (Emmy Noether Research Group SA4064/1-1, PI Sander). AACS and VR are further supported by funding from the Federal Ministry of Education and Research (BMBF) and the Baden-W{\"u}rttemberg Ministry of Science as part of the Excellence Strategy of the German Federal and State Governments. 
JIV acknowledges support from the European Research Council for the ERC Advanced Grant 101054731. LRP acknowledges support from grant PID2022-140483NB-C22 funded by MCIN/AEI/10.13039/501100011033. GH acknowledges support from grants PID2021-122397NB-C21/PID2022-136640NB-C22 funded by MCIN/AEI/FEDER/10.13039/501100011033. J.K. thanks for the support of a grant GA \u{C}R 22-34467S. This project has received funding from the European Research Council (ERC) under the European Union's Horizon 2020 research and innovation programme (grant agreement 101164755/METAL) and was supported by the Israel Science Foundation (ISF) under grant number 2434/24.
\end{acknowledgements}

\bibliographystyle{aa}
\bibliography{papers}

\begin{appendix}
\section{Dependence on the adopted RV threshold}\label{app:bin_crit}
\begin{figure}
    \centering
    \includegraphics[width=\linewidth]{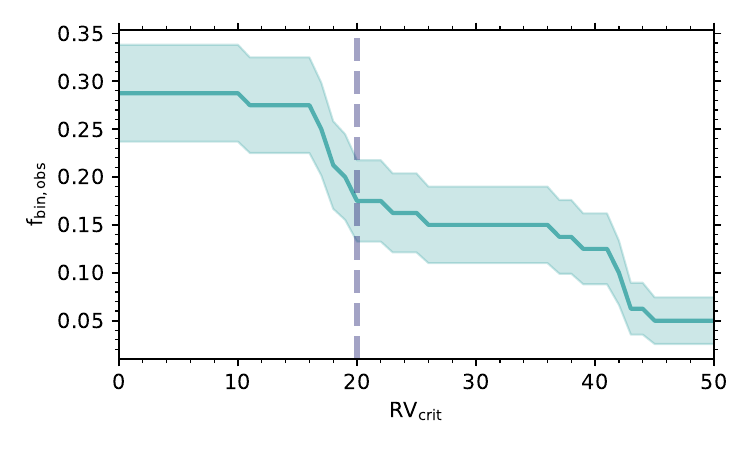}
    \caption{Observed binary fraction as a function of critical RV threshold value RV$_\mathrm{crit}$. The adopted value of 20\,\kms\ is indicated by the blue vertical line.}
    \label{fig:rvcrit}
\end{figure}

Here, we investigate the impact of the adopted critical RV threshold in the binary criteria on the observed binary fraction. For this, we vary the threshold between 1 and 50\kms, and for each threshold value, compute the observed binary fraction according to the criteria laid out in Sect.\,\ref{sec:bin_class}. Fig.\,\ref{fig:rvcrit} shows that the observed binary fraction is consistent within the errors between threshold values of approximately 16 and almost 40\,\kms, and thus not strongly sensitive to the exact value. The observed binary fraction never drops to zero as clear SB2s are flagged as such and always count in the binary fraction, even if they do not fulfill both criteria. 

\section{Orbits for OGLE-variable stars}\label{sec:ogle}
Three OBe stars are classified as photometrically variable stars in OGLE \citep{Pawlak2016}, namely BLOeM\,3-018, BLOeM\,3-031, and BLOeM\,6-034. BLOeM\,3-018 is classified as ellipsoidal variable with a period of 1.44\,d, but is not detected as binary in \bloem. We leave its analysis to the availability of additional spectroscopic epochs.

BLOeM\,3-031, and BLOeM\,6-034 are classified as ``detached or semi-detached'' system with  periods of 4.37\,d and 6.43\,d. Both are classified as binaries, in particular as SB2s, in this work. As a consistency check, we compare the OGLE orbital parameters (in particular the period P and the epoch of primary eclipse T$_\mathrm{0}$) with the measured RVs. The remaining orbital parameters (the RV semi-amplitude $K$ and the systemic velocity $\gamma$) are estimated crudely from the observed RV curve, while for simplicity the eccentricity is fixed to zero. Here, we do not perform a detailed fitting but compare if the two orbits match. 

\paragraph{BLOeM\,3-031:} This object (classified as SB2 already in \citetalias{Shenar2024} with B0:\,III: + OB\,e) shows a clear SB2 signature with a narrow-lined component that is moving with a large RV amplitude of up to 500\,\kms, measured from the strong \ion{Si}{ii} lines. Additionally, there is a double-peaked emission component that appears to be moving in anti-phase, however with a much lower amplitude than the narrow-lined star. Given the emission lines are intrinsically variable in shape and amplitude, we refrain from comparing their RVs here. The spectral signature of this target resembles the post-interaction binary \object{Sk -71$^{\circ}$ 35} recently detected in the LMC \citep[][]{Ramachandran2024}.

%\begin{figure*}
%    \centering
%    \begin{subfigure}{.5\textwidth} \centering
%    \includegraphics[width=\linewidth]{figs/BLOeM_3-031_RVcurve.pdf}
%    \newsubcap{RV curve of BLOeM\,3-031.}
%    \label{fig:rvcurve_3-031}
%    \end{subfigure}%
%    \begin{subfigure}{.5\textwidth} \centering
%    \includegraphics[width=\linewidth]{figs/BLOeM_3-031_RVextremes.pdf}
%    \newsubcap{RV extremes of BLOeM\,3-031.}
%    \label{fig:rvextr_3-031}
%   \end{subfigure}
%\end{figure*}

\begin{figure}
    \centering
\includegraphics[width=\linewidth]{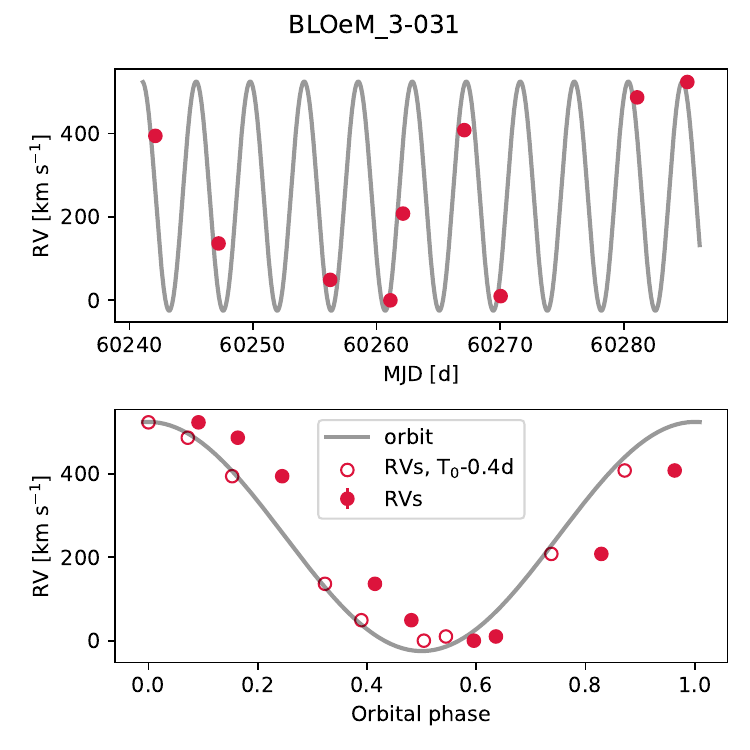}
 \caption{RV curve of BLOeM\,3-031 (filled red dots) with the photometric orbit overplotted (gray), as a function of time (upper panel) and orbital phase (lower panel). In the phase-folded curve we also include the assuming T$_\mathrm{0}$ = T$_\mathrm{0}$-0.4 days, which provide a much better fit to the data.}
    \label{fig:orbit_3-031}
\end{figure}

%The RV curve and spectra at the RV extremes of BLOeM 3-031 are shown in Figures \ref{fig:rvcurve_3-031} and \ref{fig:rvextr_3-031}.
%Note that error bars are plotted in the RV curve, which are of the order of 5\kms\ for the absorption lines, and of the order of 10\kms\ for H$\gamma$. Given the large RV amplitudes they are smaller than the symbol sizes in the figure.
Figure\,\ref{fig:orbit_3-031} shows the photometric orbit of BLOeM\,3-031 assuming $K_1=275$\,\kms and $\gamma=250$\,\kms, overplotted over the RVs measured from the \ion{Si}{ii} lines. In the phase-folded RV-curve we also include the RVs at phases shifting the photometric T$_\mathrm{0}$ by 0.4\,days, to illustrate an even better match. Such deviation would not be surprising given the short period of the system and the observing time difference of roughly 10 years between the two data sets, in which a small error in the period could lead to such an effect. This comparison demonstrates that the spectroscopic and photometric features align and BLOeM\,3-031 is indeed an interacting, semi-detached binary.

\begin{figure}
    \centering
\includegraphics[width=\linewidth]{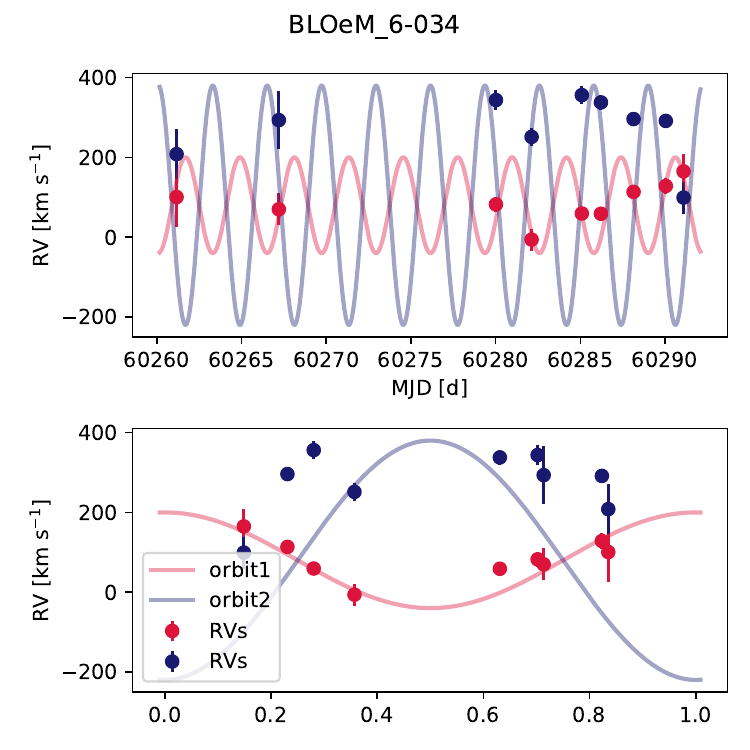}
 \caption{RV curves of BLOeM\,6-034 (red and blue dots, for the two components visible in the spectra) with the photometric orbits overplotted (in matching colors), as a function of time (upper panel) and orbital phase (lower panel).}
    \label{fig:orbit_6-034}
\end{figure}

\paragraph{BLOeM\,6-034:} BLOeM 6-034 shows the only clear SB2 feature among the sample, with two stellar components moving back and forth in particular visible in the \ion{He}{I} lines. Additionally, there seems to be a stable, third component in the emission Balmer lines that leads to line-infilling and a relatively stable H$\gamma$ profile. Given the SB2 nature of BLOeM\,6-034, for which the CCF (assuming stable line profiles) does not provide reliable RVs, we further employ double-Gaussian fitting \citep[e.g.][]{Sana2014} to determine the RVs of the object, 
%Those are shown in Figure \ref{fig:rvcurve_6-034} (the two colors indicate the primary and the secondary, 
(see Fig.\,\ref{fig:orbit_6-034}, both measured from \ion{He}{i} lines). % while two spectra at RV extremes are shown in \ref{fig:rvextr_6-034}.
Because of the complicated nature of the line profiles, we do not attempt to infer RVs of the putative emission component.

%\begin{figure*}
%    \centering
%    \begin{subfigure}{.5\textwidth} \centering
%    \includegraphics[width=\linewidth]{figs/BLOeM_6-034_RVcurve.pdf}
%    \newsubcap{RV curve of BLOeM\,6-034.}
%    \label{fig:rvcurve_6-034}
%    \end{subfigure}%
%    \begin{subfigure}{.5\textwidth} \centering
%    \includegraphics[width=\linewidth]{figs/BLOeM_6-034_RVextremes.pdf}
%    \newsubcap{RV extremes of BLOeM\,6-034.}
%    \label{fig:rvextr_6-034}
%    \end{subfigure}
%\end{figure*}

For BLOeM\,6-034, the photometric orbit does not match the spectroscopic RVs as well. This object is the only system that shows two sets of absorption lines moving in antiphase, and we estimated both RVs, however, with larger uncertainties (see Sect.\,\ref{sec:bin}). As the photometric orbit and the measured RVs did not match, we tried to get a better match using half or twice the OGLE period. Indeed, as illustrated by Fig.\,\ref{fig:orbit_6-034}, we find a better match using the OGLE T$_\mathrm{0}$, half the OGLE period ($P'=3.21$\,d), as well as $\gamma=80$\,\kms, $K_1=120$\,\kms and $K_2=300$\,\kms. While there is still no ideal match with the RVs of the secondary, the primary RVs align well with the orbit. The discrepancy could be due to problems with the RV measurements, given the highly blended spectral lines. However, the comparison and the even shorter period indicate that BLOeM\,6-034 is indeed a semi-detached binary.

\section{Comparison to catalogues of HMXBs}\label{app:xrbs}
Here, we compare the \bloem\ OBe sample with catalogues of X-ray sources. We use the catalogue of HMXBs by \citet{Antoniou2019}, which is based on their Chandra X-ray Visionary Program (XVP) and supplemented by 14 additional HMXBs identified by \citet{Haberl2016}, and the BeXRB catalogue by \citet{Coe2015}. For this comparison, we cross-match the two catalogues with the \textit{Gaia} DR3 catalogue to obtain positions, BP-RP colors and magnitudes in the same reference frame as the \bloem\ sources, resulting in 123 sources with the former, and 48 sources in the latter.

\begin{figure}
    \centering
\includegraphics[width=\linewidth]{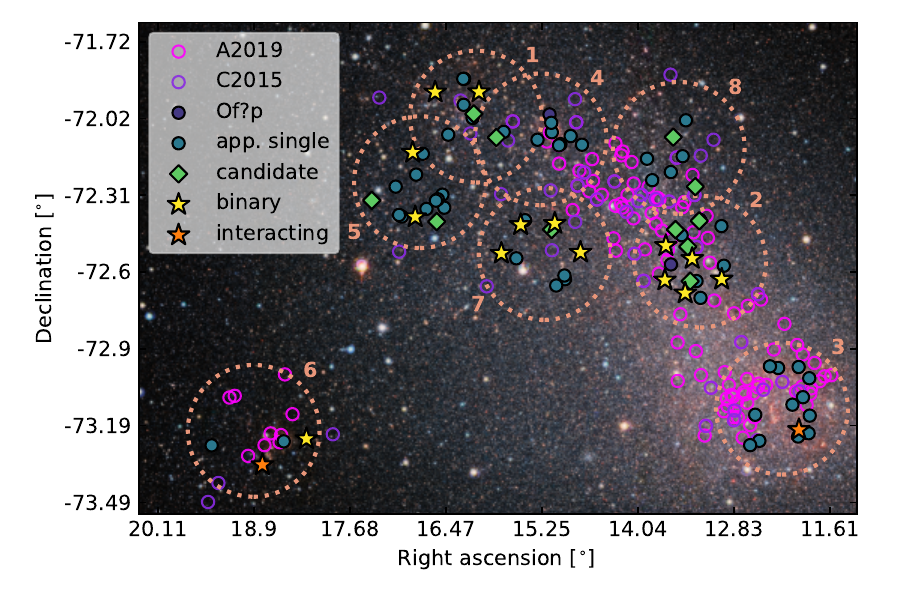}
 \caption{Same as Fig.\,\ref{fig:spatial}, with the HHMXBs of \citet{Antoniou2019} and the BeXRBs of \citet{Coe2015} indicated in purple and pink, respectively.}
    \label{fig:spatial_xrbs}
\end{figure}

\begin{figure}
    \centering
\includegraphics[width=\linewidth]{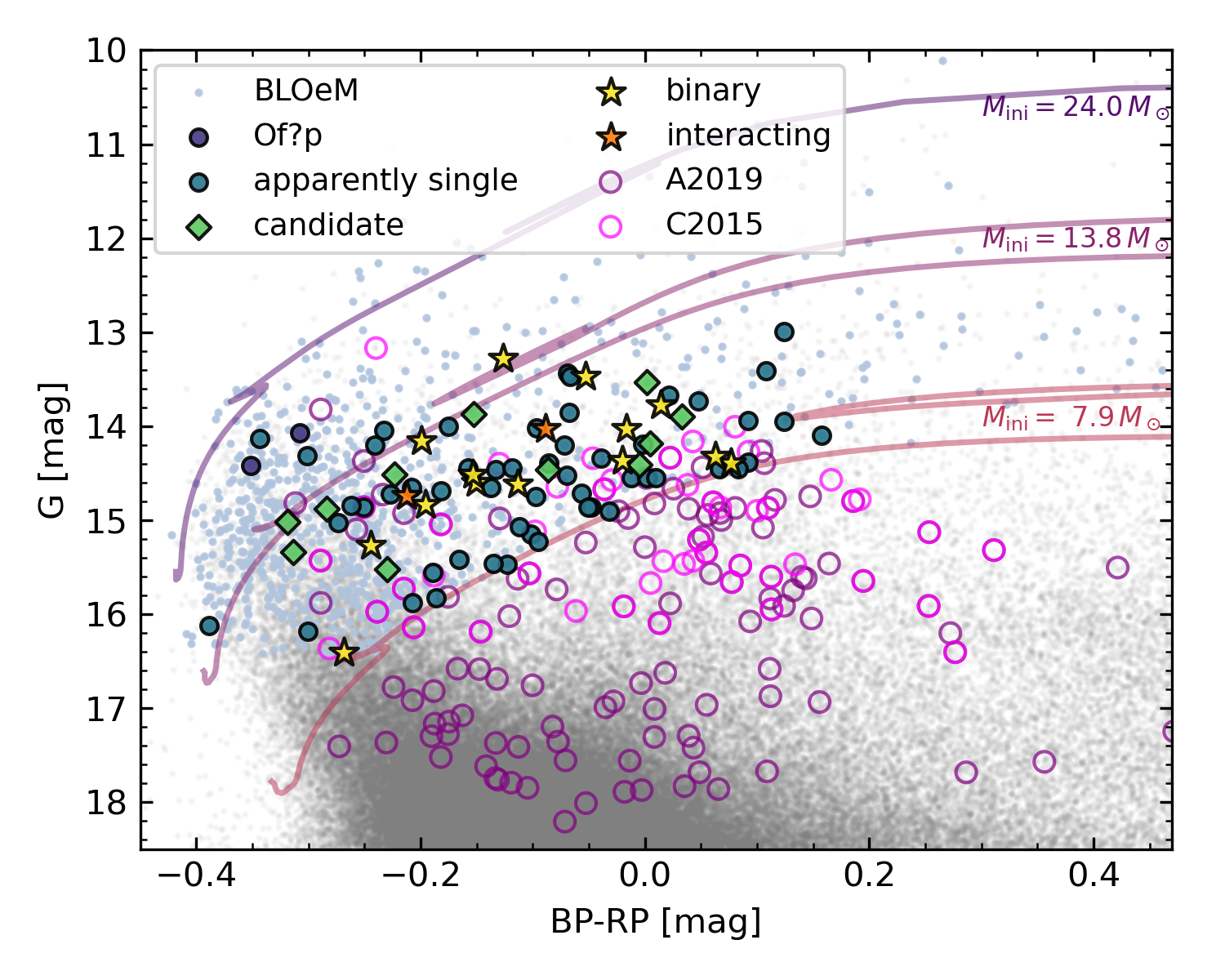}
 \caption{Same as Fig.\,\ref{fig:bin_cmd}, but with the HMXBs of \citet{Antoniou2019} and the BeXRBs of \citet{Coe2015} overlain in purple and pink, respectively. Note the adopted magnitude cut extending to fainter objects here to show the faintness of the HMXBs in particular in the \citet{Antoniou2019} catalogue.}
    \label{fig:cmd_xrbs}
\end{figure}

Overall, only two \bloem\ OBe stars coincide with HMXBs in \citet{Antoniou2019}, and there is no overlap with the BeXRBs in \citet{Coe2015}. The overall sky coverage of the XVP, and the spatial distribution of BeXRBs somewhat align with the \bloem\ fields (especially covering fields 2, 3, 4, 6 and 8, see Fig.\,\ref{fig:spatial_xrbs}). A large fraction of detected X-ray sources, however, are outside the \bloem\ fields, %and some field (i.e., 5 and 7)are barely covered.
which might be part of the reason of the low number of matches. Additionally, the BeXRBs and HMXBs in the two catalogues are significantly fainter as well as redder than the targets selected in \bloem\ (see Fig.\,\ref{fig:cmd_xrbs}). The selection based on evolutionary tracks excludes a large number of the known X-ray sources, especially the ones with redder color. Those targets could have later spectral types, or larger disks (which then lead to a stronger IR excess making them appear redder). Overall, this explains the low number of BeXRBs in the \bloem\ sample.

\section*{Appendix D: Target overview table}
Here, we provide all relevant information of the targets studied in this work. We sort them in two tables, Table\,\ref{tab:bincand} with the binaries and candidates, and Table\,\ref{tab:const} with the RV constant objects. 

\begin{sidewaystable*}
\begin{center}
\caption{\label{tab:bincand}Parameter summary of all binaries and candidate binaries. The columns give the BLOeM ID, the Gaia DR3 ID, coordinates, and G-band magnitude, and the spectral type from \citetalias{Shenar2024}. The maximum RV amplitude between two epochs is given in column $\Delta$RV$_\mathrm{max}$. The columns `C1' and `C2' indicate if the system satisfies the two binary criteria described in Sect.\,3.2. The binary status is given in status (`bin' indicates stars classified as binaries, while `cand' indicates candidates). Additional information on the type of binary and the literature are given in the last two columns.
}
\begin{tabular}{lllllrllllll} \hline \hline
BLoeM ID & Gaia DR3 ID & RA      & DEC     & G     & SpT & $\Delta$RV$_\mathrm{max}$ & C1 & C2 & status & type & literature \\
         &             & [J2000] & [J2000] & [mag] &     & [\kms]                    &    &    &    &      &  \\\hline
\object{BLOeM 1-040} & 4690525438865707904 & 15.6657917 & -71.9966111 & 13.48 & O9.7$\,$III:n$\,$e & 22.0 $\pm$ 4.9 & T & T & bin & SB1 & \\ 
\object{BLOeM 1-113} & 4690511626248698368 & 16.2209167 & -71.9886389 & 14.16 & B1$\,$II$\,$e & 42.4 $\pm$ 3.5 & T & T & bin & SB2 & \\ 
\object{BLOeM 2-021} & 4688964922657391744 & 12.7445833 & -72.7626944 & 14.39 & B2.5:$\,$II:$\,$e & 25.2 $\pm$ 3.7 & T & T & bin & SB2 & \\ 
\object{BLOeM 2-066} & 4688966739421000448 & 13.1017083 & -72.6775278 & 14.03 & O9.7$\,$III:nnn$\,$pe+ & 50.0 $\pm$ 9.5 & T & T & bin & ? &  \\ 
\object{BLOeM 2-082} & 4685960021758355584 & 13.2179167 & -72.8083889 & 14.6 & O9.2$\,$III$\,$pe & 39.9 $\pm$ 12.2 & T & T & bin & ? & HMXB$^{1,6}$ \\ 
\object{BLOeM 2-109} & 4688990692466513408 & 13.4242917 & -72.6228056 & 14.84 & B2$\,$II$\,$e & 41.7 $\pm$ 2.7 & T & T & bin & SB1 &  \\ 
\object{BLOeM 2-111} & 4685983699984262912 & 13.463875 & -72.7545556 & 14.62 & B2$\,$II:$\,$e & 17.3 $\pm$ 3.0 & T & F & bin & SB2 & \\ 
\object{BLOeM 5-069} & 4687502400374028928 & 16.5525 & -72.2153611 & 14.37 & B3$\,$II$\,$e & 42.3 $\pm$ 5.2 & T & T & bin & SB2? &  \\ 
\object{BLOeM 5-071} & 4687437254305162880 & 16.57175 & -72.4629722 & 13.28 & O8.5:$\,$Ib$\,$ + OBpe & 164.5 $\pm$ 4.9 & T & T & bin & SB2? &  \\ 
\object{BLOeM 6-001} & 4687162582543422080 & 18.123875 & -73.2916111 & 14.32 & B0.5$\,$III:$\,$e & 36.2 $\pm$ 6.6 & T & T & bin & ? &  \\ 
\object{BLOeM 7-013} & 4685988544708868864 & 14.5094583 & -72.6315556 & 14.51 & B1.5$\,$III$\,$e & 44.7 $\pm$ 5.4 & T & T & bin & SB2? &  \\ 
\object{BLOeM 7-045} & 4685993419427102848 & 14.8115 & -72.5156944 & 16.41 & B2$\,$III:$\,$e & 67.6 $\pm$ 15.9 & T & T & bin & ? &  \\ 
\object{BLOeM 7-082} & 4687491164738701056 & 15.2446667 & -72.5139444 & 13.78 & B2$\,$$\,$e+ & 42.4 $\pm$ 2.2 & T & T & bin & SB2 &  \\ 
\object{BLOeM 7-114} & 4687475118738255616 & 15.510875 & -72.6196111 & 15.27 & B2$\,$II$\,$e & 38.1 $\pm$ 5.3 & T & T & bin & - &  \\ \hline
\object{BLOeM 1-058} & 4690512313443855360 & 15.749 & -72.0798889 & 13.53 & O9.5$\,$II:$\,$pe & 34.6 $\pm$ 10.8 & F & T & cand & ? & \\ 
\object{BLOeM 2-056} & 4688981106097035008 & 12.9828333 & -72.5343333 & 13.88 & B2$\,$II:$\,$e & 17.9 $\pm$ 3.1 & T & F & cand & ? & \\ 
\object{BLOeM 2-070} & 4688962414395649792 & 13.144625 & -72.7636111 & 14.46 & B1$\,$II$\,$e & 20.0 $\pm$ 1.9 & T & F & cand & ? &  \\ 
\object{BLOeM 2-071} & 4688968049392847488 & 13.148375 & -72.6311389 & 14.51 & B1.5$\,$II:$\,$e & 16.6 $\pm$ 2.6 & T & F & cand & SB1 & \\ 
\object{BLOeM 2-097} & 4689003680451206272 & 13.2900417 & -72.5631111 & 14.41 & B5$\,$II:$\,$e & 11.3 $\pm$ 5.4 & F & F & cand & SB1 & \\ 
\object{BLOeM 4-113} & 4690502761435017984 & 15.4837083 & -72.1746389 & 15.02 & B2.5$\,$II$\,$pe & 18.4 $\pm$ 4.4 & T & F & cand & SB1 & HMXB$^{1,2,3}$ \\ 
\object{BLOeM 5-035} & 4687482780961207424 & 16.3032083 & -72.485 & 14.87 & B1.5$\,$II:$\,$e & 14.1 $\pm$ 4.8 & F & F & cand & - &  \\ 
\object{BLOeM 5-115} & 4687460481482120832 & 17.1076667 & -72.3909167 & 13.9 & O9.5:$\,$V:$\,$pe & 26.6 $\pm$ 7.6 & T & T & cand & SB2? &  \\ 
\object{BLOeM 7-051} & 4685990434494050048 & 14.85675 & -72.5381944 & 15.53 & B1$\,$II:$\,$e & 16.0 $\pm$ 8.4 & F & F & cand & - & \\ 
\object{BLOeM 8-021} & 4689032851851488000 & 13.0053333 & -72.4025 & 14.19 & O7$\,$V-IIInnn$\,$pe & 23.3 $\pm$ 22.4 & F & T & cand & ? & \\ 
\object{BLOeM 8-059} & 4689061748394242944 & 13.2452083 & -72.2081667 & 15.34 & B2$\,$II:$\,$e & 29.4 $\pm$ 7.5 & F & T & cand & ? & \\ \hline 
\object{BLOeM 3-031} & 4685836708972891648 & 11.8871667 & -73.3545278 & 14.03 & B0:$\,$III:$\,$ + OBe & 523.9 $\pm$ 4.2 & T & T & bin & SB2; IB & EB\,\textsuperscript{*} \\ 
\object{BLOeM 6-034} & 4686408313260119552 & 18.7012083 & -73.3825278 & 14.74 & B2$\,$III-II$\,$e & 61.0 $\pm$ 2.6 & F & T & bin & SB2; IB & EB\,\textsuperscript{*} \\ \hline
\end{tabular}     
\end{center}
\footnotesize{\textsuperscript{*} classification based on OGLE-III photometry from \citet{Pawlak2013}. %$\dag$ classification as \halpha\ source based on \halpha\ imaging \citep{Maravelias2017}.
X-ray source identifications are from $^1$: \citet{cscv2}, $^2$: \citet{Sturm2013} , $^3$: \citet{CXOGSGSRC}, $^4$ \citet{Laycock2010}, $^5$ \citet{Traulsen2020}, $^6$\citet{Antoniou2016}} \\
\end{sidewaystable*}

\begin{table*}
\begin{center}
\caption{\label{tab:const} Parameter summary of all RV-stable stars classified as presumably-single stars here. The columns give the BLOeM ID, the Gaia DR3 ID, coordinates, and G-band magnitude, and the spectral type from \citetalias{Shenar2024}. The maximum RV amplitude between two epochs is given in column $\Delta$RV$_\mathrm{max}$. Additional information from the literature are given in the last column.
}
\begin{tabular}{llllllrl} \hline \hline
BLoeM ID & Gaia DR3 ID & RA      & DEC     & G     & SpT & $\Delta$RV$_\mathrm{max}$ & literature \\
         &             & [J2000] & [J2000] & [mag] &     & [\kms]                     &         \\ \hline
\object{BLOeM 1-018} & 4690506025610129152 & 15.38575 & -72.1548056 & 14.52 & B2$\,$II:$\,$e & 7.0 $\pm$ 3.0 &  \\ 
\object{BLOeM 1-063} & 4690509289786901504 & 15.7671667 & -72.0950278 & 16.12 & B1.5$\,$II:$\,$e & 24.6 $\pm$ 7.4 &  \\ 
\object{BLOeM 1-073} & 4690526057340871040 & 15.8425 & -71.9436667 & 14.72 & B1$\,$II:$\,$e & 4.6 $\pm$ 2.7 &  \\ 
\object{BLOeM 1-075} & 4690526057340870272 & 15.8544167 & -71.9432778 & 13.67 & O9.2$\,$II:n$\,$pe & 16.5 $\pm$ 7.7 &  \\ 
\object{BLOeM 1-079} & 4690512828839860096 & 15.870125 & -72.0428611 & 14.02 & O9.2$\,$III:(n)$\,$pe & 18.2 $\pm$ 14.1 &  \\ 
\object{BLOeM 1-106} & 4687505801988031232 & 16.0892083 & -72.1541389 & 13.44 & B1.5$\,$$\,$e & 14.0 $\pm$ 5.2 &  \\ 
\object{BLOeM 2-017} & 4688982652297652352 & 12.7055833 & -72.55875 & 14.55 & B2$\,$II$\,$e & 8.9 $\pm$ 6.9 &  \\ 
\object{BLOeM 2-018} & 4688966090888390784 & 12.706875 & -72.7110556 & 14.13 & O6.5$\,$III:$\,$e? & 7.8 $\pm$ 2.4 &  \\ 
\object{BLOeM 2-055} & 4688984061034529408 & 12.9714167 & -72.5301944 & 14.86 & O9.7$\,$V:n$\,$e & 38.0 $\pm$ 10.1 & HMXB$^{1,3,4,5,6}$ \\ 
\object{BLOeM 2-061} & 4688961894666576768 & 13.027375 & -72.82975 & 14.84 & B2$\,$II:$\,$e & 10.7 $\pm$ 5.2 &  \\ 
\object{BLOeM 2-063} & 4688963204681907328 & 13.0646667 & -72.7644444 & 14.69 & B2$\,$III$\,$e & 7.4 $\pm$ 2.8 &  \\ 
\object{BLOeM 2-081} & 4688968461709335040 & 13.216625 & -72.5858056 & 14.45 & B2$\,$II$\,$e & 19.8 $\pm$ 6.3 &  \\ 
\object{BLOeM 2-104} & 4688987252155037568 & 13.3747917 & -72.6957222 & 14.07 & O5.5$\,$f?$\,$pe & 11.9 $\pm$ 4.3 &  \\ 
\object{BLOeM 3-011} & 4685854369881566848 & 11.715875 & -73.1590556 & 14.46 & B0.5$\,$III:$\,$e & 19.1 $\pm$ 4.5 &  \\ 
\object{BLOeM 3-016} & 4685849044122619264 & 11.7389583 & -73.3024444 & 16.18 & B2$\,$III$\,$e & 23.7 $\pm$ 10.3 &  \\ 
\object{BLOeM 3-018} & 4685836635962991104 & 11.7675833 & -73.3698056 & 14.39 & B1.5:$\,$II:$\,$e & 15.1 $\pm$ 6.2 & EB\,\textsuperscript{*} \\ 
\object{BLOeM 3-025} & 4685850242400168960 & 11.8104583 & -73.2306111 & 14.71 & B2$\,$II$\,$e & 6.9 $\pm$ 5.5 &  \\ 
\object{BLOeM 3-027} & 4685948515603957376 & 11.841 & -73.1140278 & 13.94 & B1.5$\,$II$\,$e & 8.6 $\pm$ 4.5 &  \\ 
\object{BLOeM 3-035} & 4685835884339216512 & 11.9025833 & -73.3803611 & 15.88 & B2$\,$II$\,$e & 12.0 $\pm$ 8.5 &  \\ 
\object{BLOeM 3-040} & 4685942949325301120 & 11.949375 & -73.2571667 & 14.45 & B2$\,$II:$\,$e+ & 23.5 $\pm$ 10.3 &  \\ 
\object{BLOeM 3-062} & 4685948064629440768 & 12.100625 & -73.1136389 & 14.05 & O9.7$\,$II$\,$e & 17.1 $\pm$ 3.1 &  \\ 
\object{BLOeM 3-077} & 4685947828409081088 & 12.2034167 & -73.1056944 & 14.34 & B1$\,$II$\,$e & 10.2 $\pm$ 5.0 &  \\ 
\object{BLOeM 3-095} & 4685925833826106368 & 12.3962917 & -73.3891111 & 15.83 & B2$\,$II:$\,$e & 70.0 $\pm$ 18.4 &  \\ 
\object{BLOeM 3-098} & 4685931679327948800 & 12.4280833 & -73.2883889 & 14.44 & O6$\,$V((f))n$\,$e & 34.0 $\pm$ 16.3 &  \\ 
\object{BLOeM 3-107} & 4685880277148119040 & 12.5162083 & -73.4038056 & 14.0 & O9.7$\,$II:(n)$\,$e? & 12.7 $\pm$ 4.1 &  \\ 
\object{BLOeM 4-002} & 4689016049958328576 & 14.4050833 & -72.21925 & 13.47 & B1.5:$\,$$\,$e & 11.6 $\pm$ 3.0 &  \\ 
\object{BLOeM 4-010} & 4689016221756877184 & 14.5416667 & -72.1838889 & 13.41 & B1$\,$$\,$e+ & 3.1 $\pm$ 1.2 &  \\ 
\object{BLOeM 4-022} & 4689015500202439296 & 14.6777917 & -72.2128056 & 14.86 & B2.5$\,$II:$\,$e & 27.1 $\pm$ 10.6 &  \\ 
\object{BLOeM 4-026} & 4689015328403760128 & 14.696125 & -72.2171389 & 14.55 & O9.5$\,$III$\,$pe & 14.6 $\pm$ 7.4 & HMXB$^{1,6}$ \\ 
\object{BLOeM 4-034} & 4689015736378946688 & 14.7764583 & -72.1655833 & 14.2 & B2$\,$II$\,$pe & 14.8 $\pm$ 4.7 &  \\ 
\object{BLOeM 4-035} & 4689016771512472960 & 14.7767083 & -72.1291667 & 14.86 & B3$\,$II:$\,$e & 15.4 $\pm$ 6.2 &  \\ 
\object{BLOeM 4-039} & 4690521006458654464 & 14.7915417 & -72.09675 & 14.42 & O6.5$\,$f?$\,$pe & 4.0 $\pm$ 2.5 &  \\ 
\object{BLOeM 4-071} & 4690504926098661504 & 14.9639167 & -72.1912778 & 14.31 & O9.7$\,$II:$\,$e? & 8.3 $\pm$ 2.6 &  \\ 
\object{BLOeM 5-020} & 4687486526172590720 & 16.1969583 & -72.43325 & 14.55 & B1.5$\,$II$\,$e & 11.2 $\pm$ 6.6 &  \\ 
\object{BLOeM 5-023} & 4687487247726980864 & 16.2099167 & -72.3830278 & 14.66 & B1.5$\,$III$\,$e & 5.7 $\pm$ 2.8 &  \\ 
\object{BLOeM 5-031} & 4687487007208845056 & 16.2723333 & -72.3996667 & 14.39 & B2.5:$\,$II:$\,$e & 9.5 $\pm$ 3.9 &  \\ 
\object{BLOeM 5-034} & 4687486938489379328 & 16.296875 & -72.4046389 & 14.19 & B1.5$\,$II:$\,$e & 13.0 $\pm$ 5.2 &  \\ 
\object{BLOeM 5-046} & 4687483777393718400 & 16.423125 & -72.4344444 & 13.95 & B1.5$\,$II:$\,$e & 13.3 $\pm$ 7.2 &  \\ 
\object{BLOeM 5-047} & 4687502228575362688 & 16.4250833 & -72.2226111 & 15.14 & B2.5$\,$III:$\,$e & 5.9 $\pm$ 4.9 &  \\ 
\object{BLOeM 5-066} & 4687489824707171456 & 16.5262917 & -72.3009722 & 13.74 & B2$\,$II:$\,$e+ & 10.8 $\pm$ 1.7 &  \\ 
\object{BLOeM 5-089} & 4687437494823799168 & 16.7412083 & -72.4588611 & 15.55 & B2$\,$III:$\,$e & 24.1 $\pm$ 8.5 &  \\ 
\object{BLOeM 5-092} & 4687437529183051136 & 16.7741667 & -72.4521667 & 14.2 & B1$\,$II:$\,$pe & 16.6 $\pm$ 3.0 &  \\ 
\object{BLOeM 5-094} & 4687485976416952832 & 16.7946667 & -72.343 & 15.07 & B1$\,$$\,$e & 12.1 $\pm$ 3.7 &  \\ 
\object{BLOeM 6-011} & 4687160280440948608 & 18.412 & -73.2968333 & 15.42 & B2$\,$III:$\,$e & 15.7 $\pm$ 6.1 &  \\ 
\object{BLOeM 6-113} & 4686415430023013760 & 19.3270417 & -73.2979444 & 14.44 & O$\,$$\,$nnpe & 14.3 $\pm$ 14.7 &  \\ 
\object{BLOeM 7-037} & 4685974457212254336 & 14.72675 & -72.7313056 & 15.02 & B1.5$\,$III:$\,$e & 15.7 $\pm$ 5.7 &  \\ 
\object{BLOeM 7-038} & 4685975213126426368 & 14.7310833 & -72.717 & 14.65 & B0$\,$II:$\,$e & 7.9 $\pm$ 2.7 &  \\ 
\object{BLOeM 7-047} & 4685975625443431808 & 14.8423333 & -72.7536667 & 14.75 & B0.2$\,$III:$\,$pe & 16.6 $\pm$ 6.4 &  \\ 
\object{BLOeM 7-078} & 4687491886293188608 & 15.1855 & -72.4973056 & 13.85 & B0$\,$$\,$e & 13.3 $\pm$ 4.6 &  \\ 
\object{BLOeM 7-093} & 4687477592639651072 & 15.3257083 & -72.6415 & 15.47 & B2$\,$II$\,$e & 13.1 $\pm$ 7.0 &  \\ 
\object{BLOeM 8-035} & 4689074461499247744 & 13.0714167 & -72.1463333 & 12.99 & O9.7$\,$III:$\,$e & 10.3 $\pm$ 5.0 &  \\ 
\object{BLOeM 8-043} & 4689057934463327616 & 13.1410417 & -72.2833056 & 14.91 & B2$\,$II:$\,$e+ & 12.5 $\pm$ 7.4 &  \\ 
\object{BLOeM 8-064} & 4689056628793468928 & 13.3026667 & -72.3415833 & 15.22 & B3$\,$II$\,$e & 16.3 $\pm$ 4.1 &  \\ 
\object{BLOeM 8-100} & 4689009070622046336 & 13.5451667 & -72.3691389 & 15.46 & B2$\,$II$\,$e & 17.1 $\pm$ 7.7 &  \\ 
\object{BLOeM 8-109} & 4689011754988965888 & 13.5939167 & -72.2859444 & 14.1 & B0$\,$III:$\,$pe & 22.4 $\pm$ 12.7 &  \\   \hline
\end{tabular}     
\end{center}
\footnotesize{\textsuperscript{*} classification based on OGLE-III photometry from \citet{Pawlak2013}. X-ray source identifications are from $^1$: \citet{cscv2}, $^2$: \citet{Sturm2013} , $^3$: \citet{CXOGSGSRC}, $^4$ \citet{Laycock2010}, $^5$ \citet{Traulsen2020}, $^6$\citet{Antoniou2016}} \\
\end{table*}

\section*{Appendix E: Notes on individual targets classified as binaries}\label{sec:indiv_bin}

\begin{figure*}
    \centering
    \begin{subfigure}{.5\textwidth} \centering
    \includegraphics[width=\linewidth]{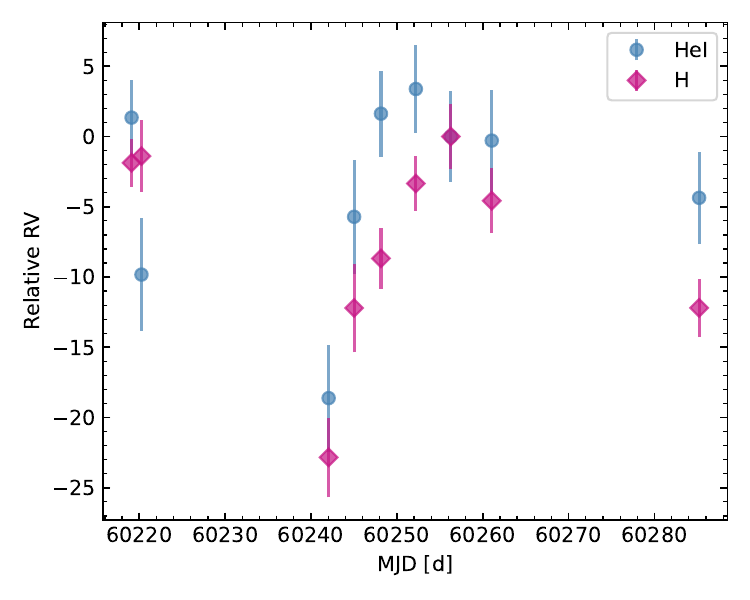}
    \newsubcap{RV curve of BLOeM\,1-040.}
    \label{fig:rvcurve_1-040}
    \end{subfigure}%
    \begin{subfigure}{.5\textwidth} \centering
    \includegraphics[width=\linewidth]{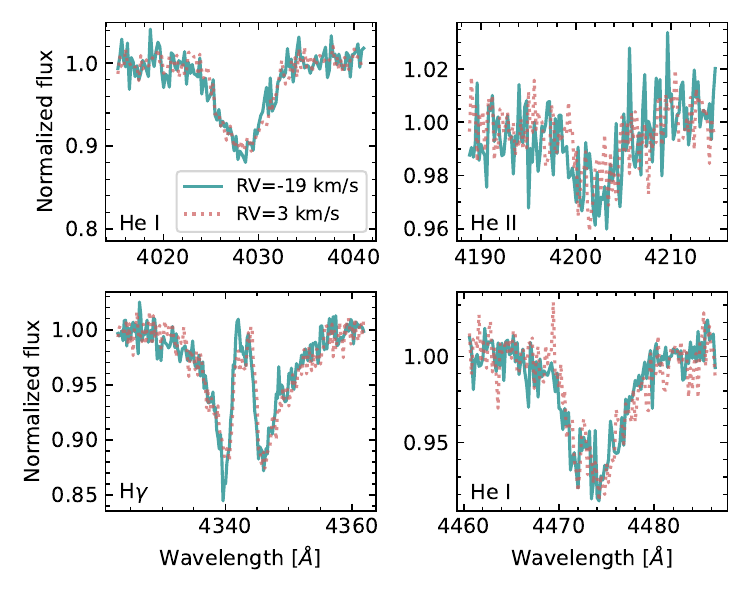}
    \newsubcap{RV extremes of BLOeM\,1-040.}
    \label{fig:rvextr_1-040}
    \end{subfigure}
\end{figure*}

\paragraph{BLOeM 1-040:} BLOeM 1-040 is an O9.7\,III:n\,e star according to \citetalias{Shenar2024}. Using all available \ion{He}{i} lines to measure RVs, we find that two epochs of the target meet both binary criteria described in Sect.\,3.2. The RV threshold of 20\,\kms\,is barely met, with a $\Delta$RV$_\mathrm{max}$ = 22\,\kms (see Fig.\,\ref{fig:rvcurve_1-040}). Given their broadening and the low S/N ratio, we refrain from measuring the RVs of the \ion{He}{ii} lines. The double-peaked emission lines follow the same trend in RVs than the \ion{He}{i} absorption lines and also indicate a binary with a period of approximately 45 days. The shift in RV measured in \ion{He}{i} and Balmer lines is within the uncertainty of the RVs measured in the reference epoch (epoch 7).

\begin{figure*}
    \centering
    \begin{subfigure}{.5\textwidth} \centering
    \includegraphics[width=\linewidth]{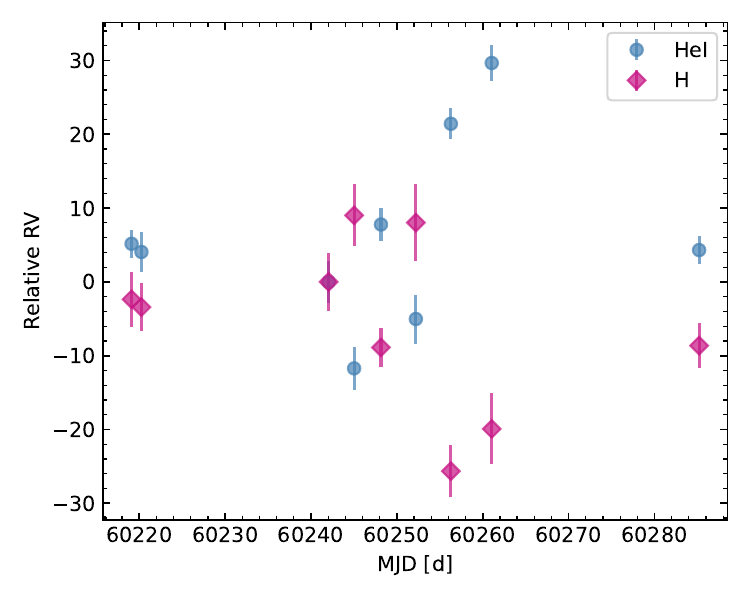}
    \newsubcap{RV curve of BLOeM\,1-113.}
    \label{fig:rvcurve_1-113}
    \end{subfigure}%
    \begin{subfigure}{.5\textwidth} \centering
    \includegraphics[width=\linewidth]{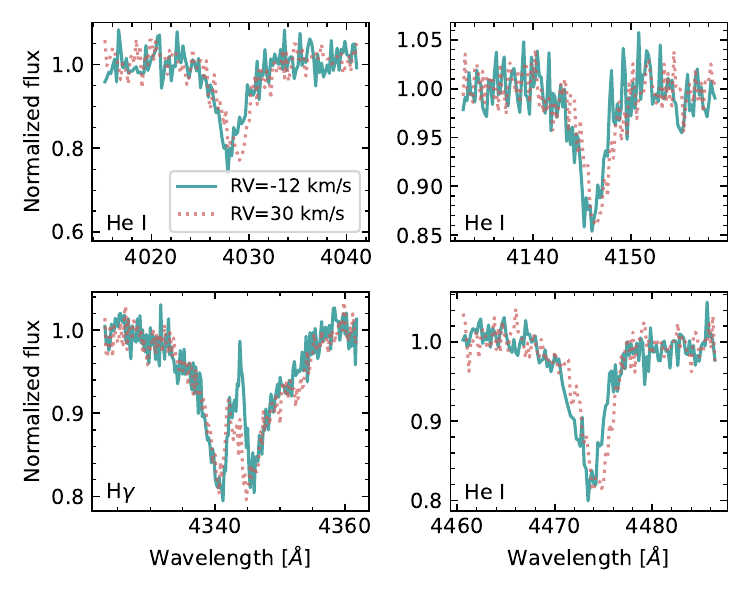}
    \newsubcap{RV extremes of BLOeM\,1-113.}
    \label{fig:rvextr_1-113}
    \end{subfigure}
\end{figure*}

\paragraph{BLOeM 1-113:} BLOeM 1-113 was classified as B1\,II\,e in \citetalias{Shenar2024}. It shows the signature of an SB2 with the narrow, non-variable, clean \ion{He}{i} absorption lines moving in apparent anti-phase with the weak emission lines, which additionally show intrinsic variability. The maximum RV amplitude $\Delta$RV$_\mathrm{max}$ is similar for both absorption and emission lines and is around 40\,\kms. The RV curve and spectra at RV extremes are shown in Figures \ref{fig:rvcurve_1-113} and \ref{fig:rvextr_1-113}.

\begin{figure*}
    \centering
    \begin{subfigure}{.5\textwidth} \centering
    \includegraphics[width=\linewidth]{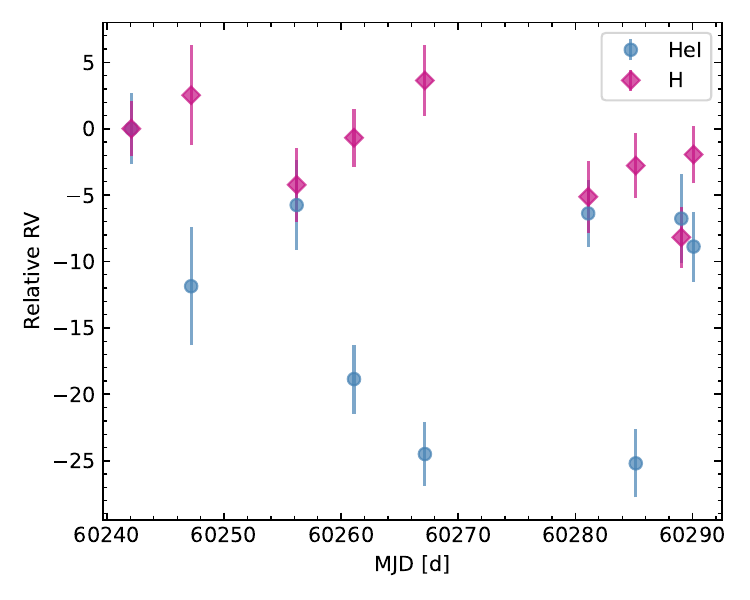}
    \newsubcap{RV curve of BLOeM\,2-021.}
    \label{fig:rvcurve_2-021}
    \end{subfigure}%
    \begin{subfigure}{.5\textwidth} \centering
    \includegraphics[width=\linewidth]{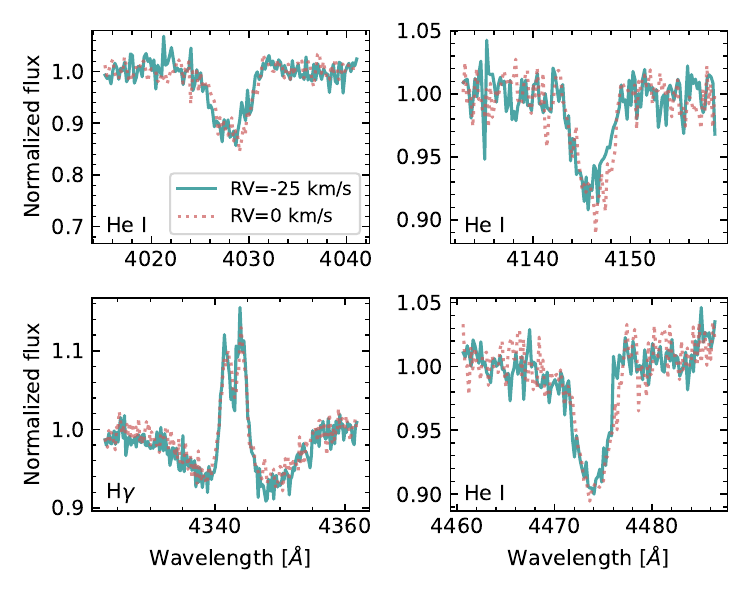}
    \newsubcap{RV extremes of BLOeM\,2-021.}
    \label{fig:rvextr_2-021}
    \end{subfigure}
\end{figure*}

\paragraph{BLOeM 2-021:} This target was assigned a spectral type of B2.5:\,II:\,e (\citetalias{Shenar2024}). Here, based on the \ion{He}{i} lines, we classify the star as binary with a $\Delta$RV$_\mathrm{max}$ of 25\,\kms\ (Fig.\,\ref{fig:rvcurve_2-021}). The emission lines, which are double-peaked and strong in particular in H$\gamma$ (Fig.\,\ref{fig:rvextr_2-021}) do not show any significant RV variations. The slight trend visible in anti-phase especially visible in epochs 3-5 might be spurious, and will be further investigated with additional epochs. 

\begin{figure*}
    \centering
    \begin{subfigure}{.5\textwidth} \centering
    \includegraphics[width=\linewidth]{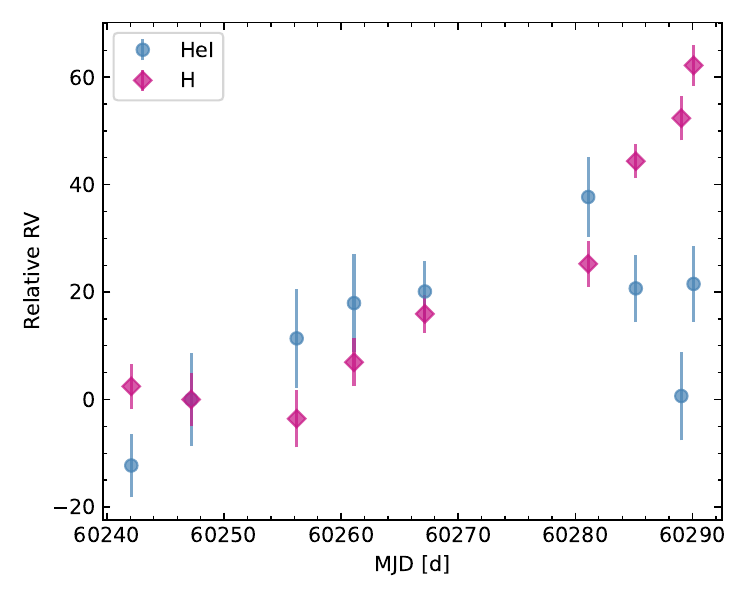}
    \newsubcap{RV curve of BLOeM\,2-066.}
    \label{fig:rvcurve_2-066}
    \end{subfigure}%
    \begin{subfigure}{.5\textwidth} \centering
    \includegraphics[width=\linewidth]{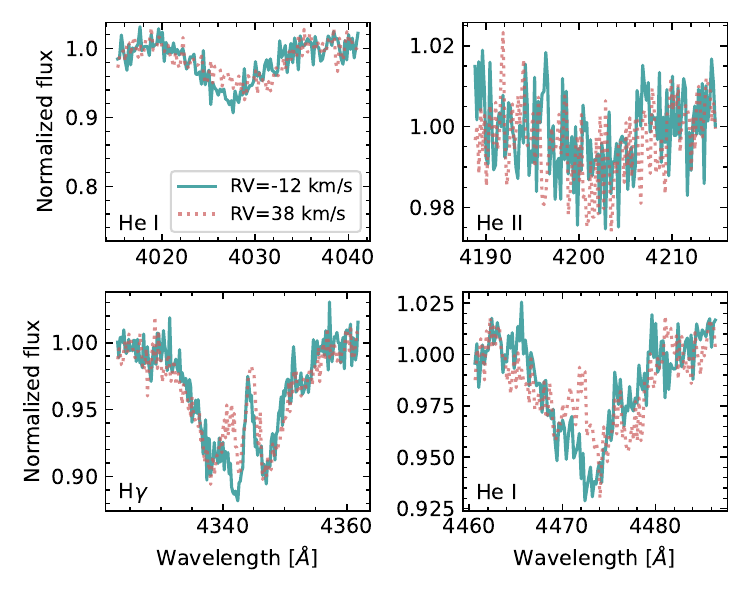}
    \newsubcap{RV extremes of BLOeM\,2-066.}
    \label{fig:rvextr_2-066}
    \end{subfigure}
\end{figure*}

\paragraph{BLOeM 2-066:} The spectral type assigned to this object in \citetalias{Shenar2024} is O9.7\,III:nnn\,pe+, indicating that the object is a rapid rotator. The spectra of BLOeM\,2-066 are on average fairly noisy. The \ion{He}{i} lines show a signature that could be interpreted as SB2, while the \ion{He}{ii} are too broad and noisy to measure reliable RVs. The emission component is a complicated superposition of several components, with the strongest clearly shifting in wavelength (see Fig.\,\ref{fig:rvcurve_2-066}). With a $\Delta$RV$_\mathrm{max}$ of 50\,\kms\ in \ion{He}{i} and almost 60\,\kms\ in H$\gamma$ and the potential SB2 signature, BLOeM 2-066 is definitely a binary. To measure the RVs more accurately, as well as to assess whether it is a genuine SB2 requires more observations. The tentative RV curve and spectra at RV extremes are shown in Figures \ref{fig:rvcurve_2-066} and \ref{fig:rvextr_2-066}, which have to be taken with a grain of salt.

\begin{figure*}
    \centering
    \begin{subfigure}{.5\textwidth} \centering
    \includegraphics[width=\linewidth]{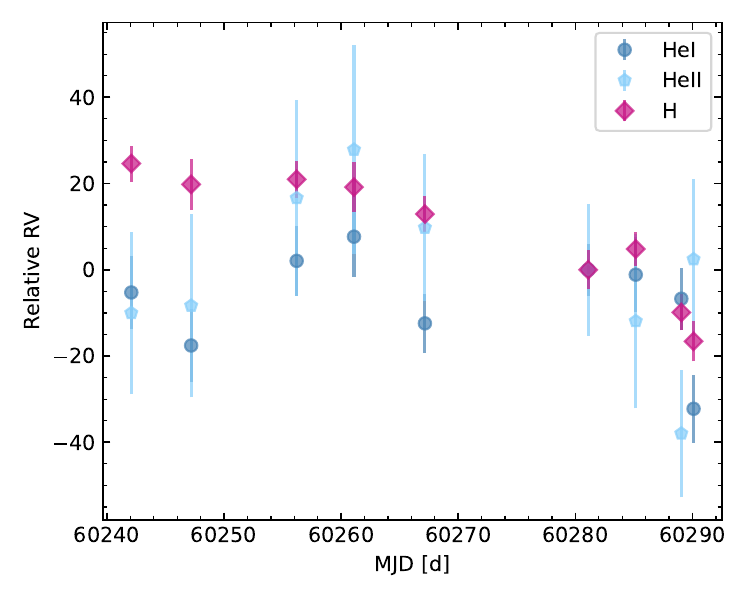}
    \newsubcap{RV curve of BLOeM\,2-082.}
    \label{fig:rvcurve_2-082}
    \end{subfigure}%
    \begin{subfigure}{.5\textwidth} \centering
    \includegraphics[width=\linewidth]{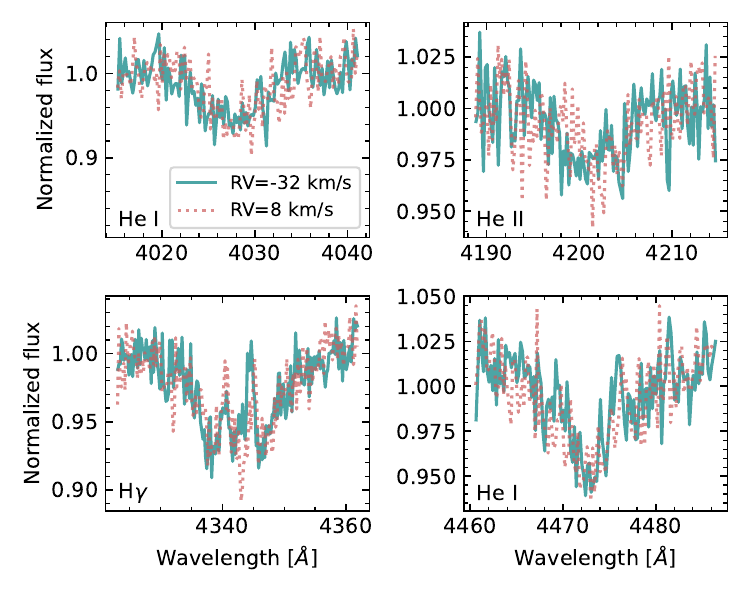}
    \newsubcap{RV extremes of BLOeM\,2-082.}
    \label{fig:rvextr_2-082}
    \end{subfigure}
\end{figure*}

\paragraph{BLOeM 2-082:} BLOeM 2-082 is one of the Oe stars in the sample with a spectral type O9.2\,III\,pe (\citetalias{Shenar2024}). Given the overall low S/N in the spectra, the determined RVs, especially from the broad \ion{He}{i} and \ion{He}{ii} lines, have large error bars (see Fig.\,\ref{fig:rvcurve_2-082}). The double-peaked Balmer emission lines, however, show a clear trend over time, with a peak-to-peak RV amplitude of almost 45\,\kms. Based on the large amplitude and the clear trend in RVs, we classify BLOeM\,2-082 as a binary. It was detected as a binary already in the RIOTS survey \citep{Lamb2016}, and coincides with an X-ray source \citep{cscv2, Antoniou2019}, making it a HMXB.

\begin{figure*}
    \centering
    \begin{subfigure}{.5\textwidth} \centering
    \includegraphics[width=\linewidth]{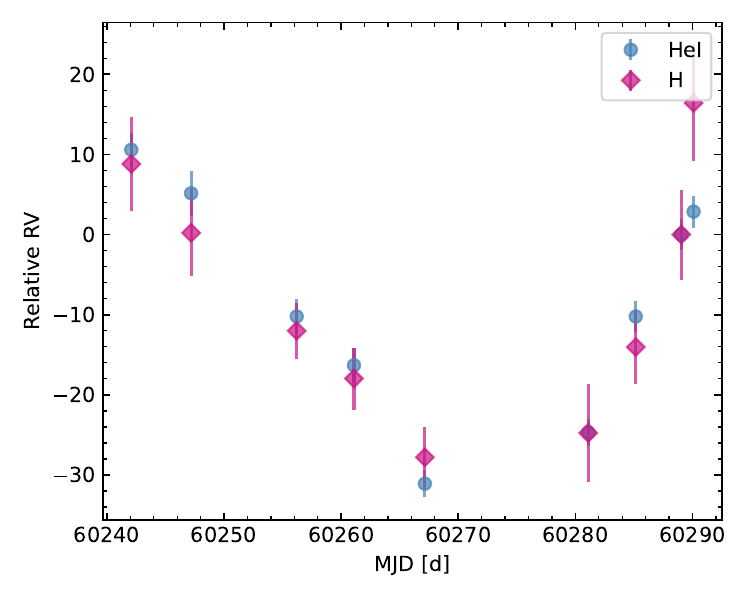}
    \newsubcap{RV curve of BLOeM\,2-109.}
    \label{fig:rvcurve_2-109}
    \end{subfigure}%
    \begin{subfigure}{.5\textwidth} \centering
    \includegraphics[width=\linewidth]{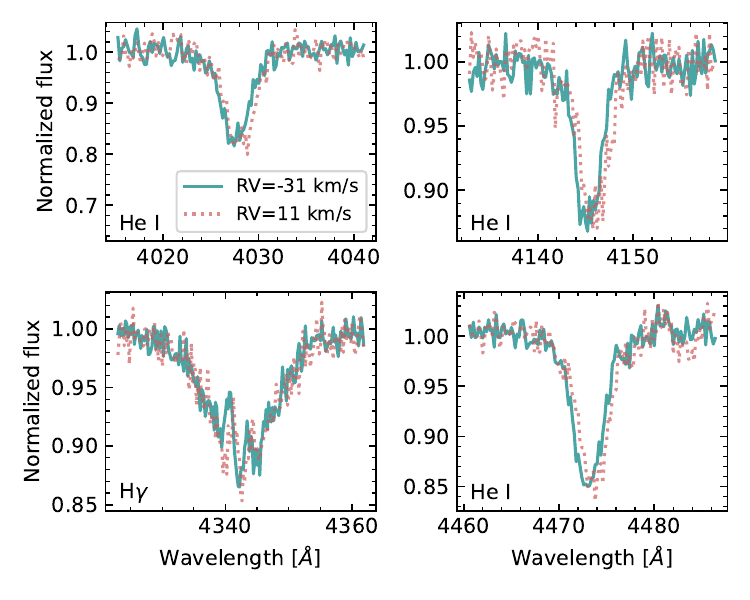}
    \newsubcap{RV extremes of BLOeM\,2-109.}
    \label{fig:rvextr_2-109}
    \end{subfigure}
\end{figure*}

\paragraph{BLOeM 2-109:} This target shows comparatively narrow \ion{He}{i} absorption lines (see Fig.\,\ref{fig:rvextr_2-109}) and was classified as B2\,II\,e (\citetalias{Shenar2024}). The RVs measured from the absorption lines show a clear sinusoidal trend with a maximum amplitude of around 40\,\kms\ (see Fig.\,\ref{fig:rvcurve_2-109}) the observed time seems to cover almost exactly half of the binary period, indicating that the period of the binary system is around 60 days. The RVs measured from the weak, double-peaked emission line in H$\gamma$ follow these measurements closely.

\begin{figure*}
    \centering
    \begin{subfigure}{.5\textwidth} \centering
    \includegraphics[width=\linewidth]{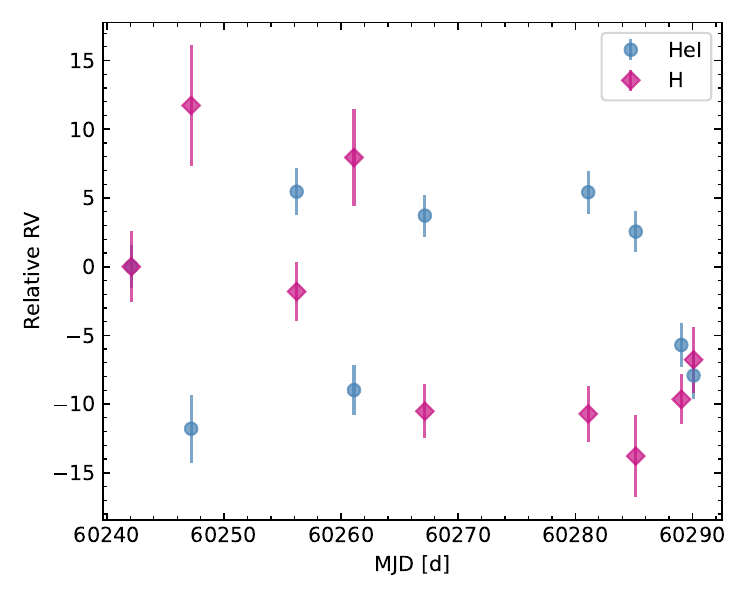}
    \newsubcap{RV curve of BLOeM\,2-111.}
    \label{fig:rvcurve_2-111}
    \end{subfigure}%
    \begin{subfigure}{.5\textwidth} \centering
    \includegraphics[width=\linewidth]{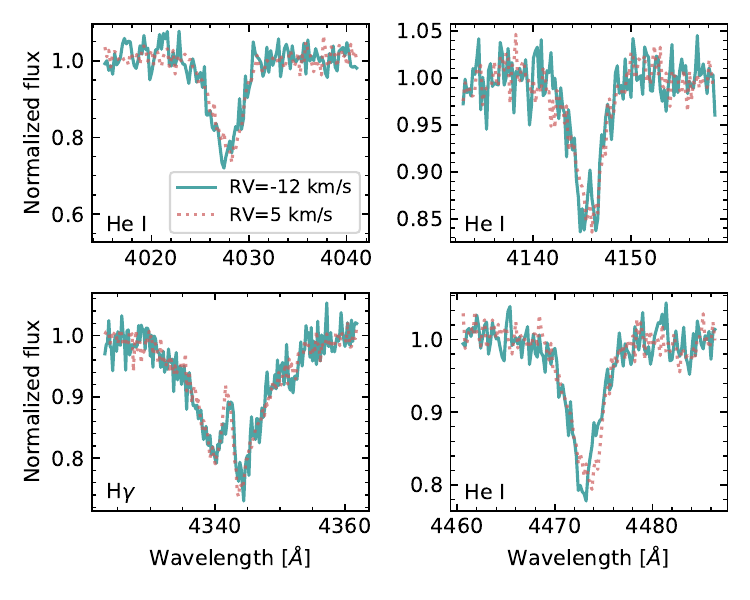}
    \newsubcap{RV extremes of BLOeM\,2-111.}
    \label{fig:rvextr_2-111}
    \end{subfigure}
\end{figure*}

\paragraph{BLOeM 2-111:} This target shows only weak emission lines with a single peak. The RVs measured from narrow \ion{He}{i} lines and the weak emission in H$\gamma$ indicate anti-phase motion, which is why we classify it as SB2. The peak-to-peak RV amplitude of 17\,\kms\ measured in the \ion{He}{i} lines is below the imposed threshold of 20\,\kms. While the one measured in H$\gamma$ is larger, it is not significant due to the larger error bars. The RV curve and spectra at RV extremes are shown in Figs.\,\ref{fig:rvcurve_2-111} and \ref{fig:rvextr_2-111}.

\begin{figure*}
    \centering
    \begin{subfigure}{.5\textwidth} \centering
    \includegraphics[width=\linewidth]{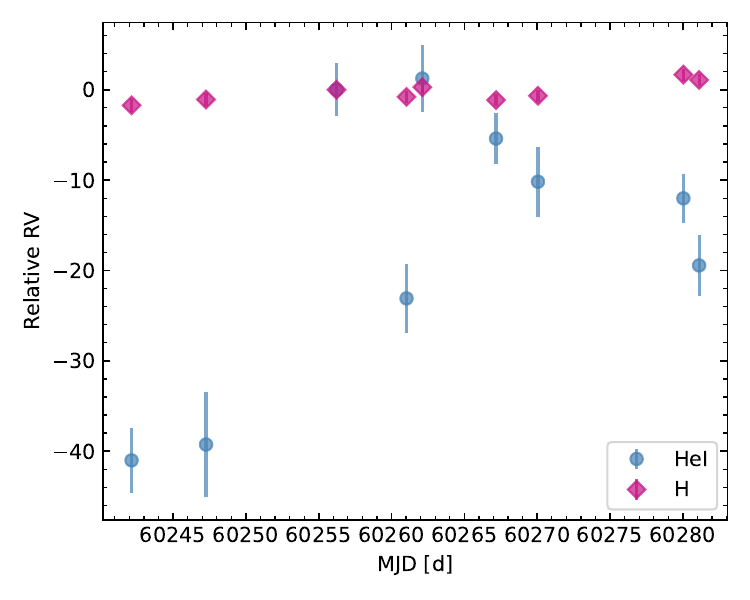}
    \newsubcap{RV curve of BLOeM\,5-069.}
    \label{fig:rvcurve_5-069}
    \end{subfigure}%
    \begin{subfigure}{.5\textwidth} \centering
    \includegraphics[width=\linewidth]{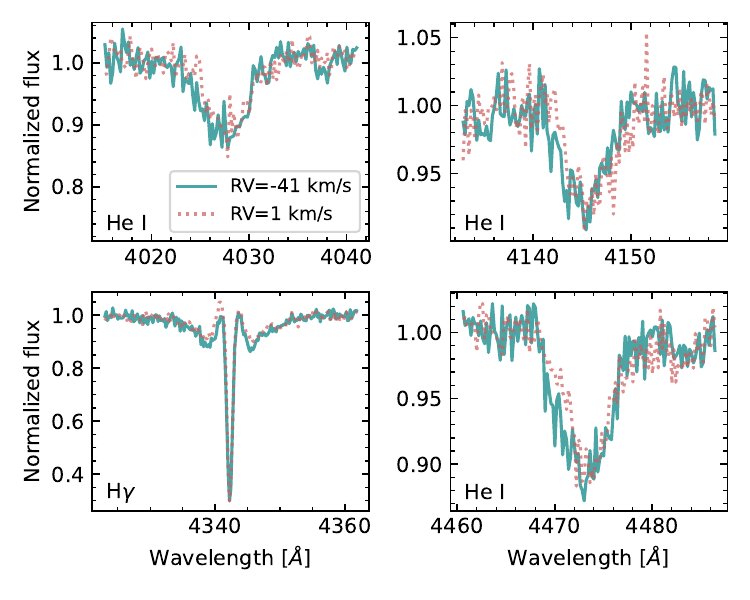}
    \newsubcap{RV extremes of BLOeM\,5-069.}
    \label{fig:rvextr_5-069}
    \end{subfigure}
\end{figure*}

\paragraph{BLOeM 5-069:} This object was classified as B3\,II\,e and shows the typical spectrum of a shell Be star (i.e., broad absorption lines, with additional emission components in the Balmer lines superimposed with narrow absorption features, see Fig.\,\ref{fig:rvextr_5-069}). Those features are interpreted as a Be star viewed almost equator-on, where the narrow absorption component arises in the disk along the line of sight. The \ion{He}{i} lines indicate BLOeM\,5-069 to be a binary, with a maximum RV amplitude of 42\,\kms. The RVs of the emission component alone are difficult to measure and dominated by the stationary strong absorption. 

\begin{figure*}
    \centering
    \begin{subfigure}{.5\textwidth} \centering
    \includegraphics[width=\linewidth]{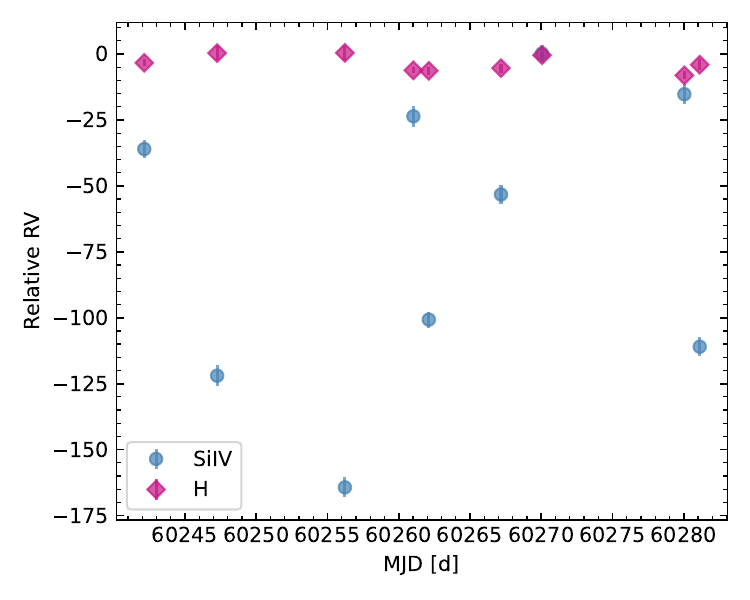}
    \newsubcap{RV curve of BLOeM\,5-071.}
    \label{fig:rvcurve_5-071}
    \end{subfigure}%
    \begin{subfigure}{.5\textwidth} \centering
    \includegraphics[width=\linewidth]{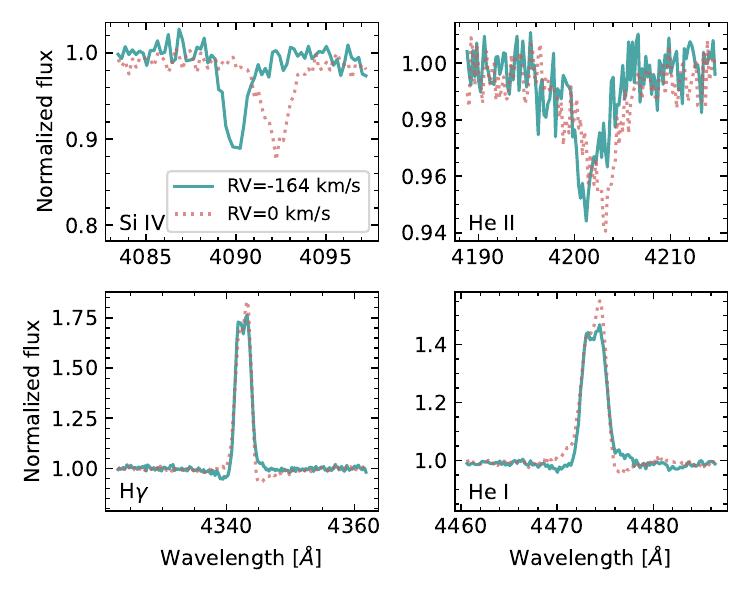}
    \newsubcap{RV extremes of BLOeM\,5-071.}
    \label{fig:rvextr_5-071}
    \end{subfigure}
\end{figure*}

\paragraph{BLOeM\,5-071:} For this particular target, classified as SB2 with an O8.5:\,Ib and an OB\,pe component in \citetalias{Shenar2024}, we use the \ion{Si}{iv} line at 4088.3\,\r{A} to estimate RVs as the line is clean and not variable. Most other lines (apart from the more noisy \ion{He}{ii} lines) are impacted by strong emission (see Fig.\,\ref{fig:rvextr_5-071}).  BLOeM 5-071 shows a similar signature as BLOeM 3-031, with strongly moving absorption lines ($\Delta$RV$_\mathrm{max}$\,=\,164\,\kms) and a fairly stable emission line component (Fig.\,\ref{fig:rvcurve_5-071}). Again, the formal RV errors are of the order of a few \kms\ but hardly visible due to the large amplitude of the RVs. The origin of the emission as well as the nature of the binary remains to be determined. 

\begin{figure*}
    \centering
    \begin{subfigure}{.5\textwidth} \centering
    \includegraphics[width=\linewidth]{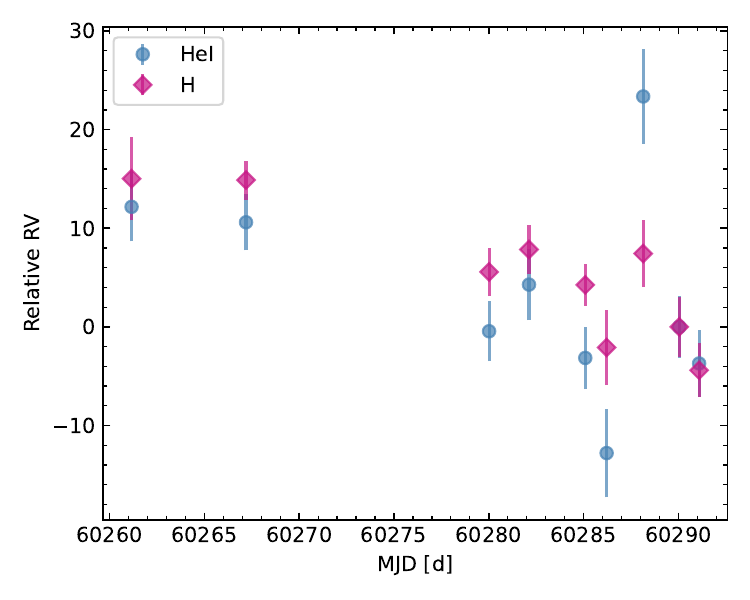}
    \newsubcap{RV curve of BLOeM\,6-001.}
    \label{fig:rvcurve_6-001}
    \end{subfigure}%
    \begin{subfigure}{.5\textwidth} \centering
    \includegraphics[width=\linewidth]{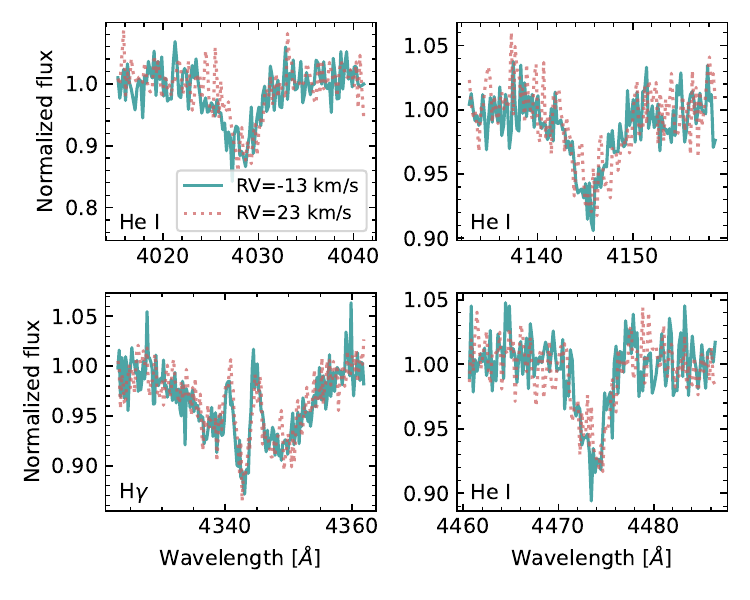}
    \newsubcap{RV extremes of BLOeM\,6-001.}
    \label{fig:rvextr_6-001}
    \end{subfigure}
\end{figure*}

\paragraph{BLOeM 6-001:} This object was classified as B0.5\,III\:,e in \citetalias{Shenar2024}. Based on the two above mentioned criteria and the RVs measured from the \ion{He}{i} lines, it is classified as binary. One epoch stands out (epoch 7, see Figs.\,\ref{fig:rvcurve_6-001} and \ref{fig:rvextr_6-001}), which also corresponds to the epoch with the lowest S/N. To ensure that the binary classification is not solely based on one spurious, potentially low-quality measurement, we re-evaluate if the criteria are met when removing this particular epoch. Given that this is true, we classify BLOeM\,6-001 as a binary. The H$\gamma$ line indicates a similar trend, with a maximum amplitude $\Delta$RV$_\mathrm{max}$ just below the threshold of 20\,\kms.

\begin{figure*}
    \centering
    \begin{subfigure}{.5\textwidth} \centering
    \includegraphics[width=\linewidth]{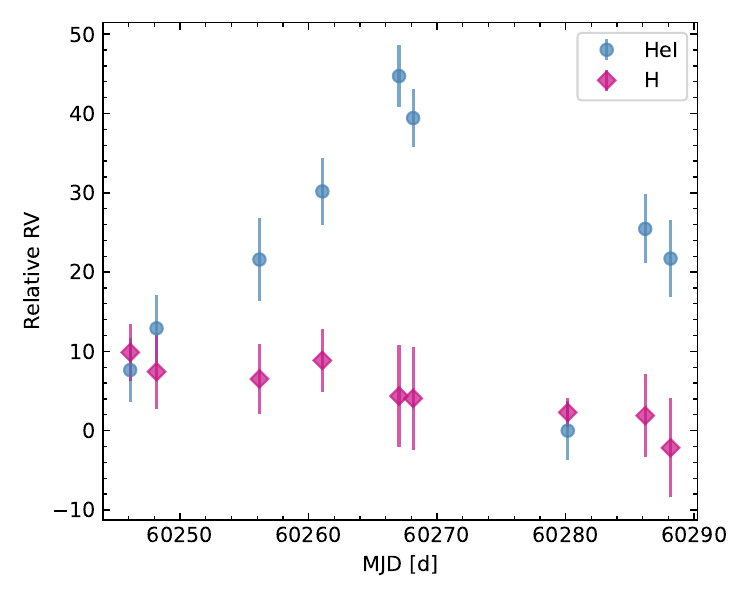}
    \newsubcap{RV curve of BLOeM\,7-013.}
    \label{fig:rvcurve_7-013}
    \end{subfigure}%
    \begin{subfigure}{.5\textwidth} \centering
    \includegraphics[width=\linewidth]{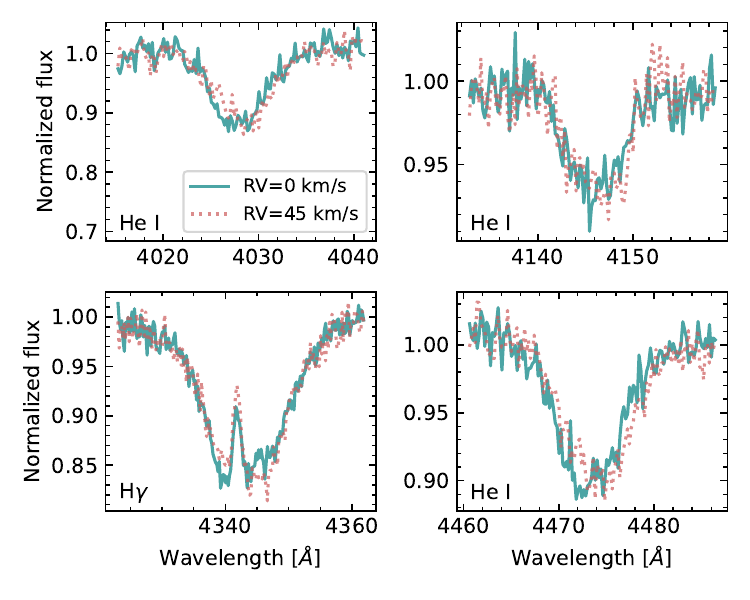}
    \newsubcap{RV extremes of BLOeM\,7-013.}
    \label{fig:rvextr_7-013}
    \end{subfigure}
\end{figure*}

\paragraph{BLOeM 7-013:} This system was classified as B1.5\,III\,e. The stable absorption lines show radial velocity variations with a maximum amplitude of 45\,\kms. The weak, single-peaked emission line shows no significant variation based on the epochs obtained so far. The emission feature is not located in the center of the line, and appears stable on the covered timescales, potentially implying that the feature stems from nebular contamination and the object is not a "true" OBe star. Figures \ref{fig:rvcurve_7-013} and \ref{fig:rvextr_7-013} show the RV curve and the spectra at the RV extremes.

\begin{figure*}
    \centering
    \begin{subfigure}{.5\textwidth} \centering
    \includegraphics[width=\linewidth]{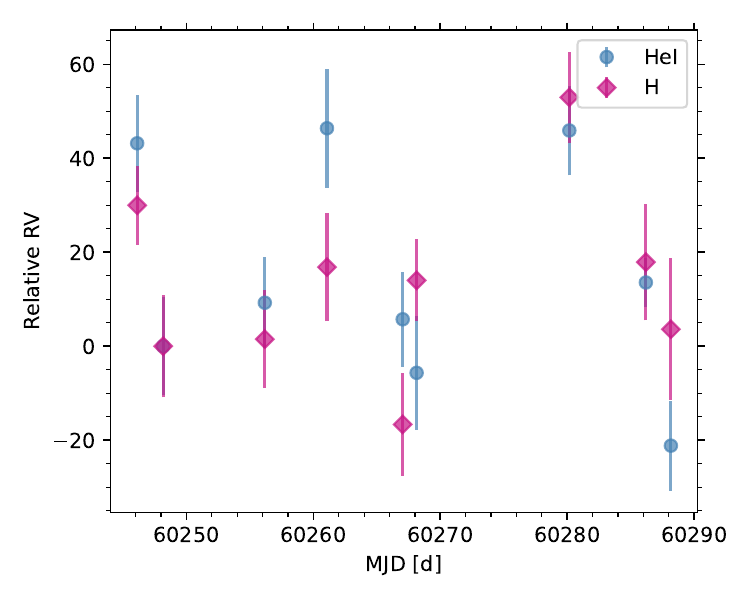}
    \newsubcap{RV curve of BLOeM\,7-045.}
    \label{fig:rvcurve_7-045}
    \end{subfigure}%
    \begin{subfigure}{.5\textwidth} \centering
    \includegraphics[width=\linewidth]{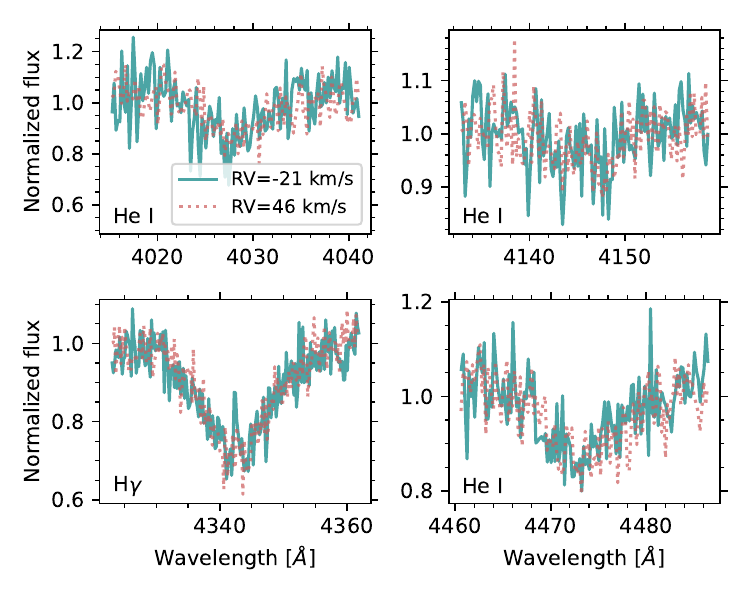}
    \newsubcap{RV extremes of BLOeM\,7-045.}
    \label{fig:rvextr_7-045}
    \end{subfigure}
\end{figure*}

\paragraph{BLOeM 7-045:} This rapidly rotating B2\,III:\,e star (see \citetalias{Shenar2024}) shows only weak emission in H$\gamma$. The broad and noisy lines make it difficult to measure RVs, and the estimated formal errors may be underestimated. Formally, the measured RVs fulfill both criteria with a peak-to-peak amplitude $>$65\,\kms, it is thus classified as binary. The RV curve as well as the spectra at RV extremes for BLOeM\,7-045 are shown in Figs.\,\ref{fig:rvcurve_7-045} and \ref{fig:rvextr_7-045}.

\paragraph{BLOeM 7-082:} This system, also called AzV\,261, was previously classified as blue supergiant and its spectroscopic and photometric variability was investigated by \citet{Kalari2018}, searching for additional luminous blue variable (LBV)-like stars. While the authors found strong photometric variability in this object, they concluded that BLOeM\,7-082 is not an LBV-like object as it does not show variations in its spectral type.\newline
In \citetalias{Shenar2024}, we classified the star as B2\,e+, based on narrow, clean \ion{He}{i} absorption lines as well as OBe-typical double-peaked emission lines. Based on the RVs measured from \ion{He}{i} lines, which show a maximum amplitude of 45\,\kms, we classify the system as binary. The strong but variable emission lines also show RV variations that appear to be in anti-phase with the absorption lines, with a similar amplitude. The binary status of this system might explain the photometric variability reported in \citet{Kalari2018}, which requires a more detailed investigation in the future.

\paragraph{BLOeM 7-114:} This target was classified as B2\,II\,e in \citetalias{Shenar2024}. Based on the RVs measured from the noisy but fairly narrow \ion{He}{i} lines (see Fig. \ref{fig:rvextr_7-114}) we classify BLOeM\,7-114 as binary. The RV curve (see Fig. \ref{fig:rvcurve_7-114}) shows RV amplitudes of up to almost 40\,\kms\ and indicates a period shorter than the three months covered by observations so far. Additional epochs will help to constrain this further.

\begin{figure*}
    \centering
    \begin{subfigure}{.5\textwidth} \centering
    \includegraphics[width=\linewidth]{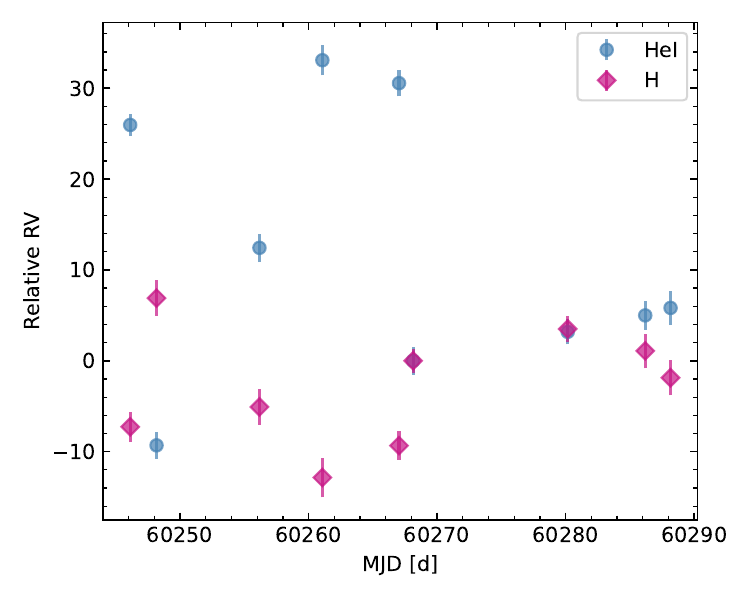}
    \newsubcap{RV curve of BLOeM\,7-082.}
    \label{fig:rvcurve_7-082}
    \end{subfigure}%
    \begin{subfigure}{.5\textwidth} \centering
    \includegraphics[width=\linewidth]{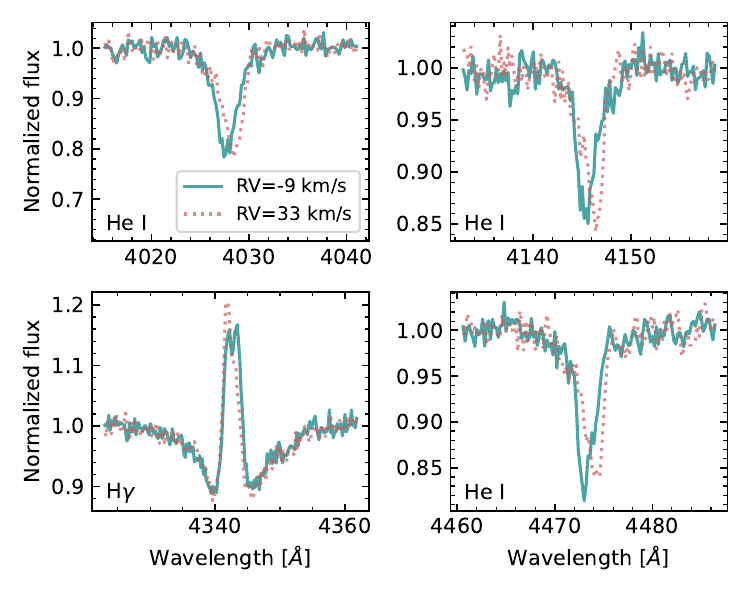}
    \newsubcap{RV extremes of BLOeM\,7-082.}
    \label{fig:rvextr_7-082}
    \end{subfigure}
\end{figure*}

\begin{figure*}
    \centering
    \begin{subfigure}{.5\textwidth} \centering
    \includegraphics[width=\linewidth]{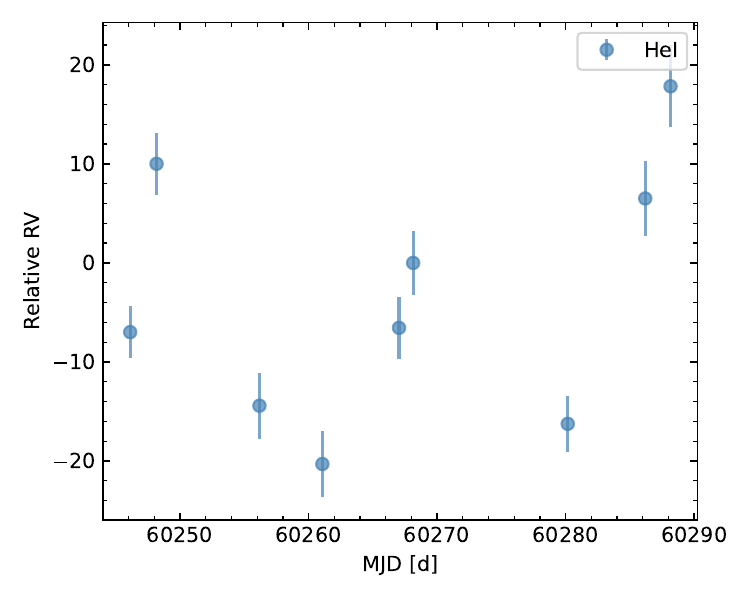}
    \newsubcap{RV curve of BLOeM\,7-114.}
    \label{fig:rvcurve_7-114}
    \end{subfigure}%
    \begin{subfigure}{.5\textwidth} \centering
    \includegraphics[width=\linewidth]{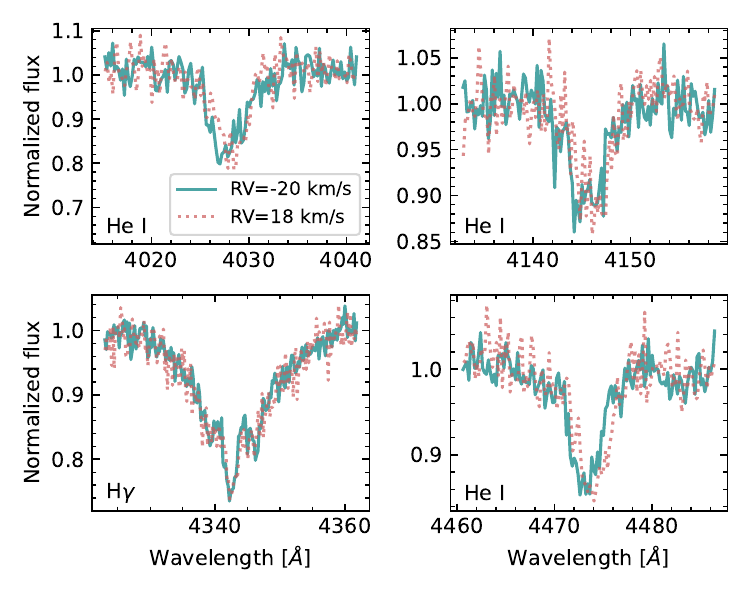}
    \newsubcap{RV extremes of BLOeM\,7-114.}
    \label{fig:rvextr_7-114}
    \end{subfigure}
\end{figure*}

\section*{Appendix F: Notes on candidate binaries}\label{sec:indiv_candidate}

\begin{figure*}
    \centering
    \begin{subfigure}{.5\textwidth} \centering
    \includegraphics[width=\linewidth]{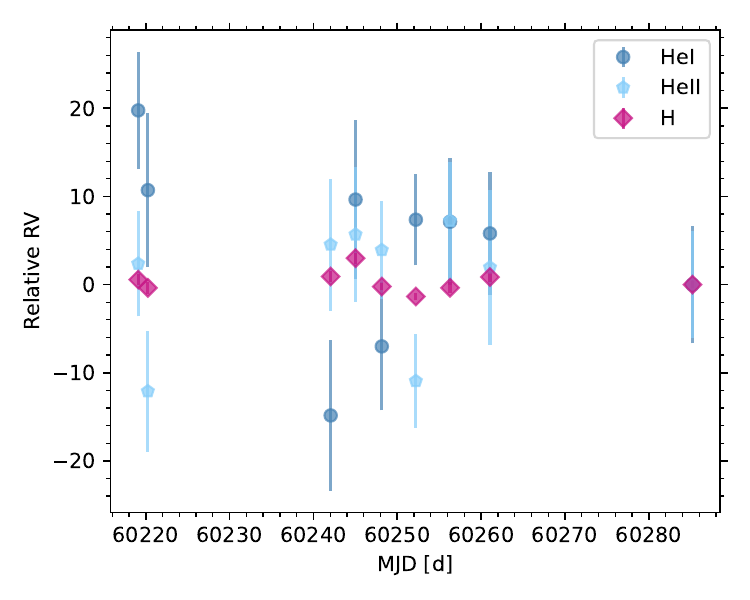}
    \newsubcap{RV curve of BLOeM\,1-058.}
    \label{fig:rvcurve_1-058}
    \end{subfigure}%
    \begin{subfigure}{.5\textwidth} \centering
    \includegraphics[width=\linewidth]{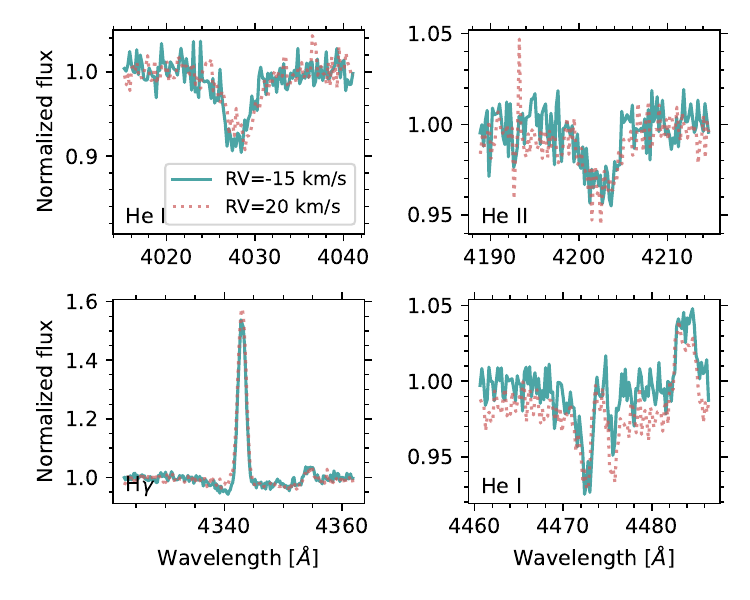}
    \newsubcap{RV extremes of BLOeM\,1-058.}
    \label{fig:rvextr_1-058}
    \end{subfigure}
\end{figure*}

\paragraph{BLOeM 1-058}: This object is one of the Oe stars, with spectral type O9.5\,II:\,pe (\citetalias{Shenar2024}). The overall spectra are quite noisy and due to the broad lines as well as strong emission lines, RVs of clean absorption lines are difficult to measure and have large error bars (see Fig.\,\ref{fig:rvextr_1-058}). While the RVs from emission lines as well as \ion{He}{ii} show no significant variability, the \ion{He}{i} absorption lines (the ones not impacted by the emission) fulfill the second binary criteria (Fig.\,\ref{fig:rvcurve_1-058}). However, given the low data quality as well as the mismatch between \ion{He}{i} and \ion{He}{ii} RVs we classify this object as candidate binary.

\begin{figure*}
    \centering
    \begin{subfigure}{.5\textwidth} \centering
    \includegraphics[width=\linewidth]{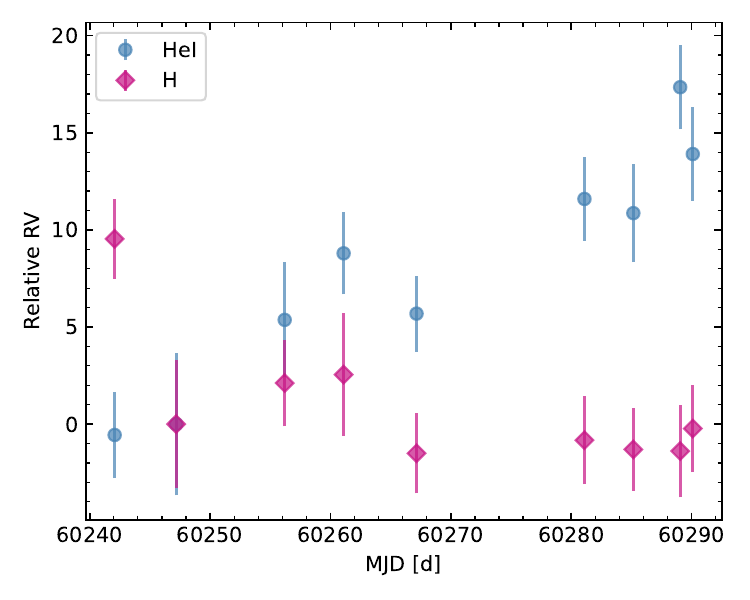}
    \newsubcap{RV curve of BLOeM\,2-056.}
    \label{fig:rvcurve_2-056}
    \end{subfigure}%
    \begin{subfigure}{.5\textwidth} \centering
    \includegraphics[width=\linewidth]{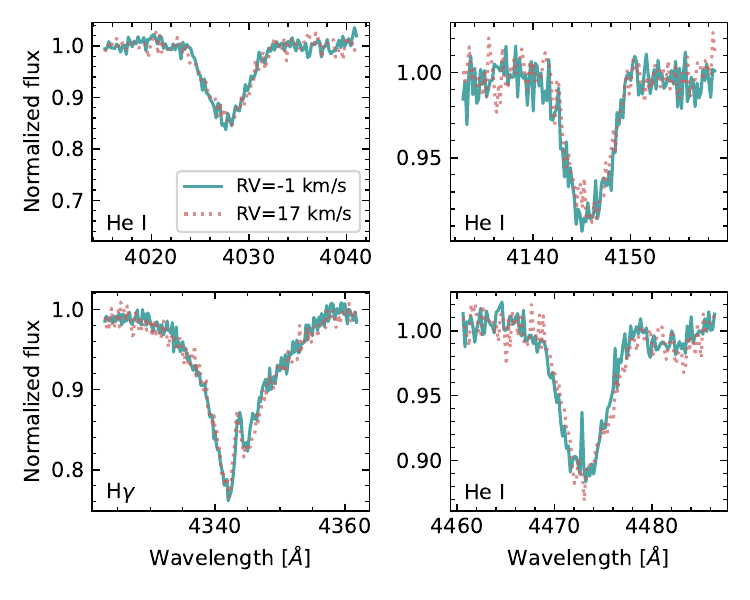}
    \newsubcap{RV extremes of BLOeM\,2-056.}
    \label{fig:rvextr_2-056}
    \end{subfigure}
\end{figure*}

\paragraph{BLOeM 2-056:} This target, a B2\,II:\,e star according to \citetalias{Shenar2024}, shows RV variations with a maximum amplitude of approximately 15\,\kms\ (see Fig.\,\ref{fig:rvextr_2-056}). As the measured RVs seem to indicate a long-term anti-phase trend between Balmer and \ion{He}{i} that is reminiscent of a binary orbit (Fig.\,\ref{fig:rvcurve_2-056}), we classify this object as candidate binary that requires additional observations to be confirmed.

\begin{figure*}
    \centering
    \begin{subfigure}{.5\textwidth} \centering
    \includegraphics[width=\linewidth]{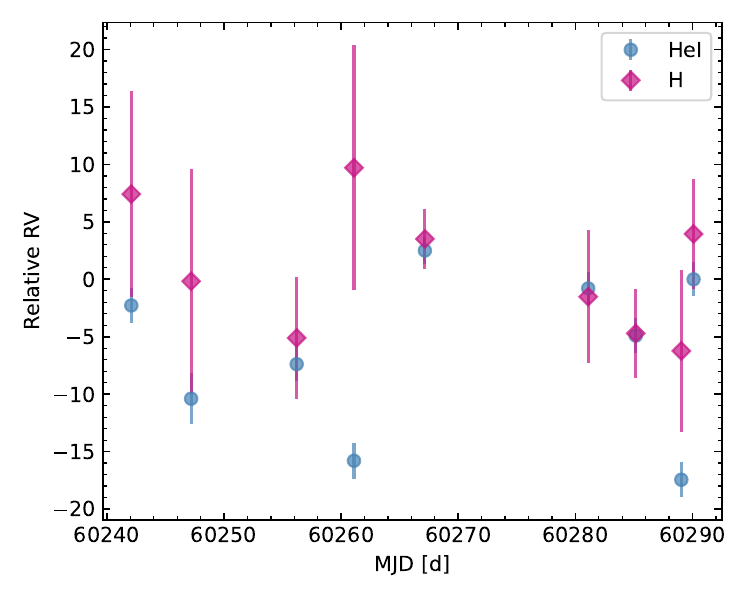}
    \newsubcap{RV curve of BLOeM\,2-070.}
    \label{fig:rvcurve_2-070}
    \end{subfigure}%
    \begin{subfigure}{.5\textwidth} \centering
    \includegraphics[width=\linewidth]{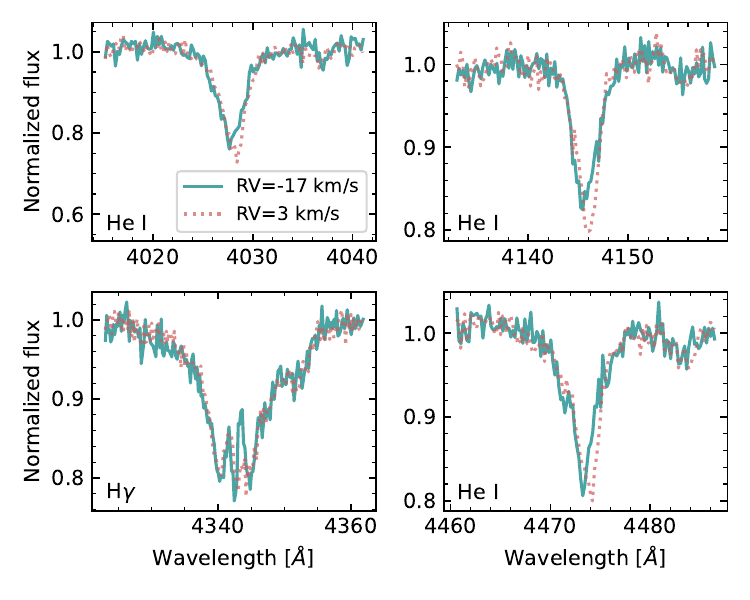}
    \newsubcap{RV extremes of BLOeM\,2-070.}
    \label{fig:rvextr_2-070}
    \end{subfigure}
\end{figure*}

\paragraph{BLOeM 2-070:} This star was classified as B1\,II\,e in \citetalias{Shenar2024}. The spectrum, in particular the \ion{He}{i} lines of the star, show variability (Fig.\,\ref{fig:rvextr_2-070}). The RVs measured from those indicate, that the target is a binary, with a maximum amplitude of 20\,\kms\ (Fig.\,\ref{fig:rvcurve_2-070}), thus just at the imposed threshold. The changes in the shape of the lines, which occur mostly in the line core, while the overall line seems stable, however, might indicate the variability observed here stems from line-profile variability (LPV) typical for pulsating OBe stars. We thus classify BLOeM\,2-070 as candidate binary.

\begin{figure*}
    \centering
    \begin{subfigure}{.5\textwidth} \centering
    \includegraphics[width=\linewidth]{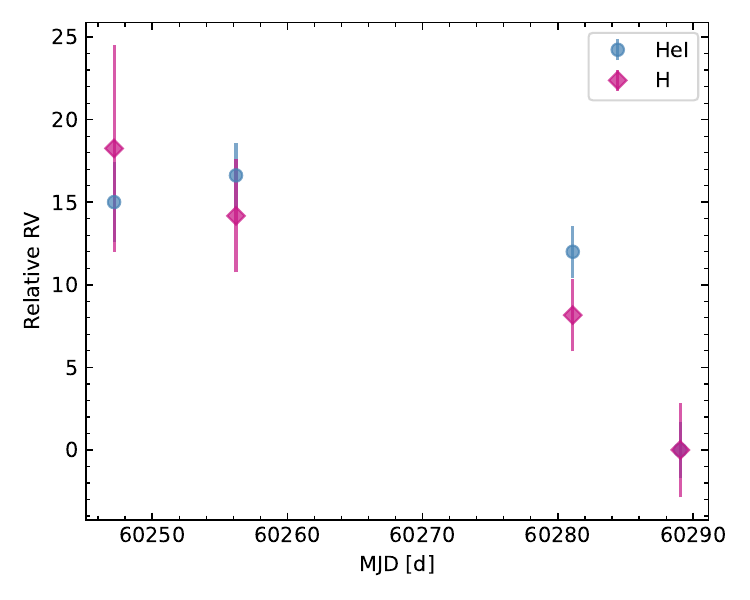}
    \newsubcap{RV curve of BLOeM\,2-071.}
    \label{fig:rvcurve_2-071}
    \end{subfigure}%
    \begin{subfigure}{.5\textwidth} \centering
    \includegraphics[width=\linewidth]{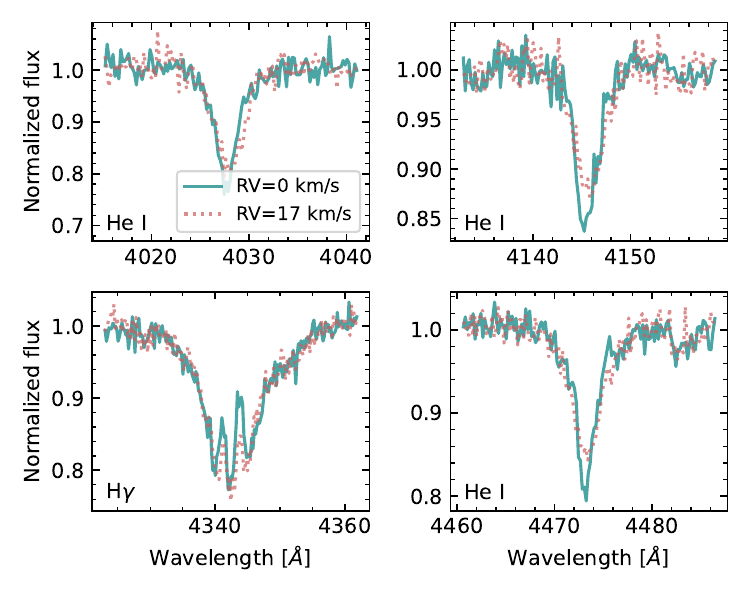}
    \newsubcap{RV extremes of BLOeM\,2-071.}
    \label{fig:rvextr_2-071}
    \end{subfigure}
\end{figure*}

\paragraph{BLOeM 2-071:} Only four epochs are available for this object. It is similar to the star discussed before, with a spectral type of B1.5\,II:\,e (\citetalias{Shenar2024}), and the \ion{He}{i} lines show variability that could either be indicative of LPVs or binarity (Fig.\,\ref{fig:rvextr_2-071}). Given that the \ion{He}{i} RVs as well as the H$\gamma$ RVs lines show the same trend, but the maximum RV amplitude does not exceed the 20\,\kms\ threshold (Fig.\,\ref{fig:rvcurve_2-071}), we classify BLOeM\,2-071 as candidate binary.

\begin{figure*}
    \centering
    \begin{subfigure}{.5\textwidth} \centering
    \includegraphics[width=\linewidth]{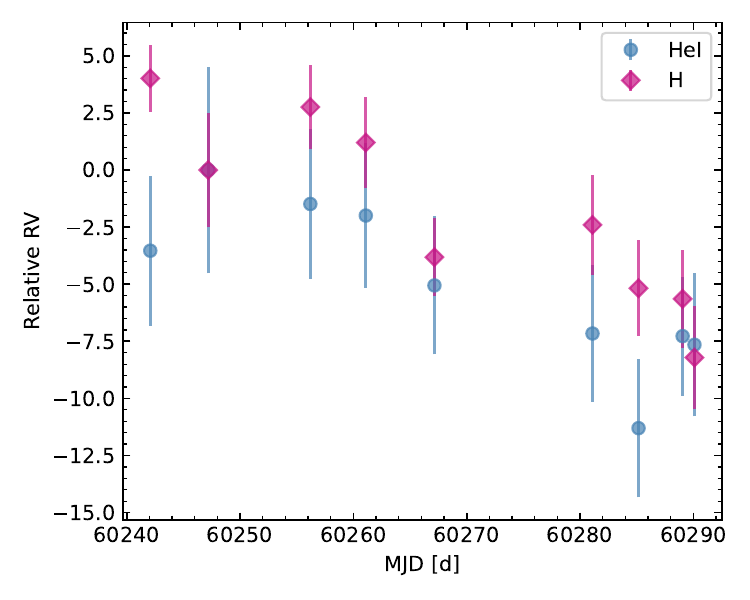}
    \newsubcap{RV curve of BLOeM\,2-097.}
    \label{fig:rvcurve_2-097}
    \end{subfigure}%
    \begin{subfigure}{.5\textwidth} \centering
    \includegraphics[width=\linewidth]{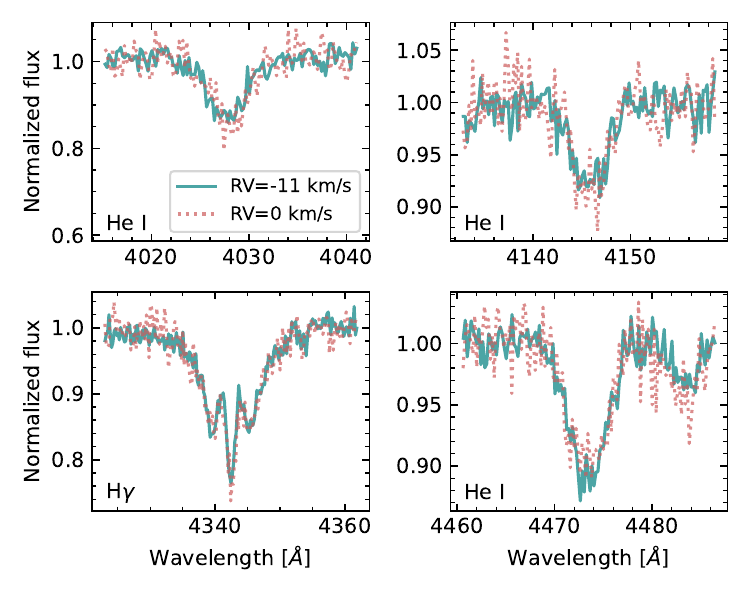}
    \newsubcap{RV extremes of BLOeM\,2-097.}
    \label{fig:rvextr_2-097}
    \end{subfigure}
\end{figure*}

\paragraph{BLOeM 2-097:} This star is the target with the latest spectral type in the sample, it was classified as B5\,II\,e. While the maximum RV amplitude $\Delta$RV$_\mathrm{max}$ measured from the \ion{He}{i} lines is only 11\,\kms\ (Fig.\,\ref{fig:rvextr_2-097}) and thus below the imposed RV threshold, the RV curves measured both from \ion{He}{i} lines and H$\gamma$ show a clear downwards trend over the covered observing period (see Fig.\,\ref{fig:rvcurve_2-097}). We thus classify the object as candidate binary, which might be confirmed by subsequent epochs.

\begin{figure*}
    \centering
    \begin{subfigure}{.5\textwidth} \centering
    \includegraphics[width=\linewidth]{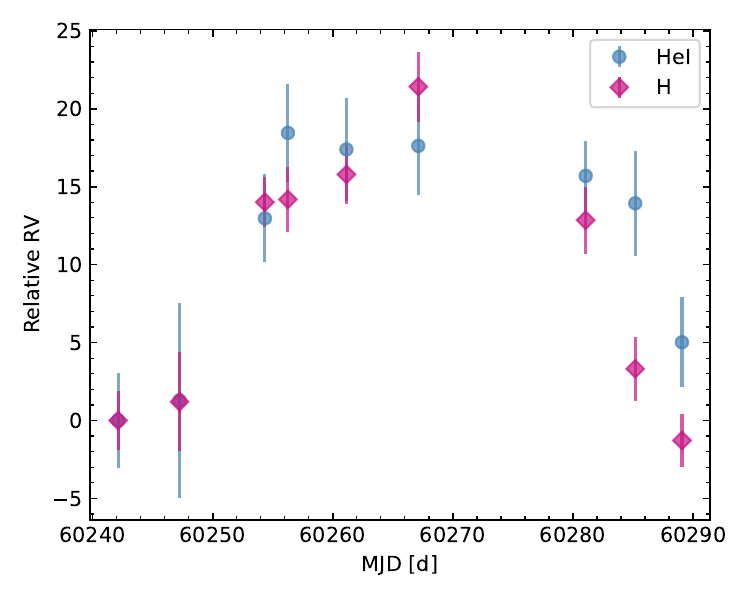}
    \newsubcap{RV curve of BLOeM\,4-113.}
    \label{fig:rvcurve_4-113}
    \end{subfigure}%
    \begin{subfigure}{.5\textwidth} \centering
    \includegraphics[width=\linewidth]{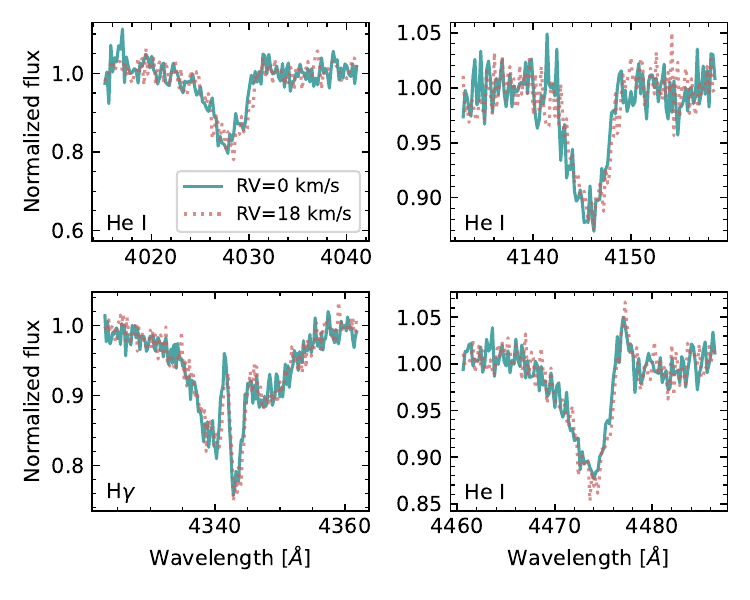}
    \newsubcap{RV extremes of BLOeM\,4-113.}
    \label{fig:rvextr_4-113}
    \end{subfigure}
\end{figure*}

\paragraph{BLOeM 4-113:} This B2.5\,II\,pe star, which shows double-peaked emission in H$\gamma$ (see Fig.\,\ref{fig:rvextr_4-113}), would by eye be classified as a binary with a period of at least 100 days (at least twice the observing span, see Fig.\,\ref{fig:rvcurve_4-113}). However, the maximum RV amplitude $\Delta$RV$_\mathrm{max}$=20\,\kms\ is just below the imposed threshold of 20\,\kms and thus the star is formally not listed as binary here. It will, however, most likely fulfill the second criterion when additional epochs are available, and is thus listed as binary candidate here. It seems to further coincide with an X-ray source \citep[see e.g.,][]{cscv2, Sturm2013}, which would make it a BeXRB.

\begin{figure*}
    \centering
    \begin{subfigure}{.5\textwidth} \centering
    \includegraphics[width=\linewidth]{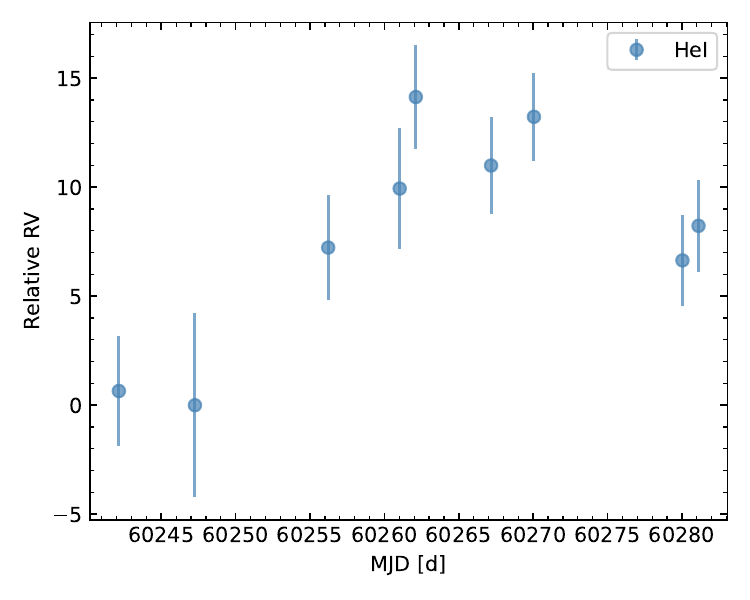}
    \newsubcap{RV curve of BLOeM\,5-035.}
    \label{fig:rvcurve_5-035}
    \end{subfigure}%
    \begin{subfigure}{.5\textwidth} \centering
    \includegraphics[width=\linewidth]{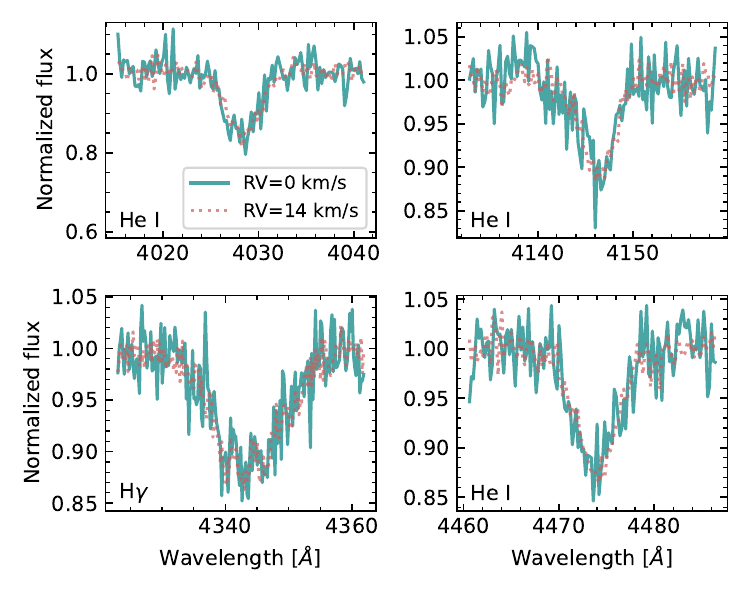}
    \newsubcap{RV extremes of BLOeM\,5-035.}
    \label{fig:rvextr_5-035}
    \end{subfigure}
\end{figure*}

\paragraph{BLOeM 5-035:} this star has a spectral type of B1.5\,II:\,e. The emission lines are quite weak and, combined with the low S/N not suitable to estimate RVs. The \ion{He}{i} RV curve of BLOeM\,5-035 shows an upwards trend over time that can be interpreted as potential binary motion (Fig.\,\ref{fig:rvcurve_5-035}). However, the measured maximum RV amplitude is only 14\,\kms\ (see Fig.\,\ref{fig:rvextr_5-035}), and the system therefore does not meet both binary criteria. Because of the trend, we however classify it as candidate binary.

\begin{figure*}
    \centering
    \begin{subfigure}{.5\textwidth} \centering
    \includegraphics[width=\linewidth]{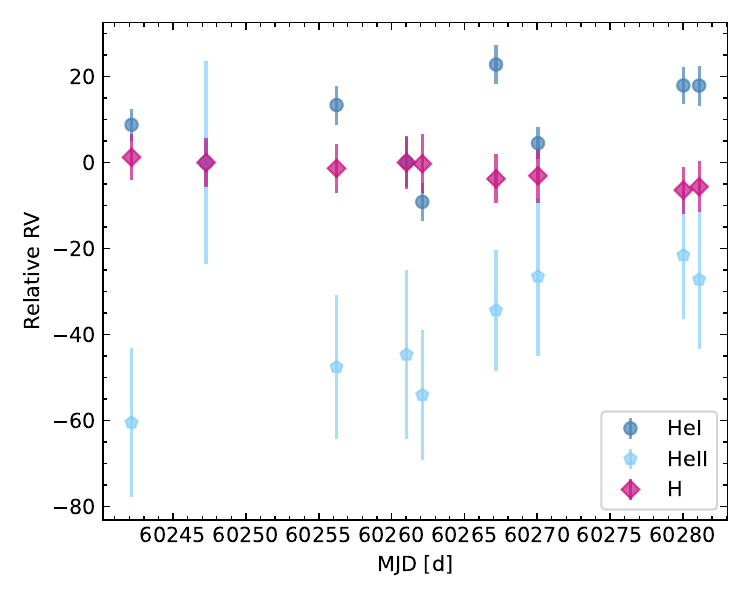}
    \newsubcap{RV curve of BLOeM\,5-115.}
    \label{fig:rvcurve_5-115}
    \end{subfigure}%
    \begin{subfigure}{.5\textwidth} \centering
    \includegraphics[width=\linewidth]{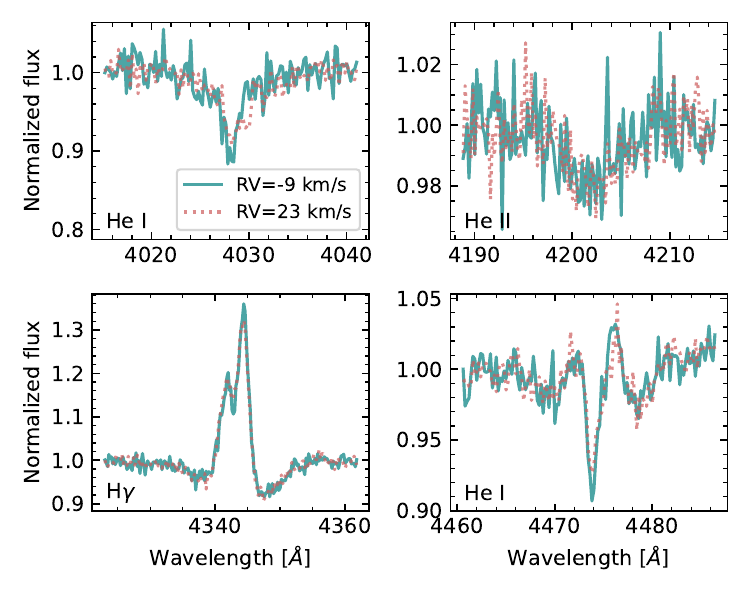}
    \newsubcap{RV extremes of BLOeM\,5-115.}
    \label{fig:rvextr_5-115}
    \end{subfigure}
\end{figure*}

\paragraph{BLOeM 5-115:} This target was also classified as Oe star in \citetalias{Shenar2024}, in particular as O9.5\,V:\,pe star. The star is formally classified as binary as the \ion{He}{i} RVs fulfill the RV criteria. The \ion{He}{ii} RVs however do not agree with them, and have large error bars (see Fig.\,\ref{fig:rvcurve_5-115}). The strong emission in Balmer lines seems to be stable. Given the binary nature of this object is not convincing, we classify it as candidate binary and await further observations to constrain the nature better.

\begin{figure*}
    \centering
    \begin{subfigure}{.5\textwidth} \centering
    \includegraphics[width=\linewidth]{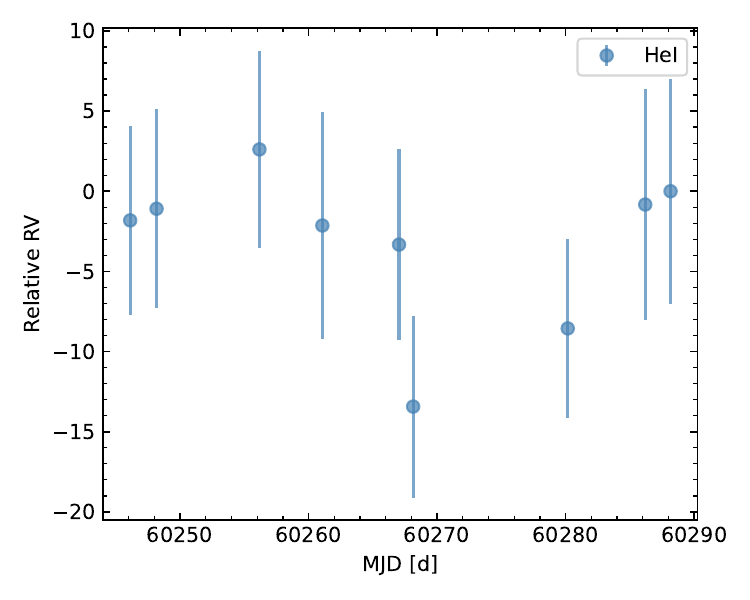}
    \newsubcap{RV curve of BLOeM\,7-051.}
    \label{fig:rvcurve_7-051}
    \end{subfigure}%
    \begin{subfigure}{.5\textwidth} \centering
    \includegraphics[width=\linewidth]{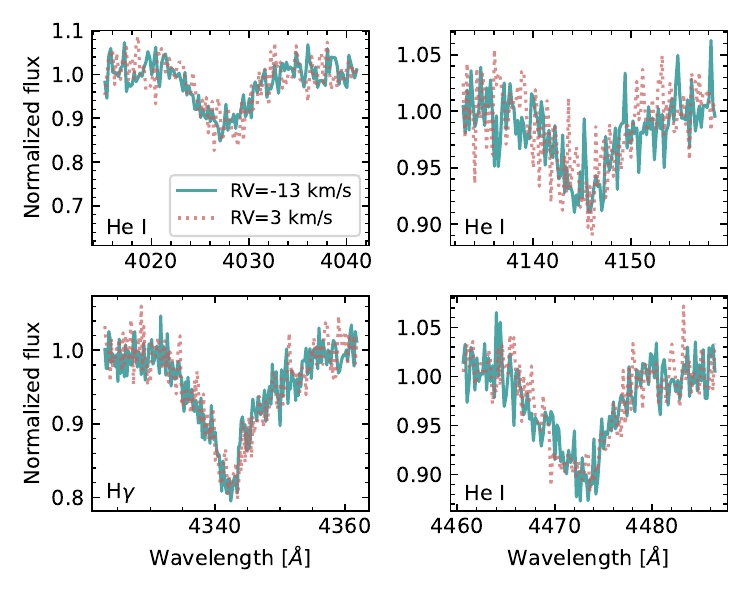}
    \newsubcap{RV extremes of BLOeM\,7-051.}
    \label{fig:rvextr_7-051}
    \end{subfigure}
\end{figure*}

\paragraph{BLOeM 7-051:} This star is also a rapidly rotating B-type star with a spectral type B1\,II:\,e (\citetalias{Shenar2024}). It shows potential weak emission infilling in the wings of H$\gamma$, too weak to measure RVs. Given the broad lines and the relatively low S/N, the RVs measured from \ion{He}{i} do not fulfill the binary criteria. The RV curve (Fig.\,\ref{fig:rvcurve_7-051}), however appears to show a sinusoidal trend. The spectra at the measured RV extremes are shown in Fig.\,\ref{fig:rvextr_7-051}.

\begin{figure*}
    \centering
    \begin{subfigure}{.5\textwidth} \centering
    \includegraphics[width=\linewidth]{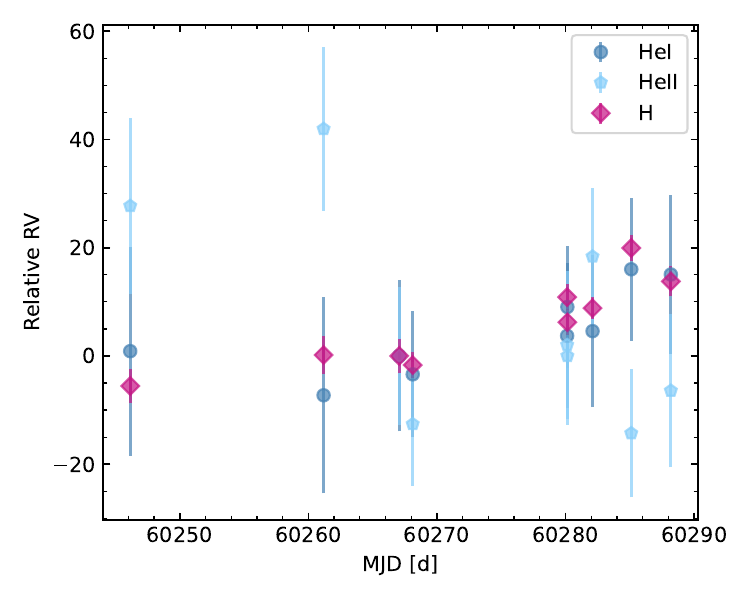}
    \newsubcap{RV curve of BLOeM\,8-021.}
    \label{fig:rvcurve_8-021}
    \end{subfigure}%
    \begin{subfigure}{.5\textwidth} \centering
    \includegraphics[width=\linewidth]{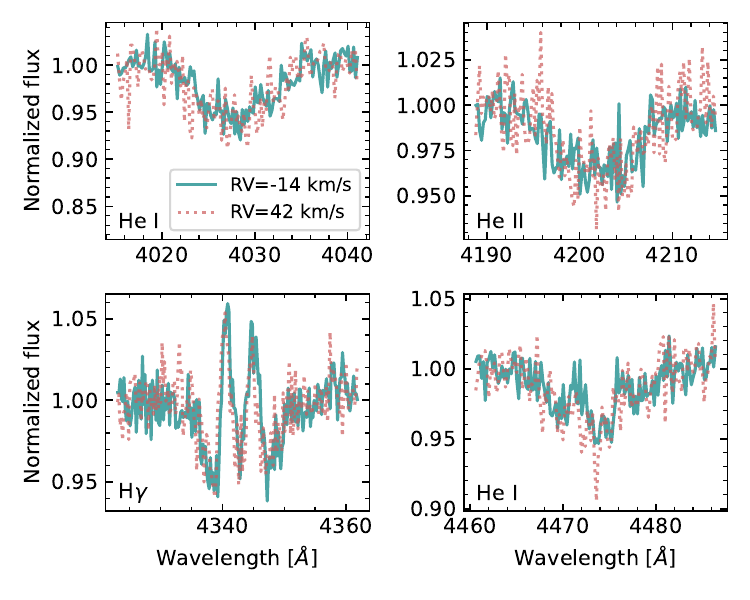}
    \newsubcap{RV extremes of BLOeM\,8-021.}
    \label{fig:rvextr_8-021}
    \end{subfigure}
\end{figure*}

\paragraph{BLOeM 8-021:} This target is an Oe star with spectral type O7\,V-IIInnn,pe (\citetalias{Shenar2024}). The spectrum of this object is very noisy and the spectral lines are broad, thus the RVs measured from absorption have large error bars (of the order of 10\,\kms). Additionally, some \ion{He}{i} lines are impacted by emission and cannot be used to estimate RVs. As the RVs measured from \ion{He}{ii} lines formally fulfill both binarity criteria, we classify this target as candidate binary, however note that no similar signature is detected in \ion{He}{i} or Balmer emission lines. The RV curve and the spectra at RV extremes for BLOeM\,8-082 are shown in Figs.\,\ref{fig:rvcurve_8-021} and \ref{fig:rvextr_8-021}.

\begin{figure*}
    \centering
    \begin{subfigure}{.5\textwidth} \centering
    \includegraphics[width=\linewidth]{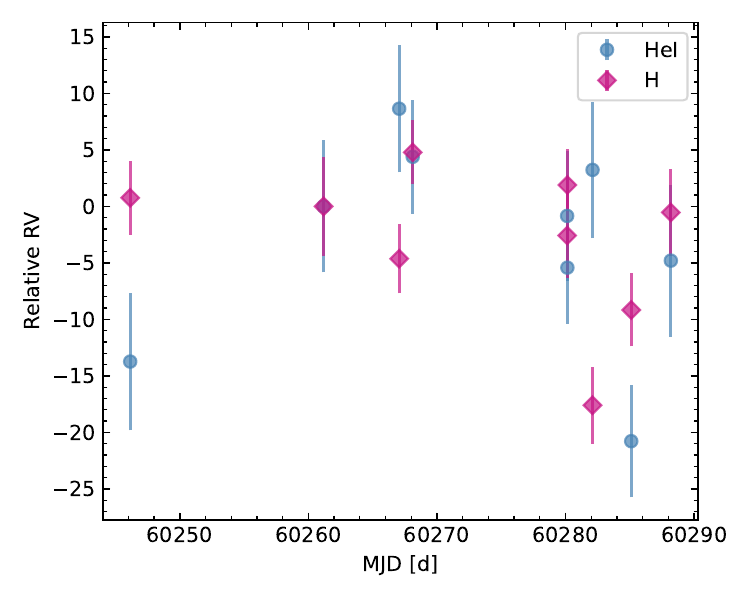}
    \newsubcap{RV curve of BLOeM\,8-059.}
    \label{fig:rvcurve_8-059}
    \end{subfigure}%
    \begin{subfigure}{.5\textwidth} \centering
    \includegraphics[width=\linewidth]{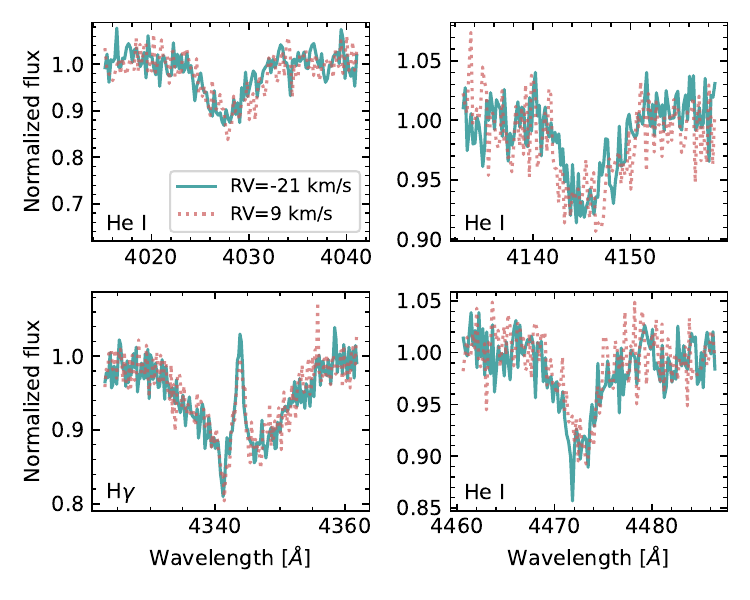}
    \newsubcap{RV extremes of BLOeM\,8-059.}
    \label{fig:rvextr_8-059}
    \end{subfigure}
\end{figure*}

\paragraph{BLOeM 8-059:} Based on the two binary criteria imposed in this work, BLOeM\,8-059 is classified as a binary from the RV measurements of the \ion{He}{i} lines (see Figs.\, \ref{fig:rvcurve_8-059} and \ref{fig:rvextr_8-059}). However, given that the S/N ratio in the spectra is low, and the measured RVs would indicate a very short period, we conservatively classify this object as binary candidate.

%{\bf \textcolor{red}{BLOeM\,5-031}}: Paul classified it as B2.5: II:e	+ early AIa in paper I. I don't see any indications for binarity though, to me it looks more like a shell star (though the lines are very deep). For now I did not include it in any statistics What do we do with it? 

\end{appendix}
\end{document}